\documentclass[11pt]{article}
\usepackage{jheppub}
\usepackage{float, extarrows, tikz-cd}
\usepackage{graphicx}
\usepackage{orcidlink}
\usepackage{slashed}
\usepackage{mathrsfs}
\usepackage{tabularx,ragged2e}
\usepackage{amssymb}
\usepackage{subcaption}
\usepackage{tikz}
\usepackage{amsmath,amssymb}
\usepackage{caption}
\usepackage{braket}
\usepackage{xcolor}
\usepackage{dsfont}
\usepackage{verbatim}
\usepackage{mathtools, xcolor,ytableau, amsfonts,tikz}
\usepackage{graphicx}

\usepackage{physics}
\newcolumntype{C}{>{\Centering\arraybackslash}X}

\numberwithin{equation}{section}

\newcommand{\Qn}{\mathcal{Q}(\mathbf{n})}
\newcommand{\Qno}{\mathcal{Q}(\mathbf{n}_1)}
\newcommand{\Qnt}{\mathcal{Q}(\mathbf{n}_2)}
\newcommand{\En}{\mathcal{E}(\mathbf{n})}
\newcommand{\Eno}{\mathcal{E}(\mathbf{n}_1)}
\newcommand{\Ent}{\mathcal{E}(\mathbf{n}_2)}
\newcommand{\mO}{\mathcal{O}}
\newcommand{\mD}{\mathcal{D}}
\newcommand{\pd}{\partial}

\newcommand{\xn}{x_{\mathbf{n}}}

\newcommand{\On}{\mathcal{O}_Q}
\newcommand{\Onb}{\bar{\mathcal{O}}_Q}
\newcommand{\cyl}{\text{cyl}}

\newcommand{\D}[1]{\mathcal{D}(\mathbf{n}_{#1})}

\author[a,b]{Gabriel Cuomo~\orcidlink{0000-0003-4668-9010},}

\author[c]{Eren Firat~\orcidlink{0009-0008-6710-6973},}

\author[c]{Filippo Nardi~\orcidlink{0009-0003-6821-278X},}

\author[d]{Lorenzo Ricci~\orcidlink{0000-0001-8704-3545}\,}

\affiliation[a]{Center for Cosmology and Particle Physics, Department of Physics, New York University, New York, NY 10003, USA}
\affiliation[b]{Department of Physics, Princeton University, Princeton, NJ  08544, USA}
\affiliation[c]{Theoretical Particle Physics Laboratory (LPTP), Institute of Physics, EPFL, Lausanne, Switzerland}
\affiliation[d]{Maryland Center for Fundamental Physics, Department of Physics, University of Maryland, College Park, MD 20742, USA}

\emailAdd{gc6696@princeton.edu}
\emailAdd{eren.firat@epfl.ch}  
\emailAdd{filippo.nardi@epfl.ch}
\emailAdd{lricci@umd.edu}

\title{Conformal Collider Physics at Large Charge}

\abstract{We study energy correlators and other event shapes in states created by operators with large global $U(1)$ charge $Q$ in Conformal Field Theories. Focusing on theories whose large charge sector is described by the superfluid Effective Field Theory (EFT), we develop a systematic framework to compute event shapes within the EFT. As formerly observed, event shapes at leading order in $1/Q$ factorize into a product of classical expectation values determined by symmetry. In contrast, the subleading contribution to energy-energy and charge-charge correlators is a nontrivial prediction of the EFT, which we compute explicitly. Our results reveal a sharp collinear enhancement of the correlation between detectors, induced by the propagation of sound. We also generalize our findings to a broad class of event shapes. }

\begin{document}

\maketitle

\newpage

\section{Introduction and summary}\label{Sec:Intro}

In Quantum Field Theory (QFT), a special class of operators known as detectors consists of integrated operators that measure fluxes in collider experiments. For instance, energy detectors are defined by
\begin{equation}\label{eq_Energy_detectors}
\En = \lim_{r \rightarrow+\infty}r^{d-2}\int_{0}^{+\infty} d t\, \mathbf{n}_i T^{0i} (t, r\mathbf{n}) \,,
\end{equation}
where $T^{\mu\nu}$ is the energy-momentum tensor. Energy correlators, expectation values of the detectors~\eqref{eq_Energy_detectors} in a momentum eigenstate, can be used to characterize the dynamics of the state produced in a collider event. More information is obtained by studying correlation functions of different detectors, generally called ``event shapes", including those associated with conserved charges.

Event shapes are inclusive yet remarkably rich observables, making them a powerful tool for reconstructing the dynamics of quantum states across different energy scales. In particular, as noted long ago~\cite{PhysRevD.17.2298,PhysRevLett.41.1585,PhysRevD.19.2018,LOUISBASHAM1979297}, energy correlators (ECs) enable a precise comparison between theoretical predictions and experimental data in collider experiments. This is because ECs are infrared safe and thus largely insensitive to the details of hadronization in processes at energies far above the QCD scale~\cite{Chen:2024nyc}. More recently, ECs have been rediscovered as promising observables at the LHC~\cite{Chen:2020vvp}, opening the door to numerous applications. These include the most precise determination to date of the strong coupling constant~\cite{CMS:2024mlf} through jet substructure, as well as a variety of probes of the Standard Model (see e.g.~\cite{Komiske:2022enw,Holguin:2022epo,Riembau:2024tom,Lee:2022uwt,Craft:2022kdo,Ricci:2022htc}).

Detectors also play a key role in many recent theoretical developments in Conformal Field Theory (CFT), where they are a special instance of the so-called light-ray operators. Most notably, light-ray operators naturally appear in the inversion formula as the objects that admit a natural analytic continuation in spin~\cite{Caron-Huot:2017vep,Kravchuk:2018htv}. They also play an important role in the average null-energy condition~\cite{Hartman:2016lgu} and several information-theoretic quantities \cite{Faulkner:2016mzt,Casini:2017roe,Ceyhan:2018zfg}. The discovery of these surprising applications of detectors has given rise to an extensive research program in CFT (see e.g.~\cite{Kologlu:2019mfz,Chang:2020qpj, Belin:2020lsr,Korchemsky:2021htm,Caron-Huot:2022eqs}), starting with the seminal work~\cite{Hofman:2008ar}, with implications extending to quantum gravity~\cite{Kologlu:2019bco,Chen:2024iuv}, celestial holography \cite{Cordova:2018ygx,Hu:2022txx} and particle physics.

Despite the recent progress, it is fair to say that event shapes are best understood for states that are \emph{close to the vacuum}, e.g. few particle states, for which there exist many available precise results (see e.g.~\cite{Belitsky:2013bja,Dixon:2018qgp,Dixon:2019uzg,Herrmann:2024yai}). It is interesting to ask what is the structure of event shapes in states that cannot be thought of as a small perturbation of the relativistic vacuum, such as those at finite density or at finite temperature,\footnote{See~\cite{Delacretaz:2018cfk} for an early study of the average null-energy condition, i.e. positivity of ECs, in thermal states.} where Poincar\'e invariance is not manifest. This is not just an academic question, but is also motivated by the possibility of using ECs to probe the dynamics of the quark-gluon plasma that is formed in heavy-ion collisions, see for instance \cite{Andres:2022ovj,Andres:2024ksi,Bossi:2024qho} and references therein. Additionally, as we will review in detail below, finite density or finite temperature states often admit universal \emph{hydrodynamic} effective field theory (EFT) descriptions. These EFTs therefore present us with a unique opportunity to study event shapes directly in strongly-coupled models, without relying on the existence of a weakly coupled or holographic limit.\footnote{Some non-perturbative results related to light-ray operators have been obtained in $\mathcal{N}=4$ SYM via integrability, see \cite{Alfimov:2014bwa,Homrich:2022cfq,Homrich:2024nwc} and ref.s therein.}

In this work, we make a step forward in the understanding of ECs and other event shapes in far-from-the-vacuum states by providing a comprehensive treatment of detectors in what is arguably the simplest nontrivial example of this sort: large charge states in CFT. More in detail, we focus on interacting $d>2$ dimensional CFTs that are invariant under an internal $U(1)$ symmetry. Let us denote by $\mO_Q$ the lowest-dimensional operator with charge $Q$ under the internal group and with $p^\mu$ the four-momentum of the state. We consider the following energy-energy correlator (EEC) for $Q\gg 1$
\begin{equation}\label{eq_EEC_intro}
    \langle\mathcal{E}(\mathbf{n}_1)\mathcal{E}(\mathbf{n}_2)\rangle_p\equiv\frac{\langle \bar{\mO}_Q(p)\mathcal{E}(\mathbf{n}_1)\mathcal{E}(\mathbf{n}_2)\mO_Q(p)\rangle}{\langle \bar{\mO}_Q(p)\mO_Q(p)\rangle}\,,
\end{equation} 
as well as other similar event shapes with two (or more) insertions of charge and other detectors. Physically, in the rest frame $p^\mu=(E,\mathbf{0})$, the state $\mO_Q(p)|0\rangle$ describes a distribution of charge $Q\gg 1$ initially localized within a large-radius ball at $x^0=0$. The charge then expands outward at the speed of light in concentric shells until it eventually reaches the detectors.

The main tool that we employ in our analysis is the conformal superfluid EFT~\cite{Hellerman:2015nra,Monin:2016jmo}, which describes correlation functions of large charge operators in a vast class of CFTs including the $O(2)$ model in $d=3$~\cite{Banerjee:2017fcx,Cuomo:2023mxg}. As we review in Sec.~\ref{Sec:LChargeRev}, the EFT consists of a single Goldstone degree of freedom that can be identified with the superfluid phonon. The EFT allows us to semiclassically compute correlation functions of large charge operators and other local operators in a systematic expansion in $1/Q$ for generic kinematics.

Unfortunately, the calculation of event shapes within EFT is not immediately straightforward. Indeed, as we will explain in detail in the main text, the Fourier transform and the integral~\eqref{eq_Energy_detectors} necessarily span regions in which the EFT breaks down. Additionally, as we review in Sec.~\ref{Sec:LChargeRev}, the EFT naturally describes correlation functions in Euclidean space or with the large charge operators inserted at \emph{imaginary} Minkowski time. Thus, a priori, we must perform a nontrivial analytic continuation of the EFT correlators to obtain Lorentzian observables. 

Despite these complications, event shapes of the sort we are interested in were previously studied in~\cite{Firat:2023lbp}. It was argued there that the semiclassical nature of the superfluid EFT implies that event shapes factorize into the product of one-point functions, e.g. for~\eqref{eq_EEC_intro}
\begin{equation}\label{eq_EEC_factorization_intro}
    \langle \Eno \Ent\rangle_p \xrightarrow[\phantom{Q \to \infty}]{Q \to \infty} \langle \Eno \rangle_p \langle \Ent \rangle_p\,,
\end{equation}
where the expectation value of a single detector is fixed by symmetries,
\begin{equation}
\langle \En \rangle_p =  \frac{(p^2)^{\frac{d}{2}} \Theta(p^2) \Theta(p^0)}{\Omega_{d-2}(n\cdot p )^{d-1}}\,.
\end{equation}
Here $\Theta$ is the Heaviside step function. In~\cite{Chicherin:2023gxt} similar results were argued to hold for generic heavy primary operators, that admit a semiclassical description as finite temperature states~\cite{Lashkari:2016vgj,Delacretaz:2020nit}. However, due to the complications mentioned above, the corrections to the factorized result~\eqref{eq_EEC_factorization_intro} were not determined.

In this work, we go beyond the leading factorized term and show that event shapes---such as the EEC---can be systematically computed in a large charge expansion in \(1/Q\), 
\begin{align}\label{Eq:CorrEnEn}
    \langle \Eno \Ent \rangle_p = \langle \Eno \rangle_p \langle \Ent \rangle_p+
    \frac{1}{\Delta_Q}\langle \Eno \Ent \rangle^{(1)}_p + \ldots\,,
\end{align}
where $\Delta_Q\sim Q^{\frac{d}{d-1}}$ denotes the scaling dimension of the large charge operator $\mO_Q$.
We compute the first connected contribution $\langle \Eno \Ent \rangle^{(1)}_p$ to EEC~\eqref{eq_EEC_intro}, the charge-charge correlator (CCC), which is analogously defined, and more general two-point event shapes. The connected contribution is not fixed by conformal symmetry and encodes non-trivial information about the state and the theory. 

Our results for the EEC and the CCC are presented in Fig.~\ref{fig:QQandEECorr}.
Notably, we find that the propagation of sound leads to an enhancement of the correlation between detectors in the collinear limit, a phenomenon that we term \emph{sound-jet} in analogy with the usual QCD jets.

To carry out this calculation, we develop a systematic framework for applying EFT to event shapes. This allows us to systematically power-count and parametrize contributions from the integration regions in which EFT breaks down in light-ray integrals. We find that such contributions are subleading to both the leading factorized term and the next-to-leading connected contribution at generic kinematics for a broad class of event shapes, including those associated with energy and charge detectors. In particular, the EFT remains valid provided the angular distance between the detectors on the celestial sphere is much larger than $1/\mu$, where $\mu\sim Q^{\frac{1}{d-1}}$ is the EFT cutoff, in the rest frame of the large charge state $p^{\mu} = (E,\mathbf{0})$.

Our results pave the way for future research in several directions. These include the study of event shapes at finite temperature or in nontrivial holographic states like black holes. Moreover, some of the techniques and ideas developed in this work have the potential for wider applicability, for instance our analysis of the light-ray integral region where EFT breaks down. We provide a detailed outlook of future directions in Sec.~\ref{Sec:Conc}. Below, we summarize our results and their derivation in greater detail.

\subsection{Summary of the paper}\label{Sec:Summ}

As mentioned above, our analysis assumes that the conformal superfluid EFT describes correlation functions of large charge operators in the CFT at hand. We review in detail this EFT and its predictions in Sec.~\ref{Sec:LChargeRev}. In short, by the state-operator correspondence, the operator $\mO_Q$ with lowest scaling dimension at fixed $U(1)$ charge $Q$ corresponds to a state with homogeneous charge density on $\mathbb{R}\times S^{d-1}$. For $Q\gg 1$, this is equivalent to the ground state at large chemical potential $\mu\sim  Q^{\frac{1}{d-1}} $. A large chemical potential generically induces a gap for all \emph{radial} modes, and at low energies, we are left with a simplified EFT description in terms of a few Goldstone degrees of freedom that account for the nonlinear realization of the spacetime and internal symmetries~\cite{Nicolis:2015sra}. In~\cite{Hellerman:2015nra,Monin:2016jmo} it was argued that the simplest and most generic option for interacting systems is that large charge operators correspond to superfluid states, that break spontaneously the $U(1)$ symmetry, and admit a low energy description in terms of a single Goldstone mode. The EFT predicts the scaling dimension $\Delta_Q\sim Q^{\frac{d}{d-1}}\gg 1$ of $\mO_Q$ and the CFT spectrum around it in a systematic expansion in $1/Q$.

The Weyl invariance of the CFT ensures that expectation values of operators in the large charge state on the cylinder admit a natural map to correlation functions in flat space. 
The map is natural in Euclidean signature, but somewhat subtle in Minkowski spacetime. In particular, expectation values in the large charge state on the Lorentzian cylinder map to Minkowski space correlators with the large charge operators inserted at \emph{imaginary} time. Technically, this is because Minkowski space admits a natural foliation in terms of the North/South quantization Hamiltonian, for which states are created by operators inserted at Euclidean time~\cite{Rychkov:2016iqz}. Physically, the imaginary time component serves to project onto a primary state, suppressing the contribution of the descendants.

Therefore, in principle, to compute EEC~\eqref{eq_EEC_intro} we need to: (i) analytically continue to Minkowski the EFT result for the four-point function with two insertions of the energy-momentum tensor to the Minkowski plane, (ii) integrate over the light-rays~\eqref{eq_Energy_detectors}, and (iii) perform the Fourier transform. However, as discussed in~\cite{Firat:2023lbp}, this procedure is technically challenging, as the Fourier transform and the light-ray integrals span several inequivalent time-orderings, and the required analytic continuations obscure the simplicity of the EFT results. In Sec.~\ref{Sec:Detectors} we argue that a more convenient strategy is to perform the Fourier transform first via a saddle-point approximation
\begin{multline}\label{Eq:ResFT}
    \int d^4x_{fi} e^{i p \cdot x_{fi}} \langle \Onb \left(x_f\right) \Eno \Ent \On\left(x_i\right) \rangle_p
     \\
    \xrightarrow[\text{point}]{\text{Saddle}} \langle \Onb \left(-i\frac{\Delta_Q p^{\mu}}{p^2}\right) \Eno \Ent \On \left(i\frac{\Delta_Q p^{\mu}}{p^2}\right) \rangle\,,
\end{multline}
where we used that $\Delta_Q\sim Q^{\frac{d}{d-1}}\gg 1$ to identify the leading exponential dependence of the correlation function with that of the two-point function $\langle \Onb \left(x_f\right)\On \left(x_i\right) \rangle$. In particular, in the rest frame $p^\mu=(E,\mathbf{0})$, the large charge operators are localized at imaginary time and thus create primary states in North-South quantization. This ensures that the correlator~\eqref{Eq:ResFT} is equivalent to an expectation value on the Lorentzian cylinder, which is precisely the natural outcome of the EFT. Therefore, using~\eqref{Eq:ResFT}, we bypass the need for analytic continuation and significantly streamline the calculation. In the main text, we also discuss how to account for corrections to the saddle-point~\eqref{Eq:ResFT} systematically. We stress that $\Delta_Q\gg 1$ directly follows from $Q \gg 1$, and hence the saddle-point expansion is not an additional approximation.

Using~\eqref{Eq:ResFT} and the Weyl invariance of the theory, we recast the ECs and other event shapes as expectation values of light-ray operators in large charge states on the Lorentzian cylinder. The relation between detectors in Minkowski space and on the Lorentzian cylinder is pictorially illustrated in Fig.~\ref{fig:LightrayIntegral}. On the cylinder, the two light-ray integrals~\eqref{eq_EEC_intro} span finite contours and are everywhere separated but at the integration endpoints, where the two operators collide and the EFT breaks down. We nonetheless estimate that the contribution from the dangerous region is very subleading as long as the angular distance between the detectors on the celestial sphere satisfies $\theta\gg 1/\mu$. For instance, in $d=3$ we find that the non-EFT contribution to EEC~\eqref{eq_EEC_intro} is suppressed at least by a factor $\sim Q^{-4}$ compared to the leading order result and is hence much smaller than the order of interest in this work.

With these preliminaries in order, it is straightforward to argue that event-shapes admit an expansion of the form~\eqref{Eq:CorrEnEn}, where the leading disconnected term is a simple consequence of the semiclassical nature of the EFT. In Sec.~\ref{Sec:EEC} we explicitly compute the first correction to EEC ($\langle \Eno \Ent \rangle^{(1)}$ in \eqref{Eq:CorrEnEn}), and CCC in the rest-frame $p^\mu=(E,\mathbf{0})$. 

The calculation of EEC and CCC at next-to-leading order is technically involved and can only be done numerically for arbitrary values of $\cos\theta = \mathbf{n}_1 \cdot \mathbf{n}_2$. We nonetheless obtained semi-analytic results in the back-to-back limit $\theta \simeq \pi$ and, most interestingly, in the collinear limit $\theta \ll 1$. In the latter the two light-ray integrals lie close to each other on the cylinder and we may resort to a flat-space approximation for the phonon propagator. Explicitly, we find for EEC in the collinear regime:
\begin{align}
\langle\Eno \Ent \rangle^{(1)} \stackrel{\theta\neq 0}{=}- \left( \frac{E}{\Omega_{d-2}}\right)^2\left[
   \frac{\sqrt{d-2} \, \Gamma \left(\frac{d}{2}+1\right)^2}{\Gamma \left(\frac{d+1}{2}\right) \Gamma \left(\frac{d+3}{2}\right)}\theta^{1-d}
   +O\left(\text{max}(1,\theta^{3-d})\right)
   \right]\,,
   \label{Eq:ShortDistanceEE_anyd_intro}
\end{align}
where we remark that we still need $\theta\gg 1/\mu$ to ensure the validity of EFT, and thus we are not sensitive to the light-ray OPE regime~\cite{Kologlu:2019mfz,Chang:2020qpj}. Remarkably,~\eqref{Eq:ShortDistanceEE_anyd_intro} predicts the formation of \textit{sound-jets} as regions of localized energy and charge at small $\theta$. In Sec.~\ref{Sec:CollLimit} we additionally discuss the compatibility of our results with Ward identities, and argue that this implies the existence of distributional contributions to the EEC localized at $\theta=0$, whose form we compute explicitly in $d=3$.

We also obtained numerical results for EEC and CCC at arbitrary values of $\theta$ in $d=3$. To this end, we found it convenient to decompose the four-point function using an unconventional form of the Goldstone propagator, formally equivalent to quantizing the EFT on $\mathbb{R}\times dS_{d-1}$. This approach is analogous to the standard procedure of selecting poles at complex momenta to compute the scattering phase shift as a function of the impact parameter $\vec{b}$ \cite{Camanho:2014apa}. Our results are shown in Fig.~\ref{fig:QQandEECorr}, where we also test the numerical predictions against the aforementioned collinear and back-to-back results, finding excellent agreement.

In Sec.~\ref{Sec:Generalized}, we extend our analysis to detectors defined by integrating generic, non necessarily conserved, local operators along light-rays at null infinity. First, we analyze detectors that are not charged under the $U(1)$ symmetry. We find that the corresponding event shapes admit an expansion analogous to~\eqref{Eq:CorrEnEn}, with the exception of detectors constructed from sufficiently light scalar operators, for which the contribution from the region in which EFT breaks down is enhanced. Interestingly, we find that EFT allows for systematic predictions in all cases, provided we parametrize the non-EFT contributions via new local insertions in the correlation function, whose form we discuss in Sec.~\ref{sec_EFT_singular} using conformal symmetry. This construction provides a further group-theoretical justification for the estimates of the non-EFT contributions discussed earlier for EEC and CCC. 

We finally discuss detectors constructed out of charged operators in Sec.~\ref{sec_charged_detectors} and argue that the corresponding light-ray integrals are dominated by the UV physics. As a consequence, their event shapes do not factorize into a homogeneous term with small corrections at large charge. We conlude in Sec.~\ref{Sec:Conc} with a detailed outlook of future directions.

\bigskip

The remainder of this paper is organized as follows. In Sec.~\ref{Sec:LChargeRev} we review the superfluid EFT and its connection with CFTs at large charge. In Sec.~\ref{Sec:Detectors}, we introduce our general setup and discuss how to compute event shapes in EFT. In Sec.~\ref{Sec:EEC}, we present the calculation of EEC and CCC at next-to-leading order. In Sec.~\ref{Sec:Generalized}, we generalize our analysis to event shapes involving generic detectors. Finally, in Sec.~\ref{Sec:Conc}, we comment on future directions. The Appendices contain several technical details of the calculations.

\vspace{0.5cm}

\emph{\textbf{Notation}: We use Latin letters \( a, b, \dots \) to denote indices on the cylinder $\mathbb{R}\times S^{d-1}$ and Greek letters \( \mu, \nu, \dots \) for indices in flat space. Spatial indices are denoted by \(i,j,\ldots \) both in flat space and on the cylinder (on $S^{d-1}$). We work in mostly minus signature. Given \( \mathbf{n} \), a unit-norm \((d-1)\)-dimensional vector, we define the associated lightlike vectors $n^{\mu} = (1,\mathbf{n})$ and $\bar{n}^{\mu} = (1,-\mathbf{n})$. We denote by $x_{\mathbf{n}}^{\mu}$ the following vector: 
\begin{align}\label{Eq:LightConeCoor}
    x_{\mathbf{n}}^{\mu} \equiv x^{+}_{\mathbf{n}} \frac{n^{\mu}}{2} + x^{-}_{\mathbf{n}} \frac{\bar{n}^{\mu}}{2}\,,
\end{align}
where $x^{+}_{\mathbf{n}}$ and $x^{-}_{\mathbf{n}}$ are lightlike coordinates with respect to $n$. Correlation functions on the cylinder are specified by an additional subscript as $\langle\ldots\rangle_{\text{\emph{cyl}}}$. Operators in flat space and on the cylinder are denoted with the same symbol.}

\section{The large charge EFT on the cylinder and in Minkowski}\label{Sec:LChargeRev}

Below we review the superfluid EFT description of large charge correlators in CFT and the Weyl map between expectation values on the cylinder and Minkowski space correlators. Readers familiar with these topics map may skip this section.

In CFT, local operators are equivalent to states for the theory quantized on the cylinder $\mathbb{R}\times S^{d-1}$. In particular, choosing the radius of the sphere to be $R$, a state corresponding to an operator with dimension $\Delta$ has energy $\Delta/R$. Expectation values on the cylinder can also be mapped to flat space correlation functions.

As explained in the introduction, in this work we are interested in $d>2$-dimensional CFTs invariant under a $U(1)$ internal symmetry. Let us denote by $\mO_Q$ the operator with lowest scaling dimension at fixed $U(1)$ charge $Q$. We assume that $\mO_Q$ is a scalar for simplicity. It is often the case that the state corresponding to the operator $\mO_Q$ with large charge $Q\gg 1 $ admits a simple EFT description, even when the theory is strongly coupled. Indeed, we expect that $\mO_Q$ corresponds to a state with homogeneous charge density on $\mathbb{R}\times S^{d-1}$. When $Q\gg 1$ there exists a parametric separation between the inverse sphere radius $R^{-1}$ and the scale associated with the charge density $\mu\sim \rho^{\frac{1}{d-1}}\sim Q^{\frac{1}{d-1}}/R$. Generic modes therefore naturally have a gap of order $\sim \mu$ above the large charge ground state and may be integrated out at low energies $\omega\ll \mu$, where we are left with a simplified description. The EFT then allows for the calculation of correlators in the large charge state in a systematic derivative expansion, whose cutoff increases with the charge.

In general, the EFT describing the large charge state depends on the model at hand. In~\cite{Hellerman:2015nra} (see also~\cite{Monin:2016jmo}) it was argued that the simplest and most generic option for interacting systems is that large charge operators correspond to superfluid states.\footnote{Other options include Fermi spheres~\cite{Komargodski:2021zzy,Dondi:2022zna,Delacretaz:2025ifh}, the axio-dilaton EFT for theories with moduli spaces~\cite{Hellerman:2017veg,Grassi:2019txd,Caetano:2023zwe,Cuomo:2024fuy}, and Reissner-N\"ordstrom extremal black holes in holographic theories.} Superfluid states break spontaneously the $U(1)$ symmetry, and admit a low energy description in terms of a single Goldstone mode. The superfluid EFT has been checked to apply in several models both at weak and strong coupling; most notably, there is compelling evidence that large charge operators in the $3d$ $O(2)$ CFT correspond to superfluid states~\cite{Banerjee:2017fcx,Alvarez-Gaume:2019biu,Badel:2019oxl,Cuomo:2023mxg}.

\subsection{The superfluid EFT}

Let us consider the Lorentzian cylinder with metric
\begin{equation}
    ds^2=dt^2-d\Omega_{d-1}^2\,,
\end{equation}
where $d\Omega_{d-1}^2$ is the metric on $S^{d-1}$ and we set the radius $R=1$ for simplicity. A superfluid state breaks spontaneously time translations and the $U(1)$ internal symmetry to a diagonal combination $H-\mu Q$, where $\mu$ may be interpreted as the chemical potential~\cite{Son:2002zn,Nicolis:2015sra}. Its action is written in terms of Goldstone field with expectation value linear in time
\begin{equation}\label{eq_EFT_bkgd}
    \chi =\mu t+\text{const}.\,.
\end{equation}
The action is constrained by Weyl and $U(1)$ invariance, and at leading order in derivatives is given by 
\begin{equation}\label{eq_large_charge_EFT}
    S=\frac{c}{d}\int d^dx\sqrt{g}(\pd\chi)^d\,,
\end{equation}
where $(\pd\chi)=\sqrt{\pd_a\chi\pd^a\chi}$. In~\eqref{eq_large_charge_EFT} $c$ is a Wilson coefficient; for instance, Monte-Carlo simulations in the $3d$ $O(2)$ model found $c\simeq 0.31$~\cite{Banerjee:2017fcx,Cuomo:2023mxg}. Higher derivative operators are suppressed by inverse powers of $(\pd\chi)^2\sim \mu^2$ compared to the leading order, see e.g.~\cite{Cuomo:2020rgt} for explicit expressions. 

Let us illustrate the calculation of the ground-state energy, corresponding to the scaling dimension of the operator $\mO_Q$. From the action~\eqref{eq_large_charge_EFT}, we find the $U(1)$ current and the energy-momentum tensor:
\begin{align}\label{current} 
J_a&= c \,\partial_a\chi(\pd\chi)^{d-2} ~,
\\
\label{emtensor} 
T_{a b}&=c \left[\pd_a\chi\pd_b\chi (\pd\chi)^{d-2} -\frac{1}{d}g_{ab}(\pd \chi)^d \right]~.
\end{align} 
Evaluating \eqref{current} on the saddle-point $\chi=\mu t$ and integrating the current over the sphere, we see that the total charge is
\begin{equation}\label{eq_charge_vs_mu}
    Q=c\,\mu^{d-1}\Omega_{d-1}\quad\implies\quad
    \mu=\left(\frac{Q}{c\,\Omega_{d-1}}\right)^{\frac{1}{d-1}}\,,
\end{equation}
where $\Omega_{d-1}=\frac{2\pi^{d/2}}{\Gamma(d/2)}$ is the volume of the $d-1$-dimensional sphere. Using this result in the energy momentum tensor we find 
\begin{equation}\label{eq_Delta_n}
\Delta_Q = 
\frac{(d-1)c\,\Omega_{d-1}}{d\left(c\,\Omega_{d-1}\right)^\frac{d}{d-1}}Q^{\frac{d}{d-1}}+\ldots\,.
\end{equation}
Higher derivative terms contribute at order $\sim Q^{\frac{d-2}{d-1}}$ to the scaling dimension~\eqref{eq_Delta_n}.

Let us now introduce the fluctuation field $\pi=\chi-\mu t$. To quadratic order, its action reads:
\begin{equation}\label{eq_cyl_action_fluct}
    S^{(2)}        =c(d-1)\mu^{d-2}\int d^dx\sqrt{g}\,\left[
    \frac{1}{2}  \dot{\pi}^2
    -\frac{1}{2(d-1)}(\pd_i\pi )^2  \right]\,,
\end{equation}
where $\dot{\pi}=\pd\pi/\pd t$ and the indices $i,j,\ldots$ run over the sphere angles (and are contracted using the appropriate spatial metric). The action~\eqref{eq_cyl_action_fluct} physically describes a \emph{phonon} mode whose speed of sound is fixed by conformal invariance: $c_s^2=1/(d-1)$. Therefore, the dispersion relation is
\begin{equation}\label{eq_freq_phonon}
    \omega_\ell=\sqrt{\frac{\ell(\ell+d-2)}{d-1}}\,,
\end{equation}
where $\ell$ is the angular momentum and $\ell(\ell+d-2)$ are the eigenvalues of the Laplacian on $S^{d-1}$. The states created by the phonon correspond to operators with the same charge $Q$ as the ground state and scaling dimension $\Delta_Q+\omega_{\ell}$. Note that the $\ell=1$ mode has $\omega_1=1$, and its excitations correspond to descendants.

From~\eqref{eq_cyl_action_fluct} we obtain the propagator of $\pi$. In the following, we will need in particular the Wightman function. This can be expressed as a sum over Gegenbauer polynomials as
\begin{equation}\label{Eq:WightCyl}
 \langle Q|\pi(t_1,\hat{m}_1)\pi(t_2,\hat{m}_2)|Q\rangle_{\cyl}= 
 \frac{1}{c(d-1)\mu^{d-2}}G_{\pi\pi}(t_1-t_2,\hat{m}_1\cdot\hat{m}_2)\,,
\end{equation}
where $|Q\rangle$ denotes the large charge ground-state, $\hat{m}_{1}$ and $\hat{m}_2$ are unit vectors that specify the positions of the operators on the sphere, and, being careful about the zero-mode,\footnote{More precisely, since the field is $2\pi$-periodic, this formula applies for correlation functions of properly quantized operators, such as $e^{\pm i\pi(t,\hat{m})}$, or of derivatives of the field.}
\begin{equation}\label{Eq:WightmanCyl}
    G_{\pi\pi}(t,x)=
    \sum_{\ell=1}^{\infty}\frac{2\ell+d-2}{(d-2)\Omega_{d-1}}\frac{e^{-i\omega_{\ell}t}}{2\omega_{\ell}}C_{\ell}^{\left(\frac{d-2}{2}\right)}\left(x\right)
    -\frac{i}{2\Omega_{d-1}}t
    \,.
\end{equation}
We remark that using the decomposition~\eqref{Eq:WightmanCyl} in correlation functions is equivalent to performing a conformal block expansion in the s-channel, where we use the OPE between the light operators and the large charge states, and the superfluid EFT controls the dimensions of the lightest exchanged operators. We refer the reader to~\cite{Jafferis:2017zna} for a detailed discussion of the conformal block decompositions at large charge. We discuss alternative convenient representations of the propagator that we will need later in App.~\ref{app_Prop}.

\subsection{From the cylinder to flat space}

By the state-operator map, the large charge ground state $|Q\rangle$ is created by the action of the corresponding primary operator on the vacuum at infinite imaginary time:
\begin{equation}\label{eq_state_op_map}
|Q\rangle=\lim_{\tau\rightarrow -\infty} e^{-\Delta_Q\tau}    \mO_Q(-i\tau,\hat{m})|0\rangle\quad\implies\quad
\langle Q|=\lim_{\tau\rightarrow \infty}e^{\Delta_Q\tau}    \langle 0| \bar{\mO}_Q(-i\tau,\hat{m})\,.
\end{equation}
The origin of~\eqref{eq_state_op_map} is simple. The action of $\mO_Q$ at a generic time on the vacuum creates a complicated linear combination of the primary and its descendant states. Pushing the field infinitely far away in imaginary time the contribution of the descendants gets exponentially suppressed, and we are left only with the lowest energy contribution: the primary state $|Q\rangle$. 

We may now exploit the Weyl invariance of the theory to relate correlation functions on the cylinder and in flat space. The transformation law for correlation functions of primary operators under Weyl rescalings is given by\footnote{Here we neglect the Weyl anomaly contribution to the partition function in even spacetime dimensions.}
\begin{equation}\label{eq_Weyl}
    \langle\mO_{1,\mu_1\ldots\mu_{J_1}}(x_1)\ldots
    \mO_{n,\nu_1\ldots\nu_{J_n}}(x_n)\rangle_g=\prod_i\Omega(x_i)^{J_i-\Delta_i}
    \langle\mO_{1,\mu_1\ldots\mu_{J_1}}(x_1)\ldots
    \mO_{n,\nu_1\ldots\nu_{J_n}}(x_n)\rangle_{\Omega^2 g}\,,
\end{equation}
where $\Delta_i$ and $J_i$ are the dimension and spin of the operators, and $g$ denotes the metric. Therefore, to obtain flat space correlators we simply need to embed the physical Minkowski spacetime inside the cylinder and use~\eqref{eq_Weyl}. The map that we will need in the following considers the wedge of the cylinder specified by (see e.g.~\cite{Kravchuk:2018htv})
\begin{equation}
    \cos t+\hat{m}^d\geq 0\;\cup\; t\in(-\pi,\pi)\,,
\end{equation}
and to each point in this region associates a point in Minkowski according to the relation
\begin{equation}\label{eq_x_explicit_Lor}
    x^\mu=\frac{X_0}{2}\left(\frac{\sin t}{\cos t+\hat{m}^d},\,
\frac{\hat{m}^i}{\cos t+\hat{m}^d}\
    \right)\,,
\end{equation}
where $X_0$ is an arbitrary scale. It is simple to check that the flat space metric is related to the cylinder one by
\begin{equation}\label{eq_cyl_Lor_metric}
    ds^2_{\text{flat}}=dx_0^2-d\mathbf{x}^2=\Omega^2(x)(dt^2-d\hat{m}^2)=\Omega^2(x)ds^2_{\text{cyl}}\,,
\end{equation}
where the Weyl factor is
\begin{equation}
    \Omega^2(x)=\frac{X_0^2}{4(\cos t+\hat{m}^d)^2}\,,
\end{equation}

Importantly for us, the primary operator insertions that specify the state~\eqref{eq_state_op_map} map to imaginary positions
\begin{equation}\label{eq_tau_insertions}
    \tau=it=\mp\infty\longrightarrow
 x^\mu=\pm\frac{i}{2}\left(X_0,\mathbf{0}\right)\equiv\pm \frac{i}{2} X^\mu\,,
\end{equation}
as it follows from~\eqref{eq_x_explicit_Lor}. 
Therefore, correlators on the Lorentzian cylinder map to correlation functions in between two large charge operators at imaginary time. Equivalently, the operator insertions at imaginary time create a primary state in North/South quantization~\cite{Rychkov:2016iqz}. As a simple example, the three-point functions with an insertion of the Noether current and the stress tensor are obtained from the cylinder matrix element as 
\begin{align}\label{eq_J_flat}
&\,\,\,\frac{\langle \bar{\mO}_Q(-\frac{i}{2} X)J^\mu(x)  \mO_Q(\frac{i}{2} X)\rangle}{\langle \bar{\mO}_Q(-\frac{i}{2} X)  \mO_Q(\frac{i}{2} X)\rangle} = \Omega(x)^{-d} \frac{\pd x^\mu}{\pd y^a}\langle Q|J^a(t,\hat{m})|Q\rangle_{\text{cyl}}
=\rho_X(x)u^\mu_X(x)\,,\\
&\begin{aligned}\label{eq_T_flat}
\frac{\langle \bar{\mO}_Q(-\frac{i}{2} X)T^{\mu\nu}(x)  \mO_Q(\frac{i}{2} X)\rangle}{\langle \bar{\mO}_Q(-\frac{i}{2} X)  \mO_Q(\frac{i}{2} X)\rangle} &= \Omega(x)^{-d-2} \frac{\pd x^\mu}{\pd y^a}\frac{\pd x^\nu}{\pd y^b}\langle Q|T^{ab}(t,\hat{m})|Q\rangle_{\text{cyl}}\\
&=\frac{d}{d-1} \varepsilon_X(x)\bigg ( u_X^\mu(x)u_X^\nu(x)-\frac{1}{d}\eta^{\mu\nu}\bigg)\,,
\end{aligned}
\end{align}
where $y^a=(\tau,\hat{m})$ collectively denotes the cylinder coordinates and we defined the Minkowski space fluid charge and energy \emph{densities} $\rho_X$ and $\varepsilon_X$, as well as the \emph{velocity} $u_X$,  as
\begin{align}
\rho_X(x)&=\frac{Q}{\Omega_{d-1}}\left[\frac{X^2}{(x-\frac{i}{2} X)^2(x+\frac{i}{2}X)^2}\right]^{\frac{d-1}{2}}\,,\hspace{0.5cm} \varepsilon_X(x) = \frac{\Delta_Q}{\Omega_{d-1}}\bigg[\frac{X^2}{\big(x+\frac{i}{2}X\big)^2\big(x-\frac{i}{2}X\big)^2}\bigg]^\frac{d}{2},\\
u^\mu_X&=\left[\frac{X^2}{(x-\frac{i}{2} X)^2(x+\frac{i}{2}X)^2}\right]^{-\frac{1}{2}}
\left[i\frac{x^\mu+\frac{i}{2}X^\mu}{(x+\frac{i}{2}X)^2}-i\frac{x^\mu-\frac{i}{2}X^\mu}{(x-\frac{i}{2}X)^2}\right]\,.
\end{align}
Note that these expressions are real. Of course,~\eqref{eq_J_flat} and~\eqref{eq_T_flat} are fixed by conformal invariance, we reproduced them here simply to illustrate the Weyl map.

\begin{figure}[t!]
    \centering
    \begin{subfigure}{0.32\textwidth}
    \centering
        \includegraphics[width=.8\linewidth]{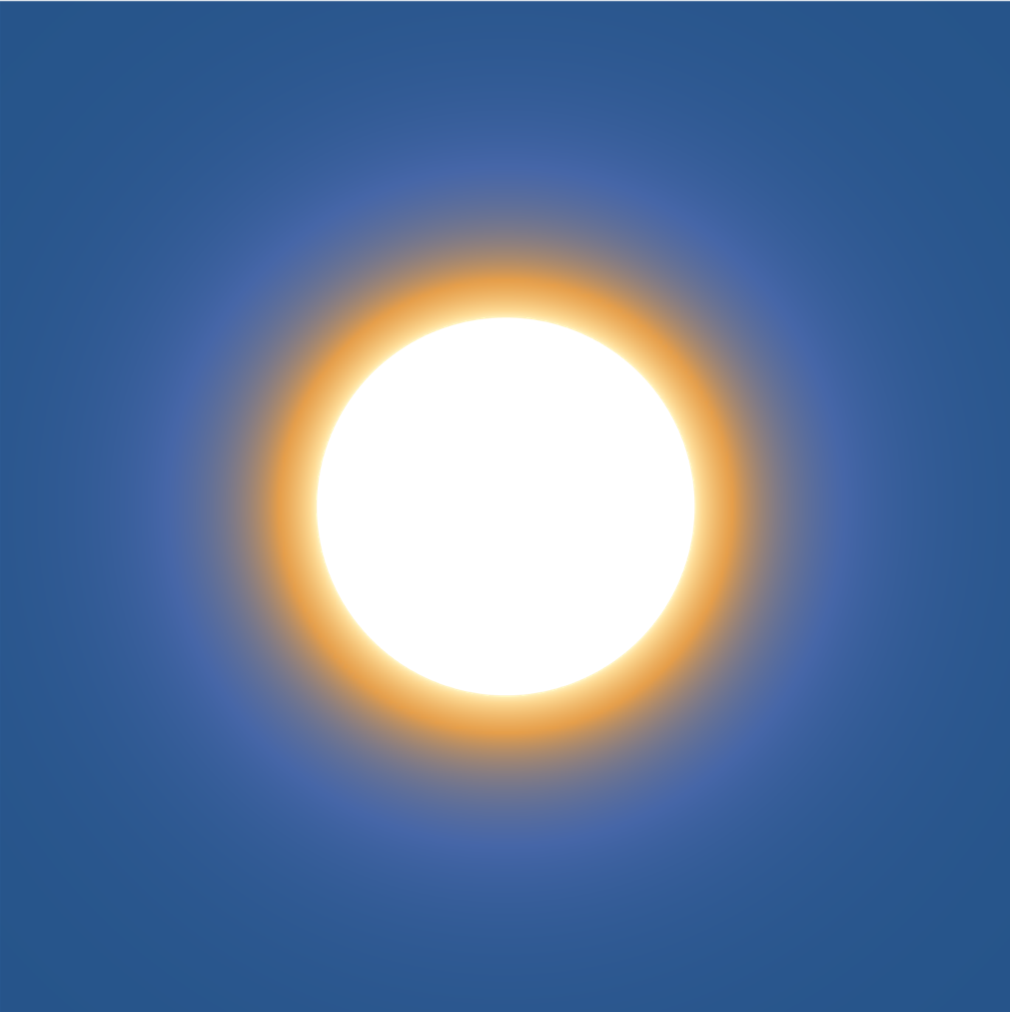}
        \caption{$x^0=0$}
    \end{subfigure}
    \begin{subfigure}{0.32\textwidth}
    \centering
        \includegraphics[width=.8\linewidth]{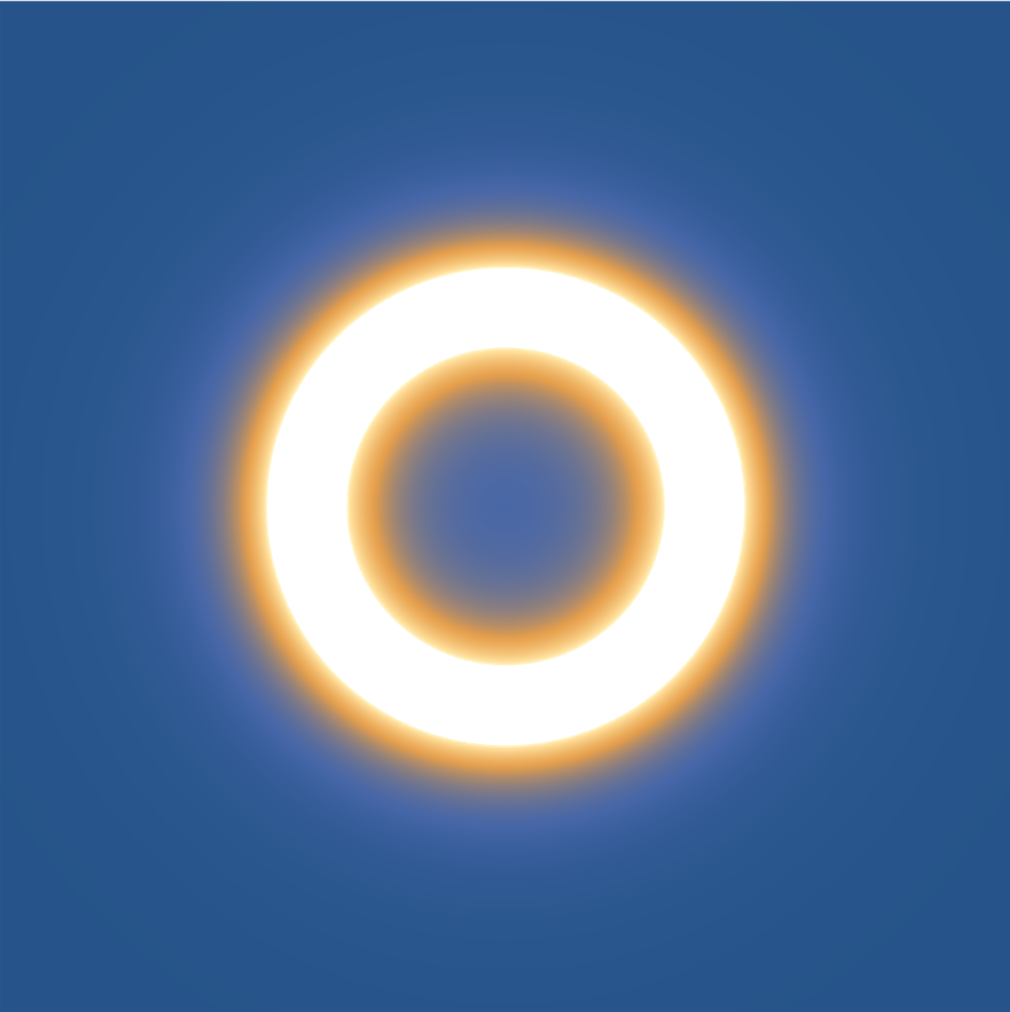}
        \caption{$x^0=2X_0$}
    \end{subfigure}
    \begin{subfigure}{0.32\textwidth}
    \centering
        \includegraphics[width=.8\linewidth]{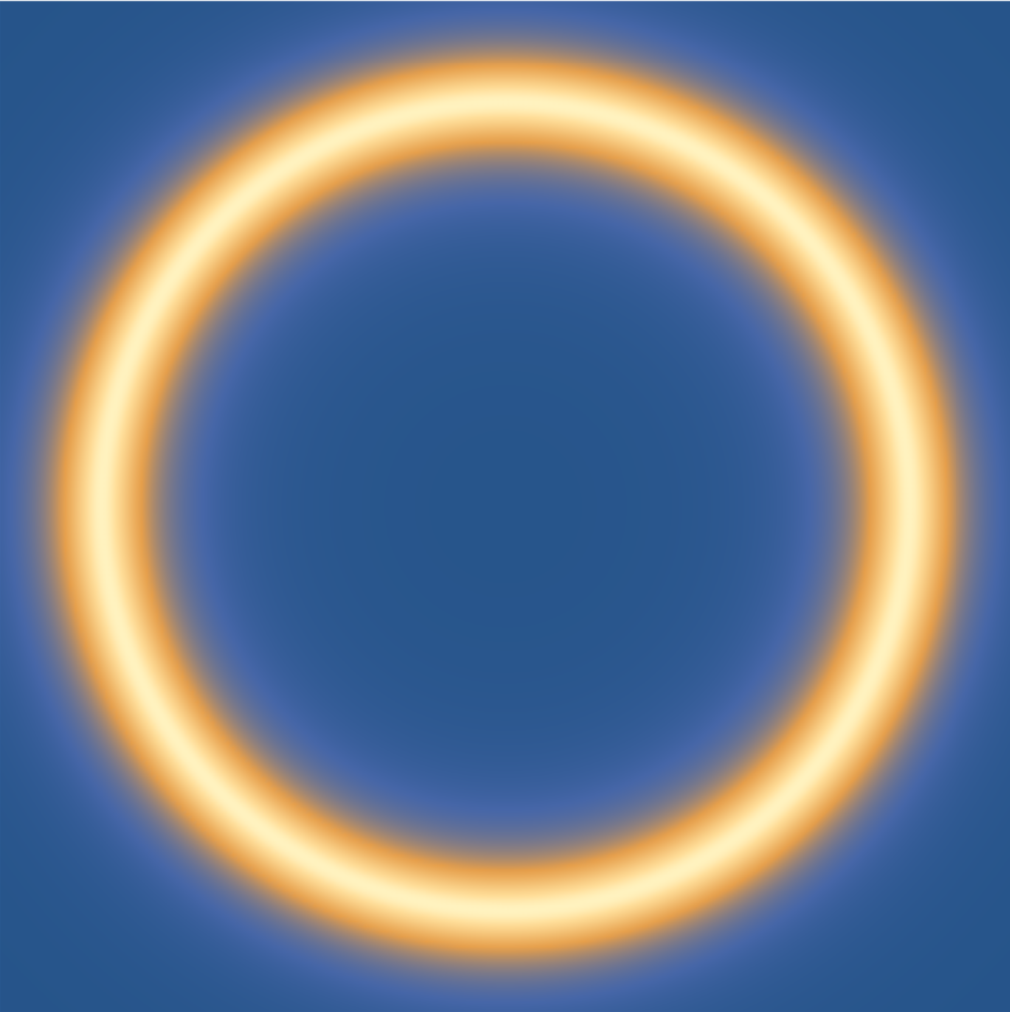}
        \caption{$x^0=4X_0$}
    \end{subfigure}
    \caption{Radial distribution of the charge density $\langle J^0(x)\rangle_X$ at different times in $d=3$.} 
    \label{fig:densityevol}
\end{figure}

The expectation values~\eqref{eq_J_flat} and~\eqref{eq_T_flat} clarify the physical meaning of the large charge state created by $\mO_Q(i X/2)$ in Minkowski space. It describes a charge distribution localized around the origin in a region of size $\sim \mu X_0$ at time $x^0=0$. By conformal invariance, the charge spreads at the speed of light around shells centered at $r\simeq x^0$, whose thickness is controlled by $\mu X_0$. Fig.  \ref{fig:densityevol} shows three snapshots of the time evolution of the charge density. Physically, as in standard hydrodynamic setups, the EFT describes correlation functions of operators in the region where the scale set by the charge flux is large compared to the separation between the insertions~\cite{Firat:2023lbp}.\footnote{Of course, the form of three-point functions is fixed by conformal invariance; the comments about the applicability of the EFT are therefore relevant for four- and higher-point correlators.} Note that the region of applicability of the EFT covers the whole spacetime for $Q\rightarrow \infty$. 

It is of course also possible to obtain correlation functions with the charged operators inserted at arbitrary points by analytic continuation. This is straightforward to do for the three-point functions~\eqref{eq_J_flat} and \eqref{eq_T_flat}, but it becomes less intuitive for higher-point correlators. Luckily, we will not need to consider such analytic continuations for our purposes.

\section{Detectors and semiclassics}\label{Sec:Detectors}

\subsection{Review of event shapes}\label{Sec:Review_EEC}

\begin{figure}
    \centering
    \begin{tikzpicture}[scale=1]
        \draw [] (-3,0) -- (0,3) -- (3,0) -- (0,-3) -- cycle;
        \draw [] (-3,0) to[out=-45,in=-135,distance=1.0cm] (3,0);
        \draw [dashed] (-3,0) to[out=45,in=135,distance=1.0cm] (3,0);
        \draw [thick,blue] (-1,-0.5) -- (0,3);
        \draw [thick,red] (1,-0.5) -- (0,3);
        \node [above] at (-1.5,-0.2) {$\mathcal{D}_{1}(\mathbf{n}_1)$};
        \node [above] at (1.5,-0.2) {$\mathcal{D}_{2}(\mathbf{n}_2)$};
        \node [above] at (1.8,1.65) {$\mathscr{I}^+$};
        \node [below] at (1.8,-1.55) {$\mathscr{I}^-$};
        \node [above] at (0,3) {$i^+$};
        \node [below] at (0,-3) {$i^-$};
        \node [right] at (3,0) {$i^0$};
    \end{tikzpicture}
    \hspace{1cm}
    \begin{tikzpicture}
\draw[thick] (0,3) ellipse (1.5 and 0.4);
\draw[thick] (-1.5,3) -- (-1.5,-3);
\draw[thick] (1.5,3) -- (1.5,-3);

\draw[thick,dashed] (1.5,-3) arc[start angle=0, end angle=180, x radius=1.5, y radius=0.4]; 
\draw[thick] (-1.5,-3) arc[start angle=180, end angle=360, x radius=1.5, y radius=0.4]; 
\draw[->,thick] (-2,-2.8)--(-2,3);
\node at (-2.3,3){$t$};
\draw[-] (-2.1,0)--(-1.9,0);
\node at (-2.3,0){$0$};
\draw[-] (-2.1,2)--(-1.9,2);
\node at (-2.3,2){$\pi$};
\draw[-] (-2.1,-2)--(-1.9,-2);
\node at (-2.4,-2){$-\pi$};
\draw[thick,blue] (-1.5, 2) to[out=-45, in=155] (1.5, 0);
\draw[dashed] (-1.5, 2) to[out=-15, in=125] (1.5, 0);
\draw[] (1.5, 0) to[out=-125, in=15] (-1.5, -2);
\draw[dashed] (1.5, 0) to[out=-165, in=55] (-1.5, -2);
\draw[-] (-1.6,-3)--(-1.4,-3);
\draw[-] (1.6,-3)--(1.4,-3);
\node at (-1.7,-3.2){$0$};
\node at (1.7,-3.2){$\pi$};
\node at (-0,-3.8){$\sigma$};
\draw[thick, ->] (-1,-3.5)to[out=-10, in=190](1, -3.5);
\node at (-0.3,0.5) {$\mathcal{D}^{\text{cyl}}$};

\node at (0,-3){$\mO_Q(t=i \infty)$};
\node at (0,3){$\bar{\mO}_Q(t=-i \infty)$};


\node [above] at (-1.2,2) {$i^+$};
\node [below] at (1.9,0.4) {$i^0$};

\end{tikzpicture}
    \caption{The detector $\mathcal{D}_{}(\mathbf{n})$ is obtained integrating a local operator on light-ray at null infinity. Left:  Penrose diagram of $\mathbb{R}^{1,2}$. Note that the entire circle at spacial infinity is actually a point $i^0$. Right: Lorenzian cylinder $\mathbb{R}\times S^2$ (each point is a circle). The source and the sink are inserted at imaginary time $t=\pm i\,\infty$ and thus create large charge states on the cylinder.}
    \label{fig:LightrayIntegral}
\end{figure}
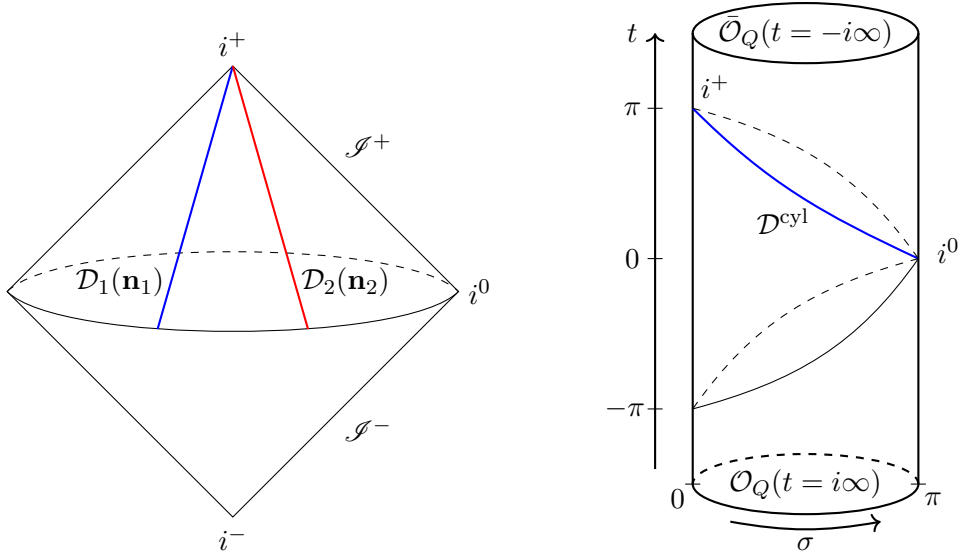

In CFTs signals move asymptotically at the speed of light,  thus detectors can be defined as integrals of local operators on light-rays at future null infinity  $\mathscr{I}^+$ (see for instance \cite{Belitsky:2013xxa}). This is depicted in Fig.~\ref{fig:LightrayIntegral} for a general detector, denoted by $\D{}$. The energy detector in \eqref{eq_Energy_detectors} can be equivalently defined as
\begin{align}
\label{energy_detector}
    \En &=2^{-d} \lim_{\xn^+ \rightarrow+\infty}(\xn^+)^{d-2}\int_{-\infty}^{+\infty} d \xn^- \bar{n}^{\mu}\bar{n}^{\nu} T_{\mu \nu} (\xn) \,,
\end{align}
and analogously for the charge detector
\begin{align} \label{charge_detector}
\Qn &=2^{1-d} \lim_{\xn^+ \rightarrow+\infty} (\xn^+)^{d-2}\int_{-\infty}^{+\infty} d \xn^- \bar{n}^{\mu}J_{\mu} (\xn)\,.
\end{align}
Here $\xn^{\pm}$ denote the components of the four-vector $x^{\mu}_{\mathbf{n}}$ in lightcone coordinates, according to our conventions in \eqref{Eq:LightConeCoor}. Even though we focus on charge and energy detectors for concreteness, the discussion in this section applies to any detector obtained integrating a local operator on a light-ray at null infinity.

Event shapes are matrix element of detectors inserted between momentum eigenstates
\begin{align}\nonumber
    \langle \mD_1(\mathbf{n}_1)\mD_2(\mathbf{n}_2)\dots \mD_k(\mathbf{n}_k)  \rangle_p &\equiv \frac{\langle{ \bar{\mO}_Q (p) \mD_1(\mathbf{n}_1)\mD_2(\mathbf{n}_2)\dots \mD_k(\mathbf{n}_k) \mO_Q(p)\rangle}}{\langle\bar{\mO}_Q (p) \mO_Q(p)\rangle}\\
    &= \frac{\int d^dx_{fi} e^{i p \cdot x_{fi}} \langle \bar{\mO}_Q (x_f) \mD_1(\mathbf{n}_1)\mD_2(\mathbf{n}_2)\dots \mD_k(\mathbf{n}_k) \mO_Q(x_i)\rangle}{\int d^dx_{fi} e^{i p \cdot x_{fi}} \langle\bar{\mO}_Q (x_f) \mO_Q(x_i)\rangle}\,,
    \label{Eq:GenEvShape}
\end{align}
where we used that detectors commute with the momentum generator $[P^{\mu}, \D{}]=0$. 
The operators $ \mO_Q(x_i) $ and $ \bar{\mO}_Q(x_f) $ are commonly referred to as the ``source'' and ``sink'', respectively. We are interested in the operators with the lowest scaling dimension at a fixed and large $ U(1) $ charge, as discussed in detail earlier. Note that the relative order of the source and sink in~\eqref{Eq:GenEvShape} is important, i.e. the event shape~\eqref{Eq:GenEvShape} is a Wightman function. The ordering between the different detectors is irrelevant instead since they commute with each other for $\mathbf{n}_1\neq\mathbf{n}_2$, as every two points in two different light-ray contours are space-like separated. We also remark that detectors annihilate the vacuum.

The definitions~\eqref{energy_detector} and~\eqref{charge_detector} may be seen as a special instance of the so-called light-ray transform in CFT~\cite{Kravchuk:2018htv}. We will  not review in detail the properties of light-ray operators. For us it suffices to say that a detector obtained integrating a primary field of dimension $\Delta$ and spin $J$ transforms as a primary of dimension $1-J$ and spin $1-\Delta$ inserted at spatial infinity under the action of the conformal group,  see e.g.~\cite{Kologlu:2019mfz}. In particular, charge and energy detectors correspond to light-ray operators with, respectively, dimension $\Delta_{\mathcal{Q}}=0$ and $\Delta_{\mathcal{E}}=-1$, and spin $J_{\mathcal{Q}}=2-d$ and $J_{\mathcal{E}}=1-d$. 

The conformal group therefore constrains the form of event shapes. In particular, one-point event shapes in scalar states are completely fixed in terms of the OPE coefficient.  Note that without loss of generality we can use Lorentz invariance to boost to the rest frame of the source/sink, $p^{\mu}= (E,\mathbf{0})$. Then the one point functions of the energy and charge detectors read
\begin{align}\label{eq_1pt_event}
    \langle \Qn \rangle = \frac{Q}{\Omega_{d-2}} \,,&& \langle \En \rangle = \frac{E}{\Omega_{d-2}}\,,
\end{align}
where we dropped the subscript $_p$ from~\eqref{Eq:GenEvShape}. The physical meaning of~\eqref{eq_1pt_event} is obvious: for a state at rest a detector at infinity measures a uniform distribution of energy and charge. 

Event shapes with two or more detectors depend on one or more cross-ratios, and are hence theory dependent. For instance, two-point charge and energy correlators read
\begin{align}\label{eq_QQ_cross_ratio}
\langle \Qno \Qnt \rangle_p &=\frac{\Theta(p^0)\Theta(p^2)(p^2)^{d-2}}{(n_1\cdot p)^{d-2}(n_2\cdot p)^{d-2}}\mathcal{G}_{\mathcal{Q}\mathcal{Q}}(\xi)\,,\\  \langle \Eno \Ent \rangle_p &=\frac{\Theta(p^0)\Theta(p^2)(p^2)^{d}}{(n_1\cdot p)^{d-1}(n_2\cdot p)^{d-1}}\mathcal{G}_{\mathcal{E}\mathcal{E}}(\xi)\,,
\label{eq_EE_cross_ratio}
\end{align}
where $\Theta$ is the step function, $\mathcal{G}_{\mathcal{Q}\mathcal{Q}}$ and $\mathcal{G}_{\mathcal{E}\mathcal{E}}$ are functions of the cross-ratio
\begin{align}
    \xi \equiv \frac{(2\,n_1\cdot n_2)(p^2)}{(2\,p\cdot n_1)(2\,p\cdot n_2)} 
    \quad \longrightarrow_{p=(E,\mathbf{0})}\quad \frac{1-\mathbf{n}_1 \cdot \mathbf{n_2}}{2}\,.
\end{align}
The variable $\xi$ encodes the interesting angular dependence on these two-point event shapes. In the rest frame of source/sink this just reduces to the angle between the detectors. In the following we shall always work in the source/sink rest frame.

Note that, in general, there is no guarantee that higher-point event shapes are well defined. Indeed, in performing the integration over the light-ray integrals, we necessarily span over dangerous configurations where two or more operators have zero or light-like separation, including the Regge and the light-cone limits \cite{Kologlu:2019bco}. Yet two and higher-point event shapes of the energy detector are always finite, and in general infinitely many other event shapes can be defined (see also \cite{Korchemsky:2021okt} for generalizations). Charge detectors not always are IR safe, but it is believed that they are well defined in the $O(2)$ CFT \cite{Caron-Huot:2022eqs}. In the following we assume that the event shapes that we are discussing are well defined.

\subsection{Event shapes in EFT}\label{Sec:FTcyl}

The semiclassical nature of the EFT implies that correlation functions of local operators in a large charge state $|Q\rangle$ admit a very simple structure: at leading order in the charge they factorize into the product of the expectation values of the individual operators. For instance, considering for concreteness two insertions of the Noether current, we find
\begin{equation}\label{eq_many_J_EFT}
    \langle Q| J_{a_1}(t_1,\hat{m}_1)J_{a_2}(t_2,\hat{m}_2) |Q\rangle_{\text{cyl}}=
    \left(\frac{Q}{\Omega_{d-1}}\right)^{2}\left[\delta_{a_1}^0\delta_{a_2}^0+O\left(n^{-\frac{d}{d-1}}\right)\right]\,,
\end{equation}
where the corrections arise from the fluctuation field $\pi$. 

We would like to relate the above factorization property of correlation functions to a similar statement for event shapes. Naively, the simplest strategy is to compute the event shape directly from the correlation function, as e.g. in~\cite{Belitsky:2013xxa}. Therefore we need to: (i) map the EFT correlator to flat space and analytically continue it to Lorentzian signature, (ii) compute the light-ray integrals, and (iii) perform the Fourier transform. Doing this for~\eqref{eq_many_J_EFT}, we indeed find
\begin{equation}\label{eq_Q_factorization_guess}
    \langle \mathcal{Q}(\mathbf{n}_1)\mathcal{Q}(\mathbf{n}_2) \rangle\approx
    \left(\frac{Q}{\Omega_{d-2}}\right)^2\,,
\end{equation}
as detailed in App.~\ref{App:Disconnected}. In words, this means that at leading order in $1/Q$, the charge distribution on the celestial sphere is completely isotropic in the rest frame of the source/sink. It was argued in~\cite{Firat:2023lbp,Chicherin:2023gxt} that also higher-point event shapes and ECs factorize in a similar fashion.

In the above discussion we neglected the corrections to the classical result in EFT. While the quantum corrections to the correlation function~\eqref{eq_many_J_EFT} are indeed suppressed for generic kinematics, it is not obvious that these remain subleading in~\eqref{eq_Q_factorization_guess} due to the several integrations that are needed to obtain the event shape. More generally, we know that~\eqref{eq_Q_factorization_guess} does not hold in certain cases and configurations, such as in the limit $\mathbf{n}_1\cdot\mathbf{n}_2 \rightarrow 1$, which is governed by the light-ray OPE (see, for instance, \cite{Kologlu:2019mfz} and references therein). It is therefore important to analyze systematically the structure of the EFT expansion for event shapes. This poses both a technical challenge and a conceptual question:
\begin{itemize}
\item At the technical level, to perform the Fourier transform we (seemingly) need to analytically continue the source and the sink away from their natural insertion points at imaginary time discussed around~\eqref{eq_tau_insertions}. This is challenging at subleading order, because the Fourier transform spans several inequivalent time orderings and especially since the propagator~\eqref{Eq:WightmanCyl} only admits closed form expressions as an infinite sum or as an implicit integral, to the best of our knowledge, see~App.~\ref{app_Prop}. Physically, as commented below~\eqref{eq_state_op_map}, the insertion of the large charge operators in Lorentzian time creates complicated linear combinations of the primary and the descendant states, obscuring the physical picture underlying the EFT.
\item At the conceptual level, when we integrate the over the light-ray contours we unavoidably encounter regions in which the EFT does not hold. Physically this is because, as $x^-_{\mathbf{n}}\rightarrow\pm\infty$, the integrated operators approach a region in which the charge flux created by the large charge operators is very diluted, and we expect the EFT to breakdown. We will characterize the breakdown of EFT more precisely in the following.
\end{itemize}
Despite these complications, it was argued in~\cite{Firat:2023lbp,Chicherin:2023gxt} that the breakdown of EFT in certain integration regions can only provide subleading corrections to the leading result~\eqref{eq_Q_factorization_guess} for generic kinematics. Below we shall confirm this conclusion, and we will further determine the scaling with the charge $Q$ of such corrections.

To overcome the complications that we just mentioned, it is convenient to compute the Fourier transform before performing the light-ray integrals. As we explain below, the integration over the distance between the source and the sink can be performed via the saddle-point expansion.
To determine the position of the saddle, let us consider the two-point function of the large charge operators:
\begin{align}
\int d^dx_{fi} e^{i p\cdot x_{fi}}
\langle\bar{\mO}_Q(x_{f})\mO_Q(x_i)\rangle&=
    \int d^dx_{fi}\frac{ e^{i p\cdot x_{fi}}}{\left[\mathbf{x}_{fi}^2-(x_{fi}^0-i\epsilon)^2\right]^{\Delta_Q}}\,.
\end{align}
In the limit $\Delta_Q\gg1$ the integral can be performed via a saddle point approximation by expressing it as
\begin{equation}
   \int d^dx_{fi} e^{\Delta_Q f(x_{fi})}\,,\qquad
   f(x_{fi})=i\frac{p\cdot x_{fi}}{\Delta_Q}-\log(-x_{fi}^2)\,.
\end{equation}
The function $f(x_{fi})$ admits a saddle for imaginary $x_{fi}$:
\begin{equation}\label{eq_saddle_pt}
    \frac{\pd f(x_{fi})}{\pd x^\mu_{fi}}=i\frac{p_\mu}{\Delta_Q}-2\frac{x_{fi,\mu}}{x^2_{fi}}=0\quad\implies\quad
    x^{\mu}_{fi}=-2i\Delta_Q\frac{p^\mu}{p^2}\,.
\end{equation}
Note that having $p^2>0$, we can think of the saddle-point as making the $i\epsilon$ finite, hence the saddle-point has the right sign as to preserve the ordering of the operators. This ensures that the integration contour may be deformed to pass through the saddle-point~\eqref{eq_saddle_pt}.
Working in the source/sink rest frame, $p^\mu=(E,\mathbf{0})$, and setting $x^{\mu}_{fi}=-2i\delta^{0}_{\mu}\Delta_Q/E+\delta x^{\mu}\sqrt{2\Delta_Q}/E $ with $\delta x^\mu$ real, we then find 
\begin{equation}\label{Eq:SaddleMom}
\begin{split}
    \int d^dx_{fi} e^{\Delta_Q f(x_{fi})}&\simeq \frac{e^{2\Delta_Q}(p^2)^{\Delta_Q}}{(2\Delta_Q)^{2\Delta_Q}}
\left(\frac{2\Delta_Q}{p^2}\right)^{d/2}\int d^d\delta x  
\,e^{-\frac{\delta x_0^2  +\delta\bf{x}^2}{2}}\left[1+O\left(\frac{\delta x^3}{\sqrt{\Delta_Q}}\right)\right]\\
&=
(2\pi)^{\frac{d}{2}}\frac{e^{2\Delta_Q}(p^2)^{\Delta_Q-d/2}}{(2\Delta_Q)^{2\Delta_Q-d/2}}\left[1+O\left(\frac{1}{\Delta_Q}\right)\right]
    \,,
    \end{split}
\end{equation}
which agrees with the expansion of the exact result reported for completeness in~\eqref{Eq:ExactFourier2Point}. 

The above logic can be applied in any correlation function with two heavy operator insertions and several light operators. Indeed, conformal invariance ensures that the dominant exponential factor \( e^{\Delta_Q f(x_{fi})} \), governed by the scaling dimensions of the source and sink, coincides with~\eqref{Eq:SaddleMom}. For instance, for the Fourier transform of the correlation function with two current insertions we find 
\begin{multline}\label{Eq:SaddlePointJJ}
    \frac{
   \int d^dx e^{i p \cdot x} \langle \bar{\mO}_{Q}(\frac{x}{2})   J^{\mu}(x_1) J^{\nu}(x_2) \mO_{Q}(-\frac{x}{2})\rangle}{\int d^dx e^{i p\cdot x} \langle \bar{\mO}_{Q}(\frac{x}{2})  \mO_{Q}(-\frac{x}{2})\rangle}
   \\=\frac{\langle \bar{\mO}_Q(-\frac{i}{2}X) J^{\mu}(x_1) J^{\nu}(x_2)  \mO_Q (\frac{i}{2}X)\rangle}{\langle \bar{\mO}_Q(-\frac{i}{2}X)   \mO_Q (\frac{i}{2}X)\rangle}
   \left[1+O\left(\frac{1}{\Delta_Q}\right) \right],
\end{multline}
where we conveniently placed the source and the sink symmetrically, and we defined
\begin{equation}\label{eq_X_def}
X^{\mu}=\left(X_0\equiv\frac{2 \Delta_Q}{E},\mathbf{0}\right)\,.
\end{equation}
Remarkably, on the saddle-point, the source and sink are at imaginary time, as in the discussion around~\eqref{eq_tau_insertions}. Therefore, using the map~\eqref{eq_x_explicit_Lor} with $X_0=2\Delta_Q/E$, we see that the correlation function~\eqref{Eq:SaddlePointJJ} is equivalent (up to the Weyl factors) to the cylinder two-point function of the current in the primary state $|Q\rangle$! We thus do not need to analytically continue the cylinder correlators predicted by the EFT to compute the event shape, bypassing the first technical complication mentioned above. As we shall explain below, the saddle-point~\eqref{Eq:SaddlePointJJ} also simplifies the analysis of the validity regime of EFT.

Let us first review in detail how to map event shapes to correlation functions of light-ray operators on the cylinder using~\eqref{Eq:SaddlePointJJ}. It is convenient to factor out one angle setting $\hat{m}^d=\cos\sigma$ in the map~\eqref{eq_x_explicit_Lor}; explicitly we consider
\begin{align}\label{Eq:ChangeOfCoordinatesCyl}
    x^{\mu} = \frac{X_0}{2} \left(\frac{\sin t}{\cos t + \cos \sigma}, \frac{\sin \sigma}{\cos t + \cos \sigma} \mathbf{n} \right)\,,
\end{align}
where $\mathbf{n}$ is a $(d-1)$-dimensional vector parametrizing $S^{d-2}$ and $0 < \sigma <\pi$.
Therefore the metric and the Weyl factor in these coordinates read
\begin{align}
    ds^2_{\text{cyl}} =dt^2 - d\sigma^2 - \sin^2 \sigma \, d\mathbf{n}^2=\frac{ds^2_{\text{flat}}}{\Omega^2(x)} \,,
    \qquad \Omega^2(x)=\frac{X_0^2}{4(\cos t+\cos\sigma)^2}\,,
\end{align}
and the lightcone coordinates are given by
\begin{align}
    \xn^{\pm} = 
    \frac{X_0}{2} \tan\left(\frac{t\pm\sigma}{2}\right)\,.
\end{align}
According to the definitions~\eqref{charge_detector} and~\eqref{energy_detector}, we see that setting $t^{\pm} = t \pm \sigma$ and using
\begin{equation}
  \frac{d x^-_{\mathbf{n}}}{dt^-}=\frac{X_0}{\left(2\cos\frac{t^-}{2}\right)^2}\,,\qquad
  \lim_{\xn^+\rightarrow\infty}\xn^+\Omega^{-1}(x)=2\cos\frac{t^-}{2}\,,
\end{equation}
the event shapes we are interested in are given by (momentarily neglecting corrections to the saddle-point)
\begin{equation}\label{eq_ev_shapes_cyl}
   \langle \mD_1(\mathbf{n}_1)\mD_2(\mathbf{n}_2)\dots \mD_k(\mathbf{n}_k)  \rangle\simeq X_0^{\sum_i(1-J_i)}\langle Q|\mD_1^{\text{cyl}}(\mathbf{n}_1)\mD_2^{\text{cyl}}(\mathbf{n}_2)\dots \mD_k^{\text{cyl}}(\mathbf{n}_k) |Q\rangle_{\text{cyl}}\,,
\end{equation}
where $J_i$ is the angular momentum of the local operator integrated in the detector and $\mD_j^{\text{cyl}}(\mathbf{n}_j)=X_0^{J-1}\mD_j(\mathbf{n}_j)$ denotes the corresponding light-ray operators on the cylinder~\cite{Kravchuk:2018htv}. For charge and energy detectors in the new coordinates these explicitly read
\begin{align}\label{Eq:ChargeOp}
    \mathcal{Q}^{\text{cyl}}(\mathbf{n})&=  \int_{- \pi}^{\pi} d t^- \left( \cos\frac{t^-}{2} \right)^{d-2}J_{-}(t^+=\pi,t^-, \mathbf{n})\,,\\
    \label{Eq:EnOp}
    \mathcal{E}^{\text{cyl}}(\mathbf{n})&=4 \int_{- \pi}^{\pi} d t^- \left(\cos\frac{t^-}{2} \right)^{d}T_{--}(t^+=\pi,t^-, \mathbf{n})\,,
    \end{align}
where $V_{-\ldots} =\frac12( V_{t\ldots}-V_{\sigma\ldots})$ as usual. We see that the light-ray integral is parametrized by $t^-\in [-\pi , \pi]$ with fixed $t^+ = \pi$ and $\mathbf{n}$ given by the angle on the celestial sphere~\cite{Hofman:2008ar}. This is illustrated on the right panel of Fig.~\ref{fig:LightrayIntegral}, where the detector spans the blue light-ray.

It is instructive to rederive the detector one-point functions~\eqref{eq_1pt_event} from~\eqref{eq_ev_shapes_cyl}. Using
\begin{equation}
    \langle Q|J_-|Q\rangle_{\cyl}=\frac{Q}{2\Omega_{d-1}}\,,\qquad
    \langle Q|T_{--}|Q\rangle_{\cyl}=
    \frac{d\Delta_Q}{4(d-1)\Omega_{d-1}}\,,
\end{equation}
from~\eqref{Eq:ChargeOp} and~\eqref{Eq:EnOp} we find
\begin{align}
    \langle \mathcal{Q}(\mathbf{n})\rangle &= \frac{Q}{2\Omega_{d-1}} \int_{- \pi}^{\pi} d t^- \left( \cos\frac{t^-}{2} \right)^{d-2} =\frac{Q}{\Omega_{d-2}}\,,\\
    \langle \mathcal{E}(\mathbf{n})\rangle &= X_0^{-1} \frac{d\Delta_Q}{(d-1)\Omega_{d-1}}\int_{- \pi}^{\pi} d t^- \left( \cos\frac{t^-}{2} \right)^{d}= \frac{E}{\Omega_{d-2}}\,,
\end{align}
where we used~\eqref{eq_X_def} in the last equation. Similarly, the factorization of event shapes follows immediately from~\eqref{eq_ev_shapes_cyl} and the semiclassical structure of the EFT predictions.

Let us finally discuss the corrections to the factorized result~\eqref{eq_Q_factorization_guess}. There are two sources of corrections: the subleading orders in the Fourier transform~\eqref{Eq:SaddlePointJJ}, and the connected contribution in~\eqref{eq_many_J_EFT}.

As for the two-point function, subleading corrections to the Fourier transform saddle-point are obtained integrating over real fluctuations $\delta x^\mu=x^\mu-X^\mu\sim O(\sqrt{\Delta_Q})$, and are thus $|\delta x|^2/|X|^2\sim 1/\Delta_Q\sim Q^{-\frac{d}{d-1}}$ suppressed. For the leading factorized term in the correlation function, this can be checked by performing the Fourier transform of the factorized term exactly as in~\cite{Firat:2023lbp}. In general, when considering subleading orders in the correlator~\eqref{eq_many_J_EFT}, the corrections to the saddle-point may be computed in terms of correlators with insertion of suitable conformal generators using
\begin{align}\label{Eq:CorrSaddle}
    \mO_Q\left(\frac{i}{2}X+\delta x\right) = \mO_Q\left(\frac{i}{2}X\right) + i\delta x^\mu \left[P_\mu,\mO_Q\left(\frac{i}{2} X\right)\right] + \ldots\,,
\end{align}
and expressing $P^\mu$ in terms of the natural conformal generators on the cylinder. 

The other correction arises integrating the connected correlator as
\begin{equation}\label{eq_JJ_connected_contribution}
   \int_{- \pi}^{\pi} d t^-_1 \left( \cos\frac{t^-_1}{2}\right)^{d-2} \int_{- \pi}^{\pi} d t^-_2 \left( \cos\frac{t^-_2}{2}\right)^{d-2} \langle Q|J_{-}(t_1^+=\pi,t_1^-,\mathbf{n}_1)J_{-}(t_2^+=\pi,t_2^-,\mathbf{n}_2)
   |Q\rangle_{\cyl}^{\text{conn}}.
\end{equation}
The connected correlator is predicted by EFT and is $\sim \mu^{-d}\sim Q^{-\frac{d}{d-1}}$ smaller than the leading order as long as the distance between the two operators is larger than the short-distance cutoff $1/\mu\sim Q^{-\frac{1}{d-1}}$. Explicitly, using~\eqref{current}, at leading order in the derivative expansion it is given by 
\begin{equation}\label{eq_current2pt_conn}
\begin{split}
\langle Q|J_{-}(t_1,\hat{m}_1)J_{-}(t_2,\hat{m}_2)|Q\rangle_{\cyl}^{(\text{conn})} &= \frac{c\, \mu^{d-2}}{4(d-1)} \prod_{i=1}^{2}\left[(d-1)\frac{\pd}{\pd t_i}-\frac{\pd}{\pd \sigma_i}\right]  G_{\pi\pi}\left(t_1-t_2,\hat{m}_1\cdot\hat{m}_2\right) \,,
\end{split}
\end{equation}
where the propagator is defined in \eqref{Eq:WightCyl} and~\eqref{Eq:WightmanCyl}. 

Therefore we need to check that the integral~\eqref{eq_JJ_connected_contribution} is not dominated by regions in which the two operators are at distances (either in space or in time) of the order of the inverse cutoff or shorter. This immediately implies that EFT only predicts the event shape for
\begin{equation}\label{eq_theta_EFT}
\theta\equiv\arccos(\mathbf{n}_1\cdot\mathbf{n}_2)\gg \mu^{-1}\,.
\end{equation}
In particular, we cannot access the light-ray OPE regime $\theta\rightarrow 0$ of the underling theory.

For generic $\mathbf{n}_1\cdot\mathbf{n}_2$ such that~\eqref{eq_theta_EFT} holds, EFT predicts the correlator everywhere within the integral but near the endpoints $t^-_1\simeq t_2^-\simeq\pm\pi$, where the operators collide and the derivative expansion breaks down. Note also that the potentially dangerous configurations $t_{1}^- = -t_2^- = \pm \pi$, where the operators become light-like separated, are safe from the EFT viewpoint since their distance lies outside the sound-cone. 

To estimate the contribution with the two operators near the same endpoint, we introduce a small cutoff at $\pm\pi \mp t_1^-=\pm\pi\mp t_2^-\sim 1/\mu$ and estimate the contribution from the dangerous region. Near $t_1^-=t_2^-=\pi$ the measure contributes as $\sim [dt^- (\pi-t^-)^{d-2}]^2\sim\mu^{-2(d-1)} $, while using the short-distance limit of the propagator the connected correlator scales as $\pd^2G_{\pi\pi}\sim \pd^2 (\pi-t^-)^{-(d-2)}\sim\mu^d$, and identically near $t_1^-=t_2^-=-\pi$. Putting these together, and accounting for the $\mu^{d-2}$ upfront~\eqref{eq_current2pt_conn}, we find
\begin{equation}\label{eq_nonEFT_JJ}
    \text{non-EFT contribution}\sim \mu^{d-2}\times \mu^{-2(d-1)}\times\mu^{d}=\mu^0\,.
\end{equation}
Therefore we conclude that the contribution from the region of the integral in which the derivative expansion breaks down is subleading not only compared to the factorized term~\eqref{eq_Q_factorization_guess} $\sim \mu^{2(d-1)}$, but also compared to the leading connected contribution $\sim \mu^{d-2}$ that arises integrating~\eqref{eq_JJ_connected_contribution}. A similar argument shows that the non-EFT contribution for EEC is smaller by a factor of $\mu^{-2(d+1)}$ compared to the leading factorized result. Note that these arguments do not fix the functional form of the non-EFT corrections, that is in general UV-dependent.

In conclusion, EFT reliably predicts charge and energy correlators, for celestial angular distances such that~\eqref{eq_theta_EFT} holds, at least up to relative order of, respectively, $\mu^{-2(d-1)}$ and $\mu^{-2(d+1)}$. In particular, EFT controls both the leading factorized result as well as the first nontrivial connected correction. It is simple to see that the relative scaling between these two is of order $\sim Q^{-\frac{d}{d-1}}\sim 1/\Delta_Q$ (accounting both for the corrections to the Fourier transform saddle-point and the connected correlator), in agreement with the naive estimate of quantum effects from the classical action $S_{class.}\sim\Delta_Q\equiv 1/\hbar_{eff.}$. 

Few comments are in order. First, as remarked before, in flat space the breakdown of EFT at the integration endpoints is associated with the dilution of the charge flux at infinity. It is remarkable that conformal invariance ensures that such breakdown of EFT can be associated with standard short-distance effects, as the map to the cylinder makes clear.

The fact that EFT only fails in a small region suggests that we might still be able to systematically compute event shapes to arbitrary subleading order, provided we include suitable additional local terms that parametrize our ignorance about the short-distance effects at $t_1^-=t_2^-=\pm \pi$. In Sec.~\ref{sec_EFT_singular} we will argue that this is indeed the case, where we will constrain such additional terms based on conformal invariance. This construction will also provide a different argument for the existence of UV contributions scaling as~\eqref{eq_nonEFT_JJ}.

Finally, as argued in~\cite{Chicherin:2023gxt}, it should be clear that the details of the large charge EFT are largely irrelevant for the factorization of event shapes. This is a general property of arbitrary \emph{semiclassical} operators, such as the large charge operators studied here, but including also, for instance, generic heavy operators, that correspond to thermal states on the cylinder~\cite{Lashkari:2016vgj,Delacretaz:2020nit}. Instead, the connected part of event shapes carries nontrivial information about the specific dynamics of the state under consideration. In the next section, we will compute the leading connected contribution of CCC and EEC for large charge operators.

\section{Energy and charge correlators}\label{Sec:EEC}

In the previous section, we discussed how to systematically compute event shapes in a large charge expansion and how they factorize at leading order. In particular, focusing, for instance, on the CCC, we argued that it admits an expansion of the form
\begin{align}\label{eq_CC_exp}
    \langle \Qno \Qnt \rangle = \langle \Qno \Qnt \rangle^{(0)}+
    \frac{1}{\Delta_Q}\langle \Qno \Qnt \rangle^{(1)}+\ldots\,,
\end{align}
where the leading result completely factorizes in terms of one-point event shapes:
\begin{align}\label{Eq:CCLO}
    \langle \Qno \Qnt \rangle^{(0)} =  \langle \Qno \rangle \langle \Qnt \rangle=\left(\frac{Q}{\Omega_{d-2}}\right)^2\,.
\end{align}
An analogous expansion holds for EEC.

In this section, we go beyond the classical approximation and present the calculation of the first correction $\langle \Qno \Qnt \rangle^{(1)}$ and $\langle \Eno \Ent \rangle^{(1)}$ to CCC and EEC. This provides the specific predictions of the superfluid EFT for the correlations in the energy and charge distributions.

As we already explained these corrections are made of two contributions
\begin{align}\label{Eq:ContQQ}
    \langle \Qno \Qnt \rangle^{(1)} =  \langle \Qno \Qnt \rangle^{(1)}_{\text{disc}}+\langle \Qno \Qnt \rangle^{(1)}_{\text{conn}} \,.
\end{align}
The first term, the disconnected contribution, arises from the corrections to the saddle point in the Fourier transform, discussed around \eqref{Eq:CorrSaddle}. This can be computed as in \cite{Firat:2023lbp}; we review the calculation in App.~\ref{App:Disconnected}. The result in the source/sink rest frame reads
\begin{align}\label{Eq:DiscCC}
    \frac{\langle \Qno \Qnt \rangle^{(1)}_{\text{disc}}}{\langle \Qno\rangle \langle \Qnt\rangle} &= - \frac{(d-2)^2}{2} \mathbf{n_1} \cdot \mathbf{n}_2\,, \\ \label{Eq:DiscEE}
    \frac{\langle \Eno \Ent \rangle^{(1)}_{\text{disc}}}{\langle \Eno\rangle \langle \Ent\rangle} &=- \frac{1+(d-1)^2\mathbf{n}_1\cdot \mathbf{n}_2}{2}\,.
\end{align}
These functions are graphically shown by the dashed lines in Fig.~\ref{fig:QQandEECorr} in 3 dimensions.

It is insightful to interpret the above equations from a physical perspective. Note that the disconnected term of the correlator~\eqref{eq_many_J_EFT} is the same both in a free and in an interacting theory. Therefore, for the sake of discussing the disconnected contribution, we can picture the large-charge state as a vast collection of free, massless particles propagating at the speed of light. In the limit of an infinite number of particles, we naturally expect a completely isotropic and dense distribution of energy and charge, as described by the leading-order result. 
At finite charge, the free particles still attempt to uniformly cover the celestial sphere, on average, but it becomes less likely to populate arbitrarily small angular regions, due to the smaller phase space available. As a result, the correlation displays an excess at large angles and a deficit at small angles, which is precisely what \eqref{Eq:DiscCC} describes.

The connected contribution is more interesting since it encodes the non-trivial dynamics of the large charge state, controlled by the fluctuations $\pi$ in the Lagrangian \eqref{eq_cyl_action_fluct}. From~\eqref{Eq:ChargeOp} and~\eqref{Eq:EnOp} we find that the connected contributions to CCC and EEC are written in terms of the EFT two-point functions as
\begin{align}\label{Eq:CCCyl}
   \frac{\langle \Qno \Qnt \rangle^{(1)}_{\text{conn}}}{\Delta_Q} &= \int_{- \pi}^\pi dt^-_1\int_{- \pi}^\pi dt^-_2 \left( \cos\frac{t^-_1}{2}\cos\frac{t^-_2}{2}\right)^{d-2}\langle Q| J_-(y_1)J_-(y_2) |Q\rangle_{\cyl}^{\text{conn}} \,, \\ \label{Eq:EECyl}
   \frac{\langle \Eno \Ent \rangle^{(1)}_{\text{conn}}}{\Delta_Q} &= \frac{16 }{X^2_0} \int_{- \pi}^\pi dt^-_1 \int_{- \pi}^\pi dt^-_2 \left( \cos\frac{t^-_1}{2}\cos\frac{t^-_2}{2}\right)^{d}\langle Q| T_{--} (y_1)T_{--} (y_2) |Q\rangle_{\cyl}^{\text{conn}} \,,
\end{align}
where $y_a=(t^+_a=\pi,t^-_a,\mathbf{n}_a)$ collectively denotes the operator coordinates.
Expanding the current and the energy-momentum tensor in terms of the goldstone field $\pi$ using~\eqref{current} and~\eqref{emtensor}, at leading order \eqref{Eq:CCCyl} and~\eqref{Eq:EECyl} are given by
\begin{align} \label{eq_QQ_conn}
     \frac{\langle \Qno \Qnt \rangle^{(1)}_{\text{conn}}}{\langle \Qno\rangle\langle \Qnt \rangle}&=
     \frac{\pi^{\frac{d-2}{2}} \Gamma \left(\frac{d+2}{2}\right)}{d^2 \Gamma \left(\frac{d-1}{2}\right)^2}
     \iint dt^-_1 dt^-_2 
     \left( \cos\frac{t^-_1}{2}\cos\frac{t^-_2}{2}\right)^{d-2}\hspace{-.6em}
     \prod_{i=1,2}\left[(d-1)\partial_{t_i}-\partial_{\sigma_i}\right]G_{\pi \pi}\,,\\ \label{eq_EE_conn}
     \frac{\langle \Eno \Ent \rangle^{(1)}_{\text{conn}}}{\langle \Eno\rangle\langle \Ent \rangle}&=
     \frac{d \,\pi^{\frac{d-2}{2}} \Gamma \left(\frac{d}{2}\right)}{8\, \Gamma \left(\frac{d+1}{2}\right)^2}
     \iint dt^-_1 dt^-_2 \left( \cos\frac{t^-_1}{2}\cos\frac{t^-_2}{2}\right)^{d}\prod_{i=1,2}\left(d\,\partial_{t_i}-2\partial_{\sigma_i}\right)G_{\pi \pi}\,,
\end{align}
where $G_{\pi \pi}$ is the Wightman 2-point function~\eqref{Eq:WightCyl} evaluated on the light-rays, i.e. for
\begin{align}\label{eq_pos_light_ray}
t_1-t_2 = \frac{t_1^- -t_2^-}{2}\,, \qquad \hat{m}_1\cdot \hat{m}_2 = \cos\left(\frac{t_1^-}{2}\right)\cos\left(\frac{t_2^-}{2}\right)\cos\theta+\sin\left(\frac{t_1^-}{2}\right)\sin\left(\frac{t_2^-}{2}\right)\,,
\end{align}
where $\cos\theta=\mathbf{n}_1\cdot \mathbf{n}_2$. Higher derivative corrections are further $1/\mu^2\sim 1/Q^{\frac{2}{d-2}}$ suppressed.

Therefore, all we have to do is to perform the integrations in~\eqref{eq_QQ_conn} and~\eqref{eq_EE_conn}. In practice, however, $G_{\pi \pi}$ is too complicated to be handled analytically for arbitrary angles. In fact, we shall also see that the obvious expansion for the propagator \eqref{Eq:WightmanCyl} is not adequate in most regimes, and one has to find alternative representations.

In the rest of the section, we compute the EEC and CCC (semi-)analytically for specific angle configurations and numerically for generic angles. For the sake of concreteness, we will only provide explicit numerical results in $d=3$ spacetime dimensions, but we will nonetheless present all derivations and formulas for general $d$.

\subsection{Collinear limit}\label{Sec:CollLimit}

Here we consider the collinear limit, i.e. $1\gg \theta \gtrsim 1/\mu$. In this limit, the dominant contribution to the light-ray integrals arises from the region in which the two operators lie close to each other: $|t_1^--t_2^-|\lesssim \theta$. This region is controlled by the short-distance expansion of the propagator, which holds for $|t_1-t_2|\sim \arccos(\hat{m}_1\cdot\hat{m}_2)\ll 1$ and is just given by the flat-space result
\begin{equation}
\begin{gathered}\label{Eq:PropSmallAngle}
G_{\pi\pi}(t_1-t_2,\hat{m}_1\cdot \hat{m}_2) \quad\underset{\hat{m}_1\rightarrow \hat{m}_2}{\overset{t_1\rightarrow t_2}{\longrightarrow}} \quad G_{\pi \pi}^{\text{(short)}}(t_1-t_2,\hat{m}_1\cdot \hat{m}_2)\,,\\
    G_{\pi\pi}^{(\text{short})}(t,\cos \sigma) =\frac{(d-1)^{\frac{d-1}{2}}}{(d-2)\Omega_{d-1}\left[-(t-i \epsilon)^2+(d-1)\,\sigma^2\right]^{\frac{d-2}{2}}}\,,
\end{gathered}
\end{equation}
described in detail in App.~\ref{Sec:ShortDistanceProp} and~\ref{App:AddColl}, where we also provide the first subleading term to~\eqref{Eq:PropSmallAngle} in $d=3$. The Goldstone propagator is as singular as the usual free massless propagator in the small-distance limit, the main difference being the sound-speed $c_s^2=1/(d-1)$ factor between the time and angular distances in the denominator.

Evaluating the event shapes within this approximation is quite straightforward since we just need to express the local correlators in terms of the short distance propagator \eqref{Eq:PropSmallAngle} and integrate over the light-rays. The qualitative features of the result are easily understood. Accounting for the two-derivatives in~\eqref{eq_QQ_conn} and~\eqref{eq_EE_conn}, the integrand $\sim \pd^2G_{\pi\pi}$ grows as $1/(\text{distance})^d$. Therefore, integrating over the relative position we conclude that the connected correlator scales as $\sim 1/\theta^{d-1}$ in the collinear regime.

For ease of presentation, in the following we only describe the final results, focusing for concreteness in $3$ dimensions. In App.~\ref{App:AddColl} we collect the technical details of the calculation.

In the frame of interest ($p = (E,\mathbf{0})$) the only nontrivial dependence is given by the angle between the detectors $\cos\theta \equiv \mathbf{n}_1 \cdot \mathbf{n}_2$. The collinear limit of EEC and CCC read
\begin{align}\label{Eq:ShortDistanceQQ}
   \langle\Qno\Qnt\rangle^{(1)}_{\text{conn}} &=- \left(\frac{Q}{2 \pi}\right)^2\left(\frac{\pi }{3}\text{Pf}\left[ \frac{1}{\theta^2} \right]+\frac{\pi}{24} \log \left(\theta ^2\right)+O\left(\theta^0\right)\right)\,,\\ 
   \label{Eq:ShortDistanceEE}
   \langle\Eno \Ent \rangle^{(1)}_{\text{conn}} &=- \left( \frac{E}{2\pi}\right)^2\left(\frac{9 \pi }{32 }
   \text{Pf}\left[ \frac{1}{\theta^2} \right]
   +\frac{93 \pi}{256}   \log \left(\theta ^2\right)+O\left(\theta^0\right)
   \right)\,.
\end{align}
The symbol $\text{Pf}$ denotes that the $1/\theta^2$ terms have to be interpreted as finite part (``Part finie'') distribution.\footnote{The finite part distribution is defined such that, when we integrate $\text{Pf}\left[ \frac{1}{\theta^2} \right]$ against a test function $f(\theta)$ which is smooth around $\theta=0$, we have
\begin{align}\label{eq_Pf_def}
    \int_{-A}^{A} d\theta \, \text{Pf}\left[\frac{1}{\theta^2}\right] f(\theta) \equiv  \lim_{\epsilon\rightarrow 0^+}\left[\int_{\epsilon}^{A} d\theta  \frac{f(\theta)}{\theta^2} +
    \int^{-\epsilon}_{-A} d\theta  \frac{f(\theta)}{\theta^2}-2\frac{f(0)}{\epsilon}\right]\,,
\end{align}
for arbitrary $A>0$. Similar distributions commonly arise in EEC calculations in theories with nontrivial mass scales, see e.g.~\cite{Ebert:2018gsn}.} As we explain in App.~\ref{App:AddColl}, this distributional result follows from the $i \epsilon$ prescription in \eqref{Eq:PropSmallAngle} which regulates the propagator at coincident points. This prescription crucially ensures that integrating~\eqref{Eq:ShortDistanceQQ} and~\eqref{Eq:ShortDistanceEE} we obtain a finite result, as required by Ward identities. Indeed, charge conservation implies that the total integral of the CCC is fixed,
\begin{align}\label{Eq:CCons}
  \int d^{d-2}  \Omega_{\mathbf{n_1}} \langle \Qno \Qnt \rangle= Q\langle  \Qnt\rangle \,, 
\end{align}
while energy and momentum conservation imply the following constraints on the EEC:
\begin{align}\label{Eq:ECons}
\int d^{d-2}  \Omega_{\mathbf{n_1}}\langle \Eno \Ent \rangle= E\,\langle  \Ent\rangle\,,&&
  \int d^{d-2}  \Omega_{\mathbf{n_1}} \mathbf{n_1}\langle \Eno \Ent\rangle=0\,.
\end{align}
These relations are completely saturated by the leading-order results. Since the disconnected terms~\eqref{Eq:DiscCC} and~\eqref{Eq:DiscEE} are manifestly finite everywhere, it is crucial that the finite part distribution makes the integration of the connected terms~\eqref{Eq:ShortDistanceQQ} and~\eqref{Eq:ShortDistanceEE} finite too. More details on how~\eqref{Eq:CCons} and~\eqref{Eq:ECons} are realized in our next-to-leading order calculation are reported in App.~\ref{App:Conservations}.

To obtain~(\ref{Eq:ShortDistanceQQ},~\ref{Eq:ShortDistanceEE}) we used the  short-distance expansion of the propagator to the first subleading order. Note that this strategy only predicts the singular terms at small angle, as the $O(1)$ part also receives seizable contributions from the integration region in which the two operators are well separated, and the full propagator has to be retained to compute it.

Before discussing the qualitative features of (\ref{Eq:ShortDistanceQQ},~\ref{Eq:ShortDistanceEE}) we stress again that EFT is valid for distances larger than the cutoff of the EFT: 
$ (t_1 - t_2)^2\gg 1/\mu^2 $, $ (\arccos \hat{m}_1 \cdot \hat{m}_2)^2\gg 1/\mu^2$. 
As the integral over the light-rays for energy and charge correlators is dominated by configurations where 
$ (t_1 - t_2)^2 \sim (\arccos \hat{m}_1 \cdot \hat{m}_2)^2\sim \theta^2 $, 
we can effectively trust our calculation only as long as $ \theta \gg 1/\mu $, as we mentioned in the previous section.

\begin{figure}[t]
    \centering  
    \begin{minipage}{0.45\textwidth}
    \centering
    \includegraphics[width=1\linewidth]{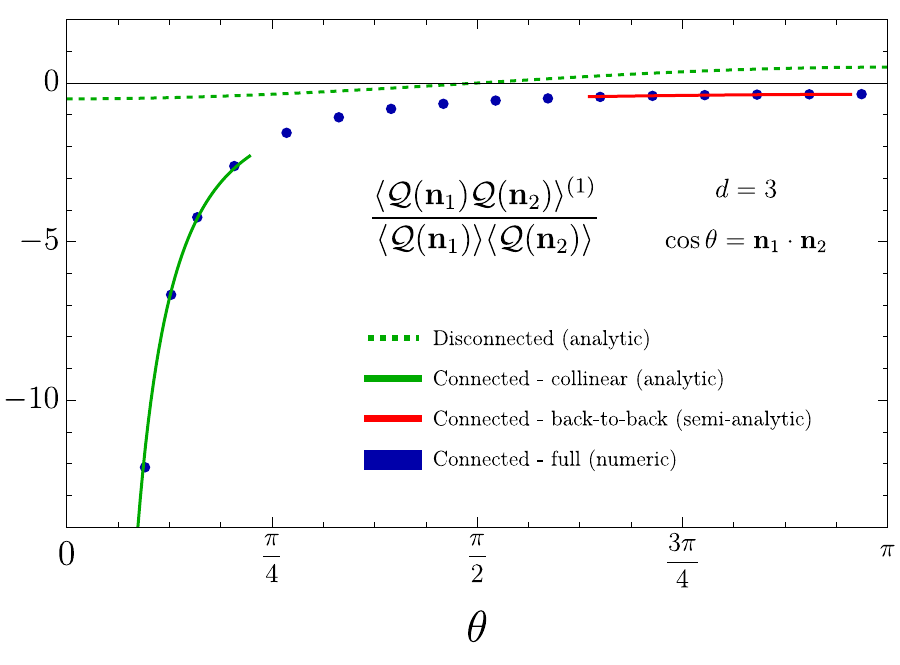}
    \end{minipage}\hspace{2em}
    \begin{minipage}{0.45\textwidth}
    \centering
    \includegraphics[width=1\linewidth]{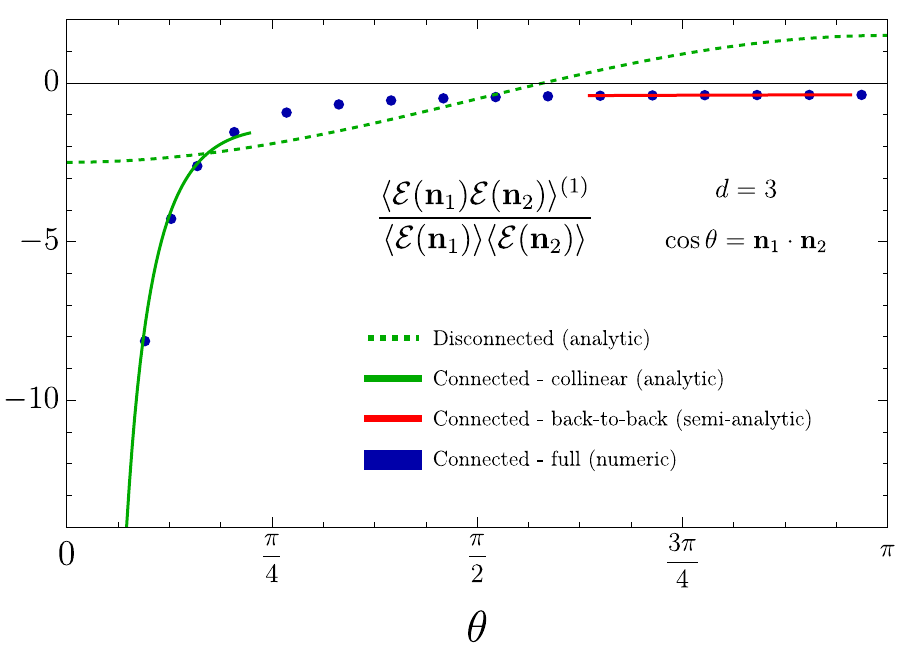}
    \end{minipage}
    \caption{Next-to-leading order charge-charge (left) and energy-energy (right) correlators as functions of the angular distance $\theta$ between the detectors. The plots show the results in $d=3$ dimensions; we added a constant shift to the collinear-limit of the connected contribution (solid green line) to improve the fit. Details on the various contributions are explained in the main text.}
    \label{fig:QQandEECorr}
\end{figure}

The results in~(\ref{Eq:ShortDistanceQQ},~\ref{Eq:ShortDistanceEE}) are shown by the continuous green line in Fig.~\ref{fig:QQandEECorr}. Remarkably, we observe that the EFT corrections predict a deficit of energy and charge over a large range \(\theta \gtrsim \delta/\mu\), with \(\delta\) sufficiently large to ensure the validity of the EFT. These corrections become more pronounced at small angles, as we approach the breakdown of the EFT. The total integral of energy and charge is fixed, implying the presence of an excess of energy and charge in a small region \(\theta \lesssim 1/\mu\), formally encoded in the finite part distribution. This phenomenon is sort of analogous to jet formation in QCD. However, the emergence of these \emph{sound-jets} is entirely governed by the sound mode of the superfluid EFT. On the other hand, the detailed structure of the energy flow inside these sound-jets lies beyond the regime of validity of the EFT.

Let us also present for generality the collinear results in arbitrary spacetime dimensions
\begin{align}\label{Eq:ShortDistanceQQ_anyd}
   \langle\Qno\Qnt\rangle^{(1)}_{\text{conn}} &\stackrel{\theta\neq 0}{=}- \left(\frac{Q}{\Omega_{d-2}}\right)^2\left[
   \frac{\sqrt{\pi } \,2^{4-d}  \Gamma \left(\frac{d}{2}+1\right) \Gamma (d)}{\sqrt{d-2} \,d^2 \Gamma \left(\frac{d-1}{2}\right)^3}\theta^{1-d}
   +O\left(\text{max}(1,\theta^{3-d})\right)\right]\,,\\ 
   \langle\Eno \Ent \rangle^{(1)}_{\text{conn}} &\stackrel{\theta\neq 0}{=}- \left( \frac{E}{\Omega_{d-2}}\right)^2\left[
   \frac{\sqrt{d-2} \, \Gamma \left(\frac{d}{2}+1\right)^2}{\Gamma \left(\frac{d+1}{2}\right) \Gamma \left(\frac{d+3}{2}\right)}\theta^{1-d}
   +O\left(\text{max}(1,\theta^{3-d})\right)
   \right]\,,
   \label{Eq:ShortDistanceEE_anyd}
\end{align}
where we focused on the leading singular term.
Comments identical to the ones above apply in this case; in particular these results only apply for $\theta\neq 0$ and imply a large concentration of charge/energy at $\theta=0$ by Ward identities.

\subsection{Back-to-back limit}\label{Sec:Backtoabcklimit}

In principle, all we have to do to compute the connected part of the event shape in general is to integrate the full propagator in~\eqref{eq_QQ_conn} and~\eqref{eq_EE_conn}. To this aim, the simplest approach seems to use the s-channel decomposition in~\eqref{Eq:WightmanCyl}, equivalent to the (detector)-(source) OPE.

This gives results in the form of a sum over $\ell$:
\begin{align}\label{eq_QQ_back}
\frac{\langle \Qno \Qnt \rangle^{(1)}_{\text{conn}}}{\langle \Qno\rangle \langle \Qnt\rangle}=
\sum_{\ell=1}^{\infty}F^{(\ell)}_{\mathcal{Q}\mathcal{Q}}(\cos\theta) \,,\qquad
\frac{\langle \Eno \Ent \rangle^{(1)}_{\text{conn}}}{\langle\Eno\rangle \langle\Ent\rangle}= \sum_{\ell=1}^{\infty}F^{(\ell)}_{\mathcal{E}\mathcal{E}}(\cos\theta)\,,
\end{align}
where 
\begin{align}
&\begin{aligned}
\label{eq_Fl_QQ}
  F^{(\ell)}_{\mathcal{Q}\mathcal{Q}}(\cos\theta)&=\frac{ \Gamma \left(\frac{d}{2}\right) \Gamma \left(\frac{d+2}{2}\right)(2 \ell+d-2)}{2 \pi  (d-2) d^2 \Gamma \left(\frac{d-1}{2}\right)^22\omega_{\ell}} \\
  &\times \iint dt_1^- dt_2^- \left(\cos\frac{t_1^-}{2}\cos\frac{t_2^-}{2}\right)^{d-2}\prod_{i=1,2}\left[(d-2)\,\partial_{t_i}-\partial_{\sigma_i}\right] e^{-i\omega_{\ell}t}C^{\left(\frac{d-2}{2}\right)}_{\ell}(\hat{m}_1\cdot\hat{m}_2)\,,
  \end{aligned} \\
  &\begin{aligned}
  \label{eq_Fl_EE}
  F^{(\ell)}_{\mathcal{E}\mathcal{E}}(\cos\theta)&=\frac{d \, \Gamma \left(\frac{d}{2}\right)^2(2\ell+d-2)}{16 \pi  (d-2) \Gamma \left(\frac{d+1}{2}\right)^22\omega_{\ell}} \\
  &\times\iint dt_1^- dt_2^-\left(\cos\frac{t_1^-}{2}\cos\frac{t_2^-}{2}\right)^d\prod_{i=1,2}\left[d\,\partial_{t_i}-2\,\partial_{\sigma_i}\right] e^{-i\omega_{\ell}t}C^{\left(\frac{d-2}{2}\right)}_{\ell}(\hat{m}_1\cdot\hat{m}_2)\,.
  \end{aligned}
\end{align}
The integrands may be written more explicitly upon taking the derivatives and evaluating the propagator at~\eqref{eq_pos_light_ray}. The resulting expressions are somewhat lengthy and we shall not report them in the main text.

For any given value of $\ell$, the integrals~\eqref{eq_Fl_QQ} and~\eqref{eq_Fl_EE} can be reduced to a sum of elementary integrals upon explicitly expanding the Gegenbauer polynomials. However, we were not able to obtain closed form results for arbitrary $\ell$. What's worst, while in principle~\eqref{eq_Fl_QQ} and~\eqref{eq_Fl_EE} yield the desired result, we observed that the sums over $\ell$ do not converge \emph{absolutely} for generic values of $\theta$; rather, the summands oscillate in a non-trivial way, and even grow with $\ell$ in absolute value for $\theta\lesssim \pi/2$.
These complications make it challenging, at best, to perform the calculation for arbitrary angles using the representation~\eqref{Eq:WightmanCyl}. 

The technical reason why the sums~\eqref{eq_Fl_QQ} and~\eqref{eq_Fl_EE} are oscillating is easy to identify. The propagator is a sum of factors that oscillate in time, each corresponding to the contribution of a given intermediate \emph{phonon} state, and the operators are integrated over various time distances. As discussed in~\cite{Kologlu:2019bco}, this implies that the conformal block decomposition in general does not converge in absolute sense, and one cannot truncate the sums~\eqref{eq_Fl_QQ} and~\eqref{eq_Fl_EE} unless the correlator is smeared against suitable test functions.

In the next section we shall provide a simple strategy to overcome these complications.
For now, we empirically observe that for $\theta$ not too far away from $\pi$ the integration of the different terms in the Gegenbauer polynomials gives rise to a destructive interference, such that $F^{(\ell)}_{\mathcal{Q}\mathcal{Q}}$ and $F^{(\ell)}_{\mathcal{E}\mathcal{E}}$ rapidly decrease in absolute value with $\ell$.
While providing a precise estimate of the convergence properties in \(\ell\) is rather complicated, in this section, we adopt a more pragmatic approach and simply use the representation in~\eqref{Eq:WightmanCyl}, truncated to some large value $\ell_{max}$, to compute EEC and CCC in the back-to-back limit. 

Therefore we Taylor expand~\eqref{eq_QQ_back} around $\theta=\pi$. In $d=3$ we find
\begin{align} \nonumber
\frac{\langle \Qno \Qnt \rangle^{(1)}_{\text{conn}}}{\langle \Qno\rangle \langle \Qnt \rangle}&=
\sum_{\ell=1}^{\infty}
F^{(\ell)}_{\mathcal{Q}\mathcal{Q}}(-1)+
\frac12(\pi-\theta)^2\sum_{\ell=1}^{\infty}
F^{(\ell)}_{\mathcal{Q}\mathcal{Q}}\,'(-1)
+O\left((\pi-\theta)^4\right)\\
&\approx 
-0.355-0.058 \,(\pi -\theta)^2+\ldots\,,
\label{eq_back_QQ}
\\
\nonumber
\frac{\langle \Eno \Ent \rangle^{(1)}_{\text{conn}}}{\langle \Eno \rangle\langle \Ent \rangle}&=
\sum_{\ell=1}^{\infty}
F^{(\ell)}_{\mathcal{E}\mathcal{E}}(-1)+
\frac12(\pi-\theta)^2\sum_{\ell=1}^{\infty}
F^{(\ell)}_{\mathcal{E}\mathcal{E}}\,'(-1)
+O\left((\pi-\theta)^4\right)\\
&\approx -0.374-0.018\, (\pi-\theta )^2+\ldots\,.
\label{eq_back_EE}
\end{align}
Some details of the calculations are given in App.~\ref{App:BacktoBack}. We see that this approach offers a semi-analytic estimate of the event shapes in the back-to-back region. These results, tested against the controlled numerical calculation presented in the next section, are shown by the red lines in Fig.~\ref{fig:QQandEECorr}.

Finally we comment that the conformal block decomposition, while it does not converge in absolute value pointwise for event shapes, does provide a convenient scheme to compute the various components of the Fourier transform of CCC and EEC. We discuss some details in App.~\ref{app_Fourier_dec}, where we also show that all the Fourier coefficients are positive as required by unitarity.\footnote{We thank A. Zhiboedov for useful comments on this point.}

\subsection{General numerical approach}\label{Sec:Num}

In  event shapes, the light-rays are always spacelike, both in the conventional sense and, more importantly for us, according to the EFT sound-cone
\begin{equation}
    (d-1)| \sigma_{12} |^2 >  |t_1 - t_2|^2.
\end{equation}
where $\sigma_{12}$ is the angular distance on $S^{d-1}$: $\cos \sigma_{12} = \hat{m}_1\cdot \hat{m}_2$.
As a result, the propagator in the event shape integrals generally describes the exchange of an \emph{off-shell} phonon. Off-shell quanta generate a potential that decays with distance. This feature, though not immediately apparent in the decomposition~\eqref{Eq:WightmanCyl}, becomes explicit when using an alternative representation of the propagator.  

This property is particularly manifest when decomposing the propagator in a different way, as explained in detail in App.~\ref{app_prop_dS}. Let us start from the Euclidean cylinder. The main idea is that, to find the propagator on the cylinder $\mathbb{R} \times S^{d-1}$, we perform a Fourier transform on $\mathbb{R}$
\begin{equation}\label{eq_prop_dS}
    G^E_{\pi \pi}(\tau, \cos\sigma) = \int_{-\infty}^{\infty} \frac{d \omega}{2 \pi} e^{i \omega \tau} F_{\omega}(\sigma)\,,
\end{equation}
and find $F_{\omega}$ by solving the equation of motion. As a result, we find that $F_{\omega}$ is the propagator on the sphere $S^{d-1}$ for a particle of mass $m^2 = (d-1) \omega^2$, see App.~\ref{app_prop_dS} for an explicit expression.\footnote{More precisely,~\eqref{eq_prop_dS} is correct up to the contribution of the zero-mode that we negleceted. This anyhow drops out once we take derivatives of the propagator, and is hence irrelevant for our purposes.} Importantly for us, $F_{\omega}$ decays exponentially for large $\omega$
\begin{equation}
    F_{\omega}(\sigma)\longrightarrow F_{\omega}^{(\text{as})}(\sigma) \sim e^{- |\omega| \sqrt{d-1}\, |\sigma|}\,,
\end{equation}
and thus, for a finite angular distance $\sigma$, we can truncate the integral in \eqref{eq_prop_dS}. 

Analogously to the relationship between radial quantization and the representation  \eqref{Eq:WightmanCyl},  \eqref{eq_prop_dS} naturally arises from quantizing the theory in $\mathbb{R} \times dS^{d-1}$ and then rotating it back to Euclidean signature in $\mathbb{R} \times S^{d-1}$.\footnote{Note however that primary states have complex momentum $i\Delta$ along $\mathbb{R}$ in $\mathbb{R}\times dS_{d-1}$, and thus the notion of unitarity is not obvious in these coordinates. Therefore, while the decomposition~\eqref{eq_prop_dS} is convenient for our purposes, we do not claim that there exists a non-perturbatively defined Hilbert space associated with such quantization procedure in a general CFT. In particular, despite some similarities, this procedure is different from angular or Rindler quantization~\cite{Agia:2022srj}, first explored in the context of event shapes in~\cite{Fox:1978vu}. } This decomposition mirrors the approach of modifying the integration contour to complex spatial momenta when computing the Fourier transform of a flat-space propagator at space-like distance. Similar complexifications of momentum are also useful in studying scattering phase shifts in the Regge limit as a function of the impact parameter~\cite{Camanho:2014apa,Kologlu:2019bco}.

We are interested in the Wightman function on the Lorentzian cylinder. This is obtained from \eqref{eq_prop_dS} by Wick-rotating the time $\tau\rightarrow i(t - i \epsilon)$. As long as we consider points that are outside the sound-cone, the integrand in \eqref{eq_prop_dS} remains exponentially suppressed for large $\omega$:
\begin{align}\label{Eq:AsDsProp}
    e^{\omega t}F_{\omega}^{(\text{as})} (\sigma) \sim e^{- |\omega|( \sqrt{d-1}\, |\sigma| -|t|)}\ll 1\,, \quad\text{for}\quad\sqrt{d-1}|\sigma| >|t|\,.
\end{align}
For any finite distance we may thus truncate the integration with a cutoff, and define an approximate propagator as
\begin{align}\label{Eq:PropApprds}
    G_{\pi \pi}^{\Lambda}(t, \cos\sigma) \simeq \int^{\Lambda}_{-\Lambda} \frac{d \omega}{2 \pi} e^{ \omega t} F_{\omega}(\sigma)+
 \int^{\infty}_{\Lambda} \frac{d \omega}{2 \pi} (e^{ \omega t}+e^{-\omega t})F_{\omega}^{(\text{as})}(\sigma)\,,
\end{align}
where the second term in this equation is evaluated analytically using the asymptotic expansion at large $\omega|\sigma|$ of the sphere propagator, see App.~\ref{app_prop_dS}. For any finite $\Lambda$, \eqref{Eq:PropApprds} is straightforward to evaluate numerically and is thus well suited for numerical integration. Derivatives of the propagator, which are needed to compute \eqref{eq_QQ_conn} and \eqref{eq_EE_conn}, can be approximated similarly.

We could now use~\eqref{Eq:PropApprds} to compute the event shapes~\eqref{eq_QQ_conn} and~\eqref{eq_EE_conn} numerically at arbitrary angular separation. However, near the endpoints $t_1^+=t_2^+=\pm \pi$ of the light-ray integrals the two operator collide and the truncated representation of the propagator~\eqref{Eq:PropApprds} is not adequate anymore. Therefore, even though the contribution near the endpoints is small due to the measure, it is better to separate the light-ray integrals in two contributions:
\begin{align} 
\frac{\langle \Qno \Qnt \rangle^{(1)}_{\text{conn}}}{\langle \Qno\rangle\langle \Qnt \rangle}&=
F^{(\text{short})}_{\mathcal{Q}\mathcal{Q}}(\cos\theta)+F^{(\text{long})}_{\mathcal{Q}\mathcal{Q}}(\cos\theta)\,, \\
\frac{\langle \Eno \Ent \rangle^{(1)}_{\text{conn}}}{\langle \Eno\rangle\langle \Ent \rangle}&= F^{(\text{short})}_{\mathcal{E}\mathcal{E}}(\cos\theta)+F^{(\text{long})}_{\mathcal{E}\mathcal{E}}(\cos\theta)\,.
\end{align}
The ``short'' term is obtained carving out a small region around the endpoints $2\pi-\eta<|t_1^-+t_2^-|$ in~\eqref{eq_QQ_conn} and~\eqref{eq_EE_conn}, with $0<\eta\ll 1$,  and replacing the propagator with its short distance approximation. From the viewpoint of Minkowski space the requirement $2\pi-\eta<|t_1^-+t_2^-|$ selects only the region near future ``$i^+$'' and space infinity ``$i^0$'' where the two light-rays touch. The ``long'' contribution is the remaining integral and is evaluated using the representation of the propagator in \eqref{Eq:PropApprds}.
The following condition on the cutoff, 
\begin{equation}
\Lambda\gtrsim \frac{4}{\eta\,|\theta|\sqrt{d-2}}\,,
\end{equation}
ensures that the exponent in \eqref{Eq:AsDsProp} is sufficiently large everywhere within the integration range.

The numerical results in $d=3$ are shown by the blue dots in Fig.~\ref{fig:QQandEECorr}. We find that $\eta=0.6$ and $\Lambda=15 $ yield precise results for $\theta\gtrsim 0.4$. Indeed, Fig.~\ref{fig:QQandEECorr} shows excellent agreement between the numerical points and the (semi-)analytical results for the collinear and back-to-back region, that we discussed before.

\section{Generalized detectors}\label{Sec:Generalized}

So far, we have considered the simplest and most physically intuitive type of detectors: those associated with energy and global charges. These are just a subset of the possible detectors that can be defined in a CFT and, more generally, in QFT. In particular, given any local operator $\mathcal{O}^{\mu_1,\ldots\mu_J}$ with scaling dimension $\Delta$ and spin $J$, such that $\Delta+J>1$, we can define a detector integrating it along a light-ray connecting spatial infinity to future infinity\footnote{The $2^{-\Delta}$ prefactor is conventional and ensures the agreement with~\eqref{charge_detector} and~\eqref{energy_detector}. This definition is equivalent to the one in \cite{Korchemsky:2021okt}.}
\begin{align}\label{Eq:GenDetector}  
    \mathcal{D}_{\mathcal{O}} (\mathbf{n}) = 2^{-\Delta} \lim_{\xn^{+}\rightarrow \infty} (\xn^+)^{\Delta - J} \int_{-\infty}^{\infty} d\xn^- \bar{n}_{\mu_1} \ldots \bar{n}_{\mu_J} \mathcal{O}^{\mu_1,\ldots\mu_J}(\xn^+ ,\xn^-)\,.
\end{align}  
Generalizations include detectors with transverse spin and, more in general, any light-ray operator, including those not associated with local operators, defines a detector \cite{Kravchuk:2018htv}. We will restrict our attention to detectors of the form~\eqref{Eq:GenDetector}.

Although generalized detectors \eqref{Eq:GenDetector} are not associated with any conserved charge, their event shapes provide valuable information about the state and, more generally, about the theory. It is therefore worthwhile to explore some of their properties and, specifically, to investigate whether they can be computed within our EFT framework. We will show that, from the EFT viewpoint, there exist ``good'' and ``bad'' detectors obtained from, respectively, neutral and charged operators. Event shapes of the former can be systematically computed within EFT (at least in some sense), while to study the latter one needs the full information on the UV theory.\footnote{Note that our definitions of ``good" and ``bad" are based solely on the EFT perspective and are not connected to the renormalization properties of the detectors~\cite{Caron-Huot:2022eqs}.}

One point event shapes involving the detector \eqref{Eq:GenDetector} are fixed by conformal invariance~\cite{Kologlu:2019mfz}
\begin{equation}\label{eq_gen_event_shape1pt}
   \langle \mathcal{D}_{\mathcal{O}}(\mathbf{n}) \rangle_p = \lambda_{\bar{\mO}_Q\mathcal{O}\,\mO_Q}  \frac{\sqrt{\pi } \,\Gamma (\Delta_Q) \Gamma \left(\Delta_Q+1-\frac{d}{2}\right) \Gamma \left(\frac{J+\Delta-1}{2}\right)}{\Gamma \left(\frac{J+\Delta }{2}\right) \Gamma \left(\frac{J-\Delta }{2}+\Delta_Q\right) \Gamma \left(\frac{\Delta+J-d}{2} +\Delta_Q\right)}\frac{(p^2)^{\frac{\Delta+J-2}2{}}}{(n\cdot p )^{\Delta -1}} \Theta(p^0) \Theta(p^2)\,,
\end{equation}
where $\lambda_{\bar{\mO}_Q\mathcal{O}\,\mO_Q}$ is the OPE coefficient of the local operator $\mathcal{O}$ defining the detector. Two point event shapes are more interesting and depend on a cross-ratio analogously to~\eqref{eq_QQ_cross_ratio} and~\eqref{eq_EE_cross_ratio}:
\begin{align}\label{Eq:GenericTwoPoint}
    \langle \mathcal{D}_{\mathcal{O}}(\mathbf{n}_1) \mathcal{D}_{\mathcal{O}}(\mathbf{n}_2) \rangle_p = \frac{(p^2)^{\Delta+J-2} \Theta(p^0) \Theta(p^2)}{(n_1\cdot p )^{\Delta -1}(n_2\cdot p )^{\Delta -1}} \mathcal{G}_{\mathcal{D_{\mathcal{O}}}\mathcal{D_{\mathcal{O}}}} (\xi) \,,
\end{align}
where, again, $\xi= \frac{(2n_1\cdot n_2)(p^2)}{(2 n_1 \cdot p)(2 n_2 \cdot p)}$ controls the dependence of the event shape with the angular distance of the detectors. Below we shall work in the source/sink rest frame. 

\subsection{Generalized detectors in EFT}\label{sec_gen_EFT}

Let us review how to represent in EFT arbitrary light operators. On general grounds, a light primary local operator of dimension $\Delta$, spin $J$, and charge $q$ flows to a local functional of the Goldstone field with the appropriate quantum numbers~\cite{Monin:2016jmo}:
\begin{equation}\label{eq_EFT_matching}
   \mO^{(q)}_{a_1\ldots a_{J}}= c_{\mO}e^{i q\chi}(\pd\chi)^{\Delta-J}\Pi_{a_1\ldots a_{J}}^{b_1\ldots b_{J}}\pd_{b_1}\chi\ldots\pd_{b_{J}}\chi+\ldots\,,
\end{equation}
where $c_{\mO}$ is an $O(1)$ Wilson coefficient, $\Pi$ is the projector onto traceless symmetric tensors and the dots stand for higher derivative terms, suppressed by a relative $ 1/\mu^2\sim Q^{-\frac{2}{d-1}}$ factor. For $q=0$, we obtain the OPE coefficient evaluating this expression on the classical profile~\eqref{eq_EFT_bkgd}:
\begin{equation}\label{eq_EFT_OPE}
    \langle Q|\mO^{(0)}_{a_1\ldots a_{J}}|Q\rangle_{\text{cyl}}=
\lambda_{\bar{\mO}_Q\mathcal{O}\,\mO_Q}\delta_{a_1}^0\ldots\delta_{a_J}^0
    \,,\qquad
\lambda_{\bar{\mO}_Q\mathcal{O}\,\mO_Q}\simeq
c_{\mO}\mu^{\Delta}\propto Q^{\frac{\Delta}{d-1}}\,.
\end{equation}
We postpone a detailed discussion of correlation functions of charge operators and the associated detectors to Sec.~\ref{sec_charged_detectors}.

To compute event shapes within EFT we will follow the same strategy as in Sec.~\ref{Sec:Detectors}. Therefore, we work on the saddle-point~\eqref{eq_saddle_pt} and map the calculation to the cylinder using \eqref{Eq:ChangeOfCoordinatesCyl}. A generalized detector on the cylinder reads
\begin{align}\label{eq_D_cyl_gen}
\mathcal{D}_{\mathcal{O}}^{\text{cyl}} (\mathbf{n})= 2^{2J-2}\int_{-\pi}^{\pi} d t^- \left(\cos \frac{t^-}{2}\right)^{\Delta + J -2} \mathcal{O}_{\underbrace{-\ldots -}_J}(t^+ = \pi, t^-,\mathbf{n})\,.
\end{align}
For instance, working in the source/sink rest frame as usual, from~\eqref{eq_EFT_OPE} and~\eqref{eq_D_cyl_gen} we obtain
\begin{equation}\label{eq_1pt_ev_shape_EFT}
\begin{split}
\langle \mathcal{D}_{\mO}(\mathbf{n})\rangle &\simeq X_0^{1-J}2^{2J-2}\int_{-\pi}^{\pi} d t^- \left(\cos \frac{t^-}{2}\right)^{\Delta + J -2}\langle Q|\mathcal{O}_{-\ldots -}|Q\rangle_{\text{cyl}} \\
&=\lambda_{\bar{\mO}_Q\mathcal{O}\,\mO_Q}\frac{E^{J-1}}{\Delta_{Q}^{J-1}}  \frac{\sqrt{\pi}\, \Gamma\left(\frac{\Delta + J -1}{2}\right)}{\Gamma\left(\frac{\Delta+J}{2}\right)}\,,
\end{split}
\end{equation}
which agrees with the exact result~\eqref{eq_gen_event_shape1pt} for $\Delta_Q\gg 1$. Note that the integral~\eqref{eq_1pt_ev_shape_EFT} only converges provided the condition $\Delta+J>1$ is satisfied.

We similarly conclude that higher-point event shape factorize into the product of one-point functions for generic kinematics as long as EFT holds, e.g.
\begin{equation}\label{eq_gen_ev_shape_factorization}
\langle \mathcal{D}_{\mO}(\mathbf{n}_1)\mathcal{D}_{\mO}(\mathbf{n}_2)\rangle\simeq \langle \mathcal{D}_{\mO}(\mathbf{n}_1)\rangle\langle\mathcal{D}_{\mO}(\mathbf{n}_2)\rangle\,.
\end{equation}
As in Sec.~\ref{Sec:FTcyl}, corrections to~\eqref{eq_gen_ev_shape_factorization} arise both from the Fourier transform and the connected correlator on the cylinder. Let us focus on the latter contribution. Expanding the neutral operator~\eqref{eq_EFT_matching} in fluctuations,
\begin{equation}
    \mO^{(0)}_{-\ldots -}\simeq 2^{-J}c_{\mO}\mu^{\Delta}\left[1+\frac{1}{\mu}\left(\Delta\dot{\pi}-J\pd_\sigma\pi\right)+\ldots
    \right]\,,
\end{equation}
we see that the leading connected contribution is given by the following integral
\begin{equation}\label{eq_connected_generalized_EFT}
X_0^{2-2J}4^{J-2} \frac{c_{\mO}^2\mu^{2\Delta-d}}{c(d-1)}\iint dt^-_1 dt^-_2 
     \left( \cos\frac{t^-_1}{2}\cos\frac{t^-_2}{2}\right)^{\Delta+J-2}
     \prod_{i=1,2}\left[\Delta\partial_{t_i}-J\partial_{\sigma_i}\right]G_{\pi \pi}\,.
\end{equation}
Even when $\theta\gtrsim 1/\mu$, as for charge and energy correlators the EFT breaks down near $t^+_1=t^+_2=\pm\pi$. Proceeding as in the discussion around~\eqref{eq_nonEFT_JJ}, we estimate that the contribution from the dangerous region scales as
\begin{equation}\label{eq_nonEFT_generalzed}
      \text{non-EFT contribution}\sim X_0^{2-2J}\times \mu^{2\Delta-d}\times\mu^{d+2-2\Delta-2J}\sim X_0^{2-2J} \times \mu^{2-2J}\,.
\end{equation}
Since the one-point function~\eqref{eq_1pt_ev_shape_EFT} scales as $X_0^{1-J}\mu^\Delta$, we see that~\eqref{eq_nonEFT_generalzed} is always smaller than the factorized result as we must have $\Delta+J>1$ for the light-ray transform to exist. However, for 
\begin{equation}\label{eq_UV_condition}
    \frac{d+2}{2}>\Delta +J\,,
\end{equation}
the UV contribution is more important than the leading connected term predicted by EFT, which scales as $X_0^{2-2J}\mu^{2\Delta-d}$ from~\eqref{eq_connected_generalized_EFT}. In fact, it is simple to check that when the condition~\eqref{eq_UV_condition} holds, the integral~\eqref{eq_connected_generalized_EFT} does not even converge near the endpoints $t_1^+=t_2^+=\pm\pi$, and thus we need a regularization procedure to compute it. Note that in unitary theories, the unitarity bounds $\Delta\geq d-2+J$ for $J\geq 1$ imply that~\eqref{eq_UV_condition} is never satisfied for detectors built out of spinning operators in $d>2$.

In the next section we will discuss the calculation of the leading correction to~\eqref{eq_gen_ev_shape_factorization}, mostly focusing on scalar operators, assuming $\Delta +J>\frac{d+2}{2}$. We will come back to the issue of UV contributions in Sec.~\ref{sec_EFT_singular}, where we will also clarify the structure of the EFT expansion in general, including when~\eqref{eq_UV_condition} holds.

\subsection{Event shapes for neutral generalized detectors}

Let us consider a two-point event shape for identical detectors with $\Delta+J>\frac{d+2}{2}$; the EFT expansion reads
\begin{align}
    \langle \mathcal{D}(\mathbf{n}_1) \mathcal{D}(\mathbf{n}_2) \rangle =  \langle \mathcal{D}(\mathbf{n}_1)\rangle\langle \mathcal{D}(\mathbf{n}_2) \rangle +
    \frac{1}{\Delta_Q} \langle \mathcal{D}(\mathbf{n}_1) \mathcal{D}(\mathbf{n}_2) \rangle ^{(1)}+\ldots\,,
\end{align}
where the leading result completely factorizes, while the first correction consists of two contributions as in~\eqref{Eq:ContQQ}
\begin{align}\label{Eq:ContDD}
    \langle\mathcal{D}(\mathbf{n}_1) \mathcal{D}(\mathbf{n}_2) \rangle^{(1)} =  \langle \mathcal{D}(\mathbf{n}_1) \mathcal{D}(\mathbf{n}_2) \rangle^{(1)}_{\text{disc}}+\langle \mathcal{D}(\mathbf{n}_1) \mathcal{D}(\mathbf{n}_2) \rangle^{(1)}_{\text{conn}} \,.
\end{align}
Here the first term is the disconnected contribution from the Fourier transform, while the second is the contribution from the connected correlator
\begin{equation}
\begin{split}
 \frac{ \langle \mathcal{D}(\mathbf{n}_1) \mathcal{D}(\mathbf{n}_2) \rangle^{(1)}_{\text{conn}}}{\langle\mathcal{D}(\mathbf{n}_1)\rangle\langle \mathcal{D}(\mathbf{n}_2) \rangle} &= \frac{\pi ^{\frac{d}{2}-1} \Gamma \left(\frac{J+\Delta }{2}\right)^2}{4 \Gamma \left(\frac{d}{2}+1\right) \Gamma \left(\frac{J+\Delta -1}{2}\right)^2}\\
 &\times
 \iint dt^-_1 dt^-_2 
     \left( \cos\frac{t^-_1}{2}\cos\frac{t^-_2}{2}\right)^{\Delta+J-2}
     \prod_{i=1,2}\left[\Delta\partial_{t_i}-J\partial_{\sigma_i}\right]G_{\pi \pi}\,.
     \end{split}
\end{equation}
The calculation is identical to the charge and energy correlators discussed in Sec.~\ref{Sec:EEC}. Below we discuss in detail some examples, mostly focusing on detectors built out of scalars in $d=3$.

\begin{figure}[t]
    \centering  
    \begin{minipage}{0.45\textwidth}
    \centering
    \includegraphics[width=1\linewidth]{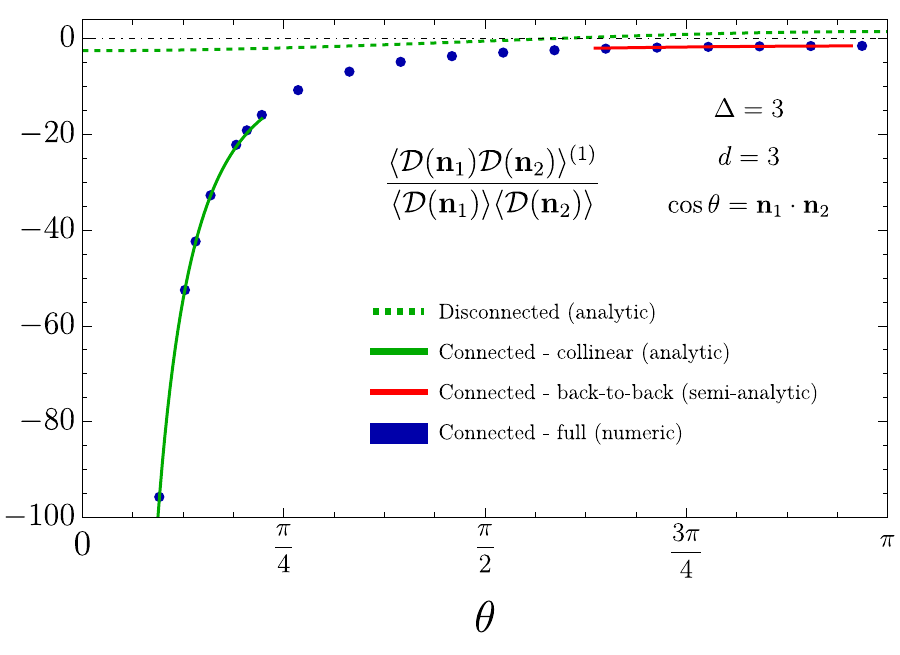}
    \end{minipage}\hspace{2em}
    \begin{minipage}{0.45\textwidth}
    \centering
    \includegraphics[width=1\linewidth]{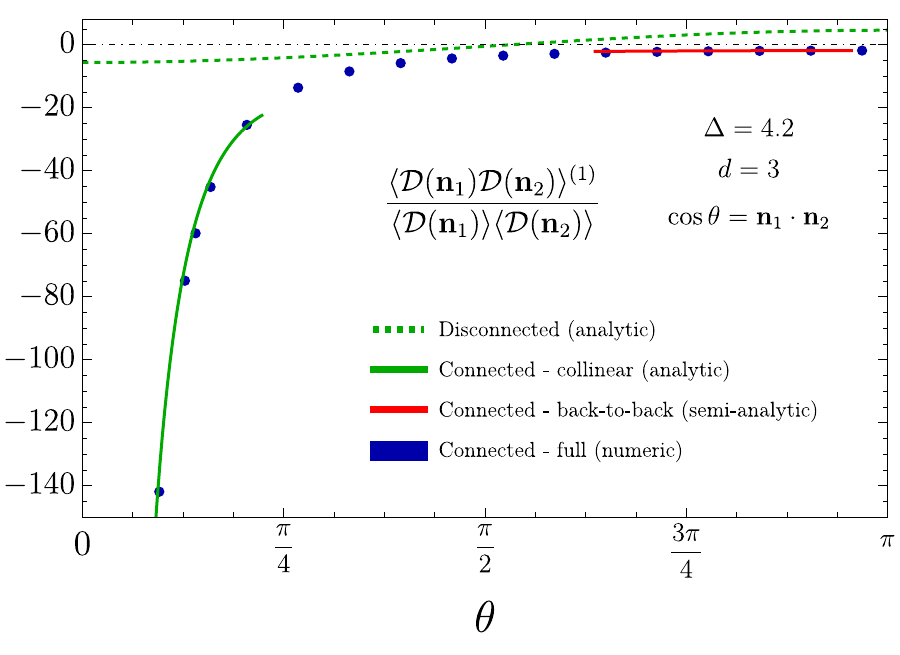}
    \end{minipage}
    \caption{Next-to-leading order two-point event shapes involving a scalar detector of dimension $\Delta=3$ (left) and $\Delta = 4.2$ (right), as functions of the angular distance $\theta$ between the detectors. The plots show the results in $d=3$ dimensions; we added a constant shift to the collinear-limit of the connected contribution (solid green line) to improve the fit. Details on the various contributions are explained in the main text.}
    \label{fig:ScalarCorr}
\end{figure}

For scalar operators in $d=3$ the first term in~\eqref{Eq:ContDD} is computed in App.~\ref{App:Disconnected} and reads
\begin{equation}
    \frac{\langle \mathcal{D}(\mathbf{n}_1) \mathcal{D}(\mathbf{n}_2) \rangle^{(1)}_{\text{disc}}}{{\langle\mathcal{D}(\mathbf{n}_1)\rangle\langle \mathcal{D}(\mathbf{n}_2) \rangle}}= -
    \frac{1+(\Delta-1)^2\mathbf{n}_1\cdot \mathbf{n}_2}{2\Delta_Q}\,.
\end{equation}
This function is graphically shown by the dashed lines in Fig.~\ref{fig:ScalarCorr} for $\Delta=3$ and $\Delta=4.2$.

In the collinear regime $1/\mu\ll\theta\ll 1$, the connected contribution follows from the short-distance limit of the propagator. For $J=0$ in $d=3$ it reads
\begin{equation}
\begin{split}
\frac{\langle \mathcal{D}(\mathbf{n}_1) \mathcal{D}(\mathbf{n}_2) \rangle^{(1)}_{\text{conn}}}{\langle \mathcal{D}(\mathbf{n}_1) \rangle\langle\mathcal{D}(\mathbf{n}_2) \rangle}\stackrel{\theta\neq 0}{=}
&-\theta^{-2}\frac{4 \Delta ^2 \Gamma \left(\Delta -\frac{5}{2}\right) \Gamma \left(\frac{\Delta }{2}\right)^2}{3 \sqrt{\pi }  \Gamma (\Delta -2) \Gamma \left(\frac{\Delta -1}{2}\right)^2}\\
&
-\log \left(\theta ^2\right)
\frac{2 \left(4 \Delta ^2-18 \Delta +21\right) \Gamma \left(\frac{\Delta }{2}+1\right)^2 \Gamma \left(\Delta -\frac{5}{2}\right) }{3 \sqrt{\pi }\, \Gamma \left(\frac{\Delta -1}{2}\right)^2 \Gamma (\Delta -1)}
+
O\left(\theta^0\right)\,.
\end{split}
\end{equation} 
Comments analogous to those in Sec.~\ref{Sec:CollLimit} apply here. Note in particular the \emph{sound-jet} singularity as $\theta\rightarrow 0$.\footnote{The universal behavior of event shapes in the collinear limit in EFT suggests that it might be possible to define some sort of effective OPE controlling this limit. We thank M. Walters for useful comments on this point.}
We checked for several integer and half-integer values of $\Delta$ that the $1/\theta^2$ term becomes the distribution $\text{Pf}\left[\frac{1}{\theta^2}\right]$ when properly regulated. We expect the same for generic $\Delta$.
We also provide the generalization of the collinear result to arbitrary $d$, $\Delta$ and $J$:
\begin{equation}\label{eq_collinear_general_d}
\begin{split}
\frac{\langle \mathcal{D}(\mathbf{n}_1) \mathcal{D}(\mathbf{n}_2) \rangle^{(1)}_{\text{conn}}}{\langle \mathcal{D}(\mathbf{n}_1) \rangle\langle\mathcal{D}(\mathbf{n}_2) \rangle}\stackrel{\theta\neq 0}{=}
   & -\theta ^{1-d}\frac{(\Delta-J )^2 \Gamma \left(\frac{d+1}{2}\right)  \Gamma \left(\frac{J+\Delta }{2}\right)^2 \Gamma \left(J+\Delta-\frac{d}{2} -1\right)}{(d-2)^{3/2} \Gamma \left(\frac{d}{2}+1\right) \Gamma \left(\frac{J+\Delta -1}{2} \right)^2 \Gamma \left(J+\Delta-\frac{d}{2} -\frac{1}{2}\right)}\\[0.6em]
    &
    +O\left(\text{max}(1,\theta^{3-d})\right)\,.
    \end{split}
\end{equation}

\begin{figure}[t]
    \centering
    \includegraphics[width=0.45\linewidth]{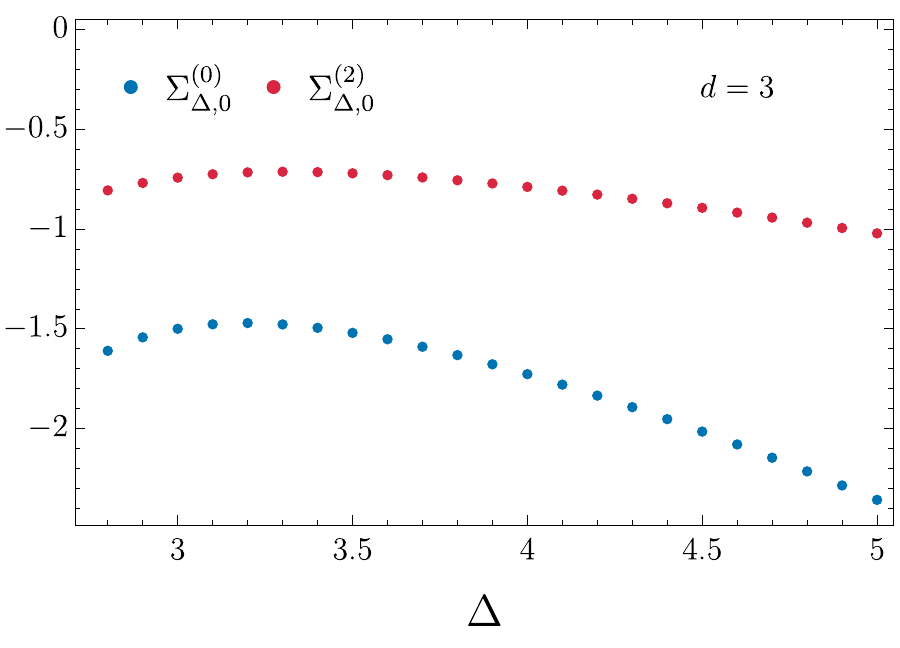}
    \caption{Two-point event shapes involving scalar detectors in the back-to-back limit. In the figure we report the two coefficients of the Taylor expansion (see \eqref{Eq:BacktoBackScalar}) for different values of $\Delta$ in $d=3$.}
    \label{fig:BacktoBackScalar}
\end{figure}

In the back-to-back regime, we obtain semi-analytic results proceeding as in Sec.~\ref{Sec:Backtoabcklimit}:
\begin{equation}\label{Eq:BacktoBackScalar}
    \frac{\langle \mathcal{D}(\mathbf{n}_1) \mathcal{D}(\mathbf{n}_2) \rangle^{(1)}_{\text{conn}}}{\langle \mathcal{D}(\mathbf{n}_1) \rangle\langle\mathcal{D}(\mathbf{n}_2) \rangle}=\Sigma^{(0)}_{\Delta,J}+\frac12(\pi-\theta)^2\Sigma^{(2)}_{\Delta,J}+O\left((\pi-\theta)^4\right)\,.
\end{equation}
We plot the results we obtained for $\Sigma^{(0)}_{\Delta,0}$ and $\Sigma^{(2)}_{\Delta,0}$ in $d=3$ for several values of $\Delta$ in Fig.~\ref{fig:BacktoBackScalar}.\footnote{For $\Delta=3$ the integrals simplify and we obtain
$\Sigma^{(0)}_{3,0}=-3/2$ exactly; we also observe $\Sigma^{(2)}_{3,0}=-3/4$ to high accuracy.} For general angles we may obtain numerical result using the representation~\eqref{eq_prop_dS} of the propagator, as in Sec.~\ref{Sec:Num}. As an illustration, we plot the results in $d=3$ for scalars with $\Delta=3$ and $\Delta=4.2$ in Fig.~\ref{fig:ScalarCorr}.

\subsection{EFT of the light-ray endpoints}\label{sec_EFT_singular}

As formerly noted, a small region of the light-ray integrals around $t_1^-=t_2^-=\pm \pi$ is not within EFT. In this section, we analyze the physics of this region further. To this aim, it will be simpler to focus directly on correlation functions of detectors on the cylinder~\eqref{eq_D_cyl_gen} in the large charge primary state. We also focus on event shapes of two identical detectors as before for concreteness. Our discussion however is general and may be straightforwardly applied to arbitrary event shapes in Minkowski space.

Let us consider an event shape with two identical neutral detectors. We have seen in Sec.~\ref{sec_gen_EFT} that the disconnected term is proportional to $\lambda_{\bar{\mO}_Q\mO\,\mO_Q}^2\sim \mu^{2\Delta}$, while the connected part~\eqref{eq_connected_generalized_EFT} naively scales as $\mu^{2\Delta-d}$. We estimated the UV contribution from the integration endpoints in~\eqref{eq_nonEFT_generalzed}, where we argued that, upon introducing short-distance cutoff at $t^-_{1}=t^-_2=\pm(\pi-\#/\mu)$, we expect a UV dominated term of the form
\begin{equation}\label{eq_gen_UV2}
    \langle Q| \mathcal{D}^{\text{cyl}}_{\mO}(\mathbf{n}_1)\mathcal{D}^{\text{cyl}}_{\mO}(\mathbf{n}_2)|Q\rangle_{\text{cyl}}\supset \mu^{2-2J}F_{\mO}(\xi)+O\left(\mu^{-2J}\right)\,,\quad\quad \xi =\frac{1-\cos \theta}{2} \,,
\end{equation}
where the function $F_{\mO}(\xi)$ receives contributions from all higher derivative terms and is thus unknown in EFT.\footnote{Estimating $F_{\mO}(\xi)$ with a cutoff in the integral and using the short-distance limit of the propagator~\eqref{Eq:PropSmallAngle}, we expect that it grows as $\theta^{1-d}$ in the collinear limit as the EFT result.} The UV component dominates over the connected EFT result for $\Delta+J<(d+2)/2$. In the limiting case $\Delta+J=(d+2)/2$ we further expect a logarithmic enhancement $\sim\log\mu$ of the UV contribution~\eqref{eq_gen_UV2}, as it may easily be checked estimating $F_{\mO}$ with a cutoff and using the short-distance limit of the propagator~\eqref{Eq:PropSmallAngle}.

Since the region in which EFT breaks down is small, we can parametrize, and renormalize, the UV contribution to the event shape in terms of local counterterms at the light-ray endpoints $(t=\pi,\sigma=0)$ and $(t=0,\sigma=\pi)$. The structure of these counterterms follows from the transformation properties of light-ray operators, of which the detectors under considerations are special instances. Below, we briefly review the ingredients that we need for our analysis following~\cite{Kravchuk:2018htv}. 

We remind the reader that in index-free notation a spin $J$ operator is encoded in a homogeneous polynomial of degree $J$ as
\begin{equation}
    \mO(y;z)=\mO^{a_1\ldots a_J}(y)z_{a_1}\ldots z_{a_J}\,,
\end{equation}
where $z$ is a null vector: $z^2=0$. Using embedding coordinates for the sphere, a null vector on $\mathbb{R}\times S^{d-1}$ can be written as $(z^0,\vec{z}\,)$, where $(z^0)^2=\vec{z}^{\,2}$ and $\vec{z}\cdot\hat{m}=0$, i.e. $\vec{z}$ is tangent to the sphere. In this language, fractional spin representations of the rotation group simply correspond to homogeneous functions of $z$ with fractional degree. Local operators cannot have fractional spin, but such representations may occur for nonlocal operators such as the ones we are interested in.

Using index-free notation, a light-ray operator on the cylinder reads~\cite{Kravchuk:2018htv}
\begin{multline}\label{eq_lightray_cyl}
    \mathbf{L}[\mO](t,\hat{m};z^0,\vec{z})= 2^{J-1}(z^0)^{1-\Delta-J}\\
    \times\int_0^\pi d\kappa (\sin\kappa)^{\Delta+J-2}\mO(t+\kappa,\hat{m}\cos\kappa+\vec{z}/z^0 \sin\kappa;z^0,\vec{z}\cos\kappa-z^0\hat{m}\sin\kappa)\,,
\end{multline}
where we fixed the prefactor to agree with our conventions~\eqref{eq_D_cyl_gen}.
The nontrivial claim is that this integral yields an operator that transforms as a primary of the conformal group with dimension $1-J$ and spin $1-\Delta$, at either of the two integration endpoints, which are related by the action of the center of the Lorentzian conformal group $\mathcal{T}$~\cite{Kravchuk:2018htv}. 

It is simple to check that the detectors we are interested in are obtained setting $t=0$ and $\sigma=\pi$, i.e. the north pole, and taking a null vector with $z^0=1$ and $\vec{z}=(\mathbf{n},0)$ in~\eqref{eq_lightray_cyl}.\footnote{In flat space, detectors are light-ray operators at spatial infinity:
\begin{equation}
    \mathcal{D}_{\mO}^{\text{flat}}(\mathbf{n})=\lim_{x\rightarrow\infty}(-x^2)^{2\Delta}
    \mathbf{L}[\mO](x;z)\,,\qquad
    \mathbf{L}[\mO](x;z)=2^{-\Delta}\int_{-\infty}^{\infty}d\alpha\,\mO\left(x-\frac{z}{\alpha};z\right)\,.
\end{equation}} Therefore the product of two detectors is nothing but the product of two light-ray operators \emph{at coincident points}:
\begin{equation}\label{eq_det=lightray}
     \mathcal{D}^{\text{cyl}}_{\mO}(\mathbf{n}_1)\mathcal{D}^{\text{cyl}}_{\mO}(\mathbf{n}_2)
     =
\mathbf{L}[\mO](y_0;z_1)
\mathbf{L}[\mO](y_0;z_2)
     \,,
\end{equation}
where $y_0=(t=0,\sigma=\pi)$ and $z_1$, $z_2$ denote the appropriate null polarization vectors. In light of the former discussion, the product $\mathcal{D}^{\text{cyl}}_{\mO}(\mathbf{n}_1)\mathcal{D}^{\text{cyl}}_{\mO}(\mathbf{n}_2)$ transforms as a primary of dimension $2-2J$ with \emph{two spins} $1-\Delta$, corresponding to the two polarization vectors. This simply means that the product $\mathbf{L}[\mO](y;z_1)\mathbf{L}[\mO](y;z_2)$ is homogeneous in both $z_1$ and $z_2$ with degree $1-\Delta$. Note that a similar argument determines the spin of the light-ray operators in the OPE of two detectors~\cite{Hofman:2008ar,Kologlu:2019mfz}.

We are now ready to provide a complete EFT treatment of detectors. We claim that the product of two detectors in the EFT flows to the light-ray integrals that we have considered so far, \emph{plus} additional local EFT operators that encode the short-distance contributions from the points in which the integrated operators collide:
\begin{equation}\label{eq_DD_counterterm}
\mathcal{D}^{\cyl}_{\mO}(\mathbf{n}_1)\mathcal{D}^{\cyl}_{\mO}(\mathbf{n}_2)\vert_{\text{UV}}\stackrel{\text{RG}}{\longrightarrow}
\mathcal{D}^{\cyl}_{\mO}(\mathbf{n}_1)\mathcal{D}^{\cyl}_{\mO}(\mathbf{n}_2)\vert_{\text{EFT}}+
\mO_{ct}(y_0;z_1,z_2)+\mO_{ct}(\mathcal{T}y_0;z_1,z_2)\,,
\end{equation}
where $y_0=(t=0,\sigma=\pi)$, $\mathcal{T}y_0=(t=\pi,\sigma=0)$, while the polarization vectors are $z_1^0=1$ and $\vec{z}_1=(\mathbf{n}_1,0)$ and similarly for $z_2$. The explicit form of the local \emph{counterterms} is easily obtained by requiring that they transform as the product of two light-ray operators under the conformal group. At leading order in derivatives, this fixes the form of $\mO_{ct}$ up to a function of a cross-ratio
\begin{align}\label{eq_counterterm}
\mO_{ct}(y;z_1,z_2)=\frac{(\pd\chi)^{2\Delta-2J} }{[(z_1\cdot\pd\chi)(z_2\cdot\pd\chi)]^{\Delta-1}}F_{\mO}\left(\frac{(\pd\chi)^2 (2z_1\cdot z_1)}{(2z_1\cdot\pd\chi)(2z_2\cdot\pd\chi)}\right)+\ldots\,,\quad \,.
\end{align}
where the function $F_{\mO}$ is arbitrary and the dots stand for further corrections, which are suppressed by relative factors of $1/(\pd\chi)^2\sim 1/\mu^2$. The homogeneity properties of~\eqref{eq_counterterm} in $z_1$ and $z_2$ are obvious. Additionally, \eqref{eq_counterterm} is easily checked to transform homogeneously under Weyl rescalings, and this in turn implies that $\mO_{ct}$ also transforms as a primary under conformal transformations. 

Using $z_1\cdot z_2=1-\cos\theta$ and that $z_1\cdot\pd\chi=z_2\cdot\pd\chi=\mu$ on the classical profile~\eqref{eq_EFT_bkgd}, the expectation value of~\eqref{eq_counterterm},
\begin{equation}
\langle Q| \mO_{ct}(y;z_1,z_2)|Q\rangle_{\text{cyl}}= \mu^{2-2J} F_{\mO}(\xi)\,, \quad \quad \xi = \frac{1-\cos \theta}{2}\,,
\end{equation}
yields a correction to the event shape in agreement with the cutoff estimate~\eqref{eq_gen_UV2}. In other words, the counterterm~\eqref{eq_counterterm} eliminates the need for a cutoff to account for the new UV contributions.

In summary, whenever $\Delta+J>(d+2)/2$ we do not need to introduce a cutoff to compute the light-ray integrals. When instead~\eqref{eq_UV_condition} is satisfied, the light-ray integrals need to be regulated. This may be done in a mass-independent scheme, such as analytic continuation in $\Delta$. In both cases, we systematically parameterize in EFT the new UV contributions from the integration endpoints with the local counterterm~\eqref{eq_counterterm}, without the need for explicit cutoff or hand-wavy arguments. Note that when computing the integral~\eqref{eq_connected_generalized_EFT} by analytic continuation in $\Delta$ we encounter poles when $\Delta+J=(d+2)/2$, as the collinear result~\eqref{eq_collinear_general_d} explicitly shows. These are renormalized in a standard fashion by the new terms in~\eqref{eq_DD_counterterm}, leading to a $\log\mu$ enhancement of the result, again in agreement with the former physical estimates.

So far we have discussed generic detectors. The physically most interesting case of energy and charge detectors however is special. Indeed, Ward identities impose further restrictions on the form of~\eqref{eq_counterterm}. For instance, the function $F_{\mO}(\xi)$ must integrate to zero over $\theta$ as discussed in Sec.~\ref{Sec:CollLimit}. We did not study systematically all the constraints that the counterterms must satisfy, nor if such constraints admit nontrivial solutions for charge and energy detectors. Note also that our discussion implies that one should consider additional similarly constructed UV counterterms for event shapes with three or more detectors, that arise from the points where three or more operators collide. 

\subsection{Breakdown of factorization for charged detectors}\label{sec_charged_detectors}

From~\eqref{eq_EFT_matching}, we see that charged operators in EFT are proportional to a fast oscillating factor $e^{i q\mu t}$ on the EFT background solution~\eqref{eq_EFT_bkgd}. For instance, a correlation function with two charged operator insertions on the large charge ground state reads
\begin{equation}\label{eq_charged_correlator}
\begin{split}
    \langle Q|\bar{\mO}^{(q)}_{a_1\ldots a_J}(t_1,\hat{m}_1)
    \mO^{(q)}_{b_1\ldots b_J}(t_2,\hat{m}_2)    |Q\rangle_{\text{cyl}}&=|\lambda_{\bar{\mO}_{Q+q}\mO\,\mO_Q}|^2e^{iq\mu(t_2-t_1)}\delta_{a_1}^0\ldots\delta_{a_J}^0\delta_{b_1}^0\ldots\delta_{b_J}^0\\
    &\times
    \left[1+\frac{q^2}{c(d-1)\mu^{d-2}}G_{\pi\pi}+\ldots  \right]   \,,
    \end{split}
\end{equation}
where $\bar{\mO}^{(q)}$ is the conjugate of $\mO^{(q)}$ and we defined the OPE coefficient\footnote{Tadpole diagrams provide a correction to this OPE coefficient $\propto q^2\mu^{\Delta -(d-2)}$ \cite{Cuomo:2020rgt}.}
\begin{equation}
\langle Q+q|\mO^{(q)}_{a_1\ldots a_J} |Q\rangle_{\text{cyl}}=  \delta_{a_1}^0\ldots\delta_{a_J}^0 \lambda_{\bar{\mO}_{Q+q}\mO\,\mO_Q}\,,\qquad
\lambda_{\bar{\mO}_{Q+q}\mO\,\mO_Q}=c_{\mO}\mu^{\Delta}+\ldots\,.
\end{equation}
The origin of the oscillating factor is clear: by charge conservation, the lowest energy state exchanged between the two light insertions in~\eqref{eq_charged_correlator} corresponds to the operator $\mO_{Q+q}$, whose gap is $\Delta_{Q+q}-\Delta_Q\simeq q\,\mu\sim q\, Q^{\frac{1}{d-2}}$. The connected term in parenthesis therefore describes the exchange of states with gap slightly above $\Delta_{Q+q}-\Delta_Q$.

In this section, we consider event shapes associated with charged operators. For simplicity, we shall work directly on the cylinder as in the previous section and focus on the event shape that is obtained integrating~\eqref{eq_charged_correlator}:
\begin{equation}\label{eq_charged_ev_shape}
    \langle Q|\mathcal{D}^{\text{cyl}}_{\bar{\mO}^{(q)}}(\mathbf{n}_1)  \mathcal{D}^{\text{cyl}}_{\mO^{(q)}}(\mathbf{n}_2)|Q\rangle_{\text{cyl}}=
    2^{4J-4}\iint d t^-_1dt^-_2 \left(\cos \frac{t^-_1}{2}\cos\frac{t^-_2}{2}\right)^{\Delta+J  -2} 
    \hspace{-1em}\langle Q|
    \bar{\mO}^{(q)}_{-\ldots-}
    \mO^{(q)}_{-\ldots-}  
    |Q\rangle_{\text{cyl}} \,,
\end{equation}
where we suppressed the operators coordinates for brevity.
When we integrate~\eqref{eq_charged_ev_shape} using~\eqref{eq_charged_correlator}, the fast oscillating factor $e^{iq\mu(t_2^- -t_1^-)/2}$ makes the integral unavoidably peaked around frequencies $\sim \mu$ away from the EFT window. Therefore, event shapes associated with charged operators are dominated by the UV even at leading order. Below we illustrate this point more in detail, showing that if one tries to compute the event shape within EFT one finds that the light-ray integrals are peaked around the singular endpoints. Consequently, there is no parametric separation between the leading order and the subleading term. In particular, unlike for neutral operators, the result \textit{does not factorize} into a term proportional to the square of the OPE coefficient with only small corrections at large charge. Intuitively, the absence of factorization can be understood on physical grounds: the integrated operators have non-zero charge, thus the detectors are necessarily correlated with each other by charge conservation.\footnote{It is however also important that the chemical potential is large: $\mu\gg 1$. In free theories or CFTs with moduli spaces, the chemical potential is $O(1)$~\cite{Cuomo:2024fuy} and our arguments do not apply.} The EFT nonetheless predicts some subleading terms in the expansion.

Let us consider first the integration of the leading term in~\eqref{eq_charged_correlator}
\begin{equation}\label{eq_charged_ev_shape_LO}
    4^{J-2}|\lambda_{\bar{\mO}_{Q+q}\mO\,\mO_Q}|^2
    \int_{-\pi}^\pi d t^-_1
    \left(\cos \frac{t^-_1}{2}\right)^{\Delta+J  -2} e^{-iq\mu t_1^-/2}
    \int_{-\pi}^\pi dt^-_2\left(\cos \frac{t^-_2}{2}\right)^{\Delta+J  -2}
    e^{iq\mu t_2^-/2}\,.
\end{equation}
The above integrals may be performed exactly, but it is instructive to evaluate them perturbatively in $1/\mu$. Consider for instance the integration over $t_2^-$. Assuming $q>0$ for concreteness, we can close the contour at $i\infty$ and recast our original problem in terms of two integrals in the complex plane over $t^-_2\in (\pi,\pi+i\infty)$ and $t^-_2\in (-\pi,-\pi+i\infty)$
\begin{equation}\label{eq_charged_int1}
\begin{split}
\int_{-\pi}^\pi dt^-_2
    e^{iq\mu t_2^-/2}\left(\cos \frac{t^-_2}{2}\right)^{\Delta+J  -2}&=
   - \int_{\pi}^{\pi+i\infty} dt^-_2e^{iq\mu t_2^-/2}\left(\cos \frac{t^-_2}{2}\right)^{\Delta  -2}-\text{c.c.}\\
&= -2\sin\left(\frac{\pi}{2}(q\mu -\Delta-J)\right)\int_{0}^{\infty} d\tau e^{-q\mu \tau/2}\left(\sinh\frac{\tau}{2}\right)^{\Delta+J-2}\,,
   \end{split}
\end{equation}
where in the last line we change variables via $t_2^-=\pi+i\tau$. We therefore see that for $q\mu\gg 0$ the integral is localized around $\tau\simeq 0$, i.e. $t_2^-=\pm \pi$. Rescaling $\tau\rightarrow \tau/q\mu$ and expanding the integrand for large $q\mu$ we find that~\eqref{eq_charged_ev_shape_LO} gives
\begin{equation}\label{eq_charged_LO}
4^{J }\Gamma (\Delta+J -1)^2  \sin ^2\left(\frac{\pi}{2}   (q\mu -\Delta -J)\right)\frac{|\lambda_{\bar{\mO}_{Q+q}\mO\,\mO_q}|^2}{(q\,\mu )^{2 \Delta +2J-2}}  \left[1+O\left(\frac{1}{\mu^2}\right)\right]\propto \mu^{2-2J}\,,
\end{equation}
where to determine the scaling with $\mu$ we used that $\lambda_{\bar{\mO}_{Q+q}\mO\,\mO_Q}\simeq c_{\mO}\mu^{\Delta}$. 

Differently than for neutral operators, the result~\eqref{eq_charged_LO} is proportional to $\mu^{2-2J}$ independently of the scaling dimension of the operator. In light of the discussion in the previous section, we recognize that~\eqref{eq_charged_LO} is of the same order as the UV contribution, described by the counterterm~\eqref{eq_counterterm} in EFT. We conclude therefore that the naive leading order EFT result is not physical, and we can only predict the scaling $\propto \mu^{2-2J}$ of the event shape via the arguments in~\ref{sec_EFT_singular}. The origin of this conclusion is clear from the manipulations in~\eqref{eq_charged_int1}: the fast oscillating phase makes the integrals peak at the integration endpoints $t^1_-=\pm \pi$ and $t^2_-=\pm \pi$, where the measure suppresses the scaling with $\mu$. In particular, the light-ray integrals receive equal contributions from the UV dominated region $t^1_-\simeq t^2_-\simeq \pm \pi$, where the operators collide, and from $t^1_-\simeq-t^2_-\simeq \pm \pi$, where the two operators are at large separation and EFT applies (recall that there are no lightcone singularities in EFT). 

It should be clear that a similar mechanism is at place when we integrate the first subleading order in~\eqref{eq_charged_correlator}. By the same manipulations as in~\eqref{eq_charged_int1} the result consists of two contributions. The first is concentrated in the UV region $t^1_-\simeq t^2_-\simeq \pm \pi$\footnote{For some values of $\theta$ there occur singularities on the countour $\tau_1,\,\tau_2\in\mathbb{R}$, corresponding to the operators becoming lightlike in $dS_{d-1}\times\mathbb{R}$. These can be avoided by slightly modifying the contour and are inessential for our arguments.} 
\begin{equation}
\begin{split}
\frac{4^{J-2}|\lambda_{\bar{\mO}_{Q+q}\mO\,\mO_Q}|^2q^2}{c(d-1)\mu^{d-2}}&\left\{
\int_{0}^{\infty} d\tau_1\int_{0}^{\infty} d\tau_2\, e^{-\frac{1}{2} q\mu (\tau_1+\tau_2)} \left[\sinh \left(\frac{\tau_1}{2}\right) \sinh \left(\frac{\tau_2}{2}\right)\right]^{\Delta +J-2}\right.\\[0.5em]
&
\left.\times
\left[G_{\pi\pi}\vert_{t_1^-=\pi-i\tau_1,t_2^-=\pi+i\tau_2}
+G_{\pi\pi}\vert_{t_1^-=-\pi-i\tau_1,t_2^-=-\pi+i\tau_2}
\right]\right\}\,.
\end{split}
\end{equation}
and also contributes at order $\mu^{2-2J}$, as it can be seen using the short-distance expansion of the propagator~\eqref{Eq:PropSmallAngle}. Note that, based on the arguments in the previous section, we expect that this UV contribution admits an expansion in inverse powers of $\mu^2$. The other contribution instead is focused around $t^1_-\simeq -t^2_-\simeq \pm \pi$ and reads
\begin{equation}\label{eq_charged_NLO2}
\begin{split}
 -&\frac{4^{J-2}|\lambda_{\bar{\mO}_{Q+q}\mO\,\mO_Q}|^2q^2}{c(d-1)\mu^{d-2}} 
 \left\{
\int_{0}^{\infty} d\tau_1\int_{0}^{\infty} d\tau_2\, e^{-\frac{1}{2} q\mu (\tau_1+\tau_2)} \left[\sinh \left(\frac{\tau_1}{2}\right) \sinh \left(\frac{\tau_2}{2}\right)\right]^{\Delta +J-2}\right.\\[0.5em]
&
\left.\times
\left[
e^{i \pi  (\Delta +J-q\mu )}
G_{\pi\pi}\vert_{t_1^-=\pi-i\tau_1,t_2^-=-\pi+i\tau_2}
+e^{-i \pi  (\Delta +J-q\mu )}
G_{\pi\pi}\vert_{t_1^-=-\pi-i\tau_1,t_2^-=\pi+i\tau_2}
    \right]\right\}\,.
    \end{split}
\end{equation}
This last contribution is physical and calculable within EFT. Using that the propagator is regular at $t_1^-=-t^-_2=\pm\pi$, we see that~\eqref{eq_charged_NLO2} scales as $\mu^{2-2J-(d-2)}$ and thus, differently than in~\eqref{eq_charged_ev_shape_LO}, is not contaminated by the UV dominated contribution for generic $d$. Therefore, the result of~\eqref{eq_charged_NLO2} is a well-defined prediction of EFT despite being subleading. For instance, expanding the integrand in~\eqref{eq_charged_NLO2} for small $\tau_1$ and $\tau_2$ and putting everything together, in $d=3$ we find the following result
\begin{align}\label{eq_charged_d3}
\langle Q|\mathcal{D}^{\text{cyl}}_{\bar{\mO}^{(q)}}(\mathbf{n}_1)
    \mathcal{D}^{\text{cyl}}_{\mO^{(q)}}(\mathbf{n}_2)|Q\rangle_{\text{cyl}}& =\mu^{2-2J} F_{\mO}(\xi)\\
    -\mu^{1-2J}\frac{c_{\mO}^2q^{4-2 \Delta -2 J}}{c\, 4^{1-J}}& \Gamma (J+\Delta -1)^2  \cos \left(\pi  (\Delta +J-q\mu )\right)
    G_{\pi\pi}\left(\pi,-1\right) +O\left(\mu^{-2J}\right)  \,,
    \nonumber
\end{align}
where the function $F_{\mO}(\xi)$ is not predicted by EFT, while the subleading contribution is homogeneous and arises from~\eqref{eq_charged_NLO2}. In~\eqref{eq_charged_d3} the propagator at $t_{12}=\sigma_{12}=\pi$ is equal to\footnote{To compute $G_{\pi\pi}\left(\pi,-1\right)$ we isolated the leading oscillatory piece of the sum~\eqref{Eq:WightmanCyl}, evaluated its sum analytically using $P_{\ell}(-1)=(-1)^{\ell}$, and then computed the sum of the remaining difference, which converges in absolute sense, numerically.}
\begin{equation}
G_{\pi\pi}\left(\pi,-1\right)=G_{\pi\pi}\left(-\pi,-1\right)\stackrel{d=3}{\simeq}
0.0712 \,.
\end{equation}

In conclusion, event shapes associated with charged operators are dominated by the microphysics, and do not factorize into a classical homogeneous term with small corrections. The EFT nonetheless predicts some subleading terms in the expansion, as in~\eqref{eq_charged_d3}.

\section{Outlook}\label{Sec:Conc}

In this work, we studied event shapes for large charge states in CFTs. We obtained explicit results for EEC and CCC and systematized the application of EFT to general event shapes. Our results are summarized in the Introduction in Sec.~\ref{Sec:Summ}. Below, we comment on future research directions.

First, it would be interesting to understand how our EFT predictions emerge from a controlled, UV-complete model. A natural candidate is the Wilson-Fisher $O(2)$ fixed point in $4-\varepsilon$ dimensions, where one can access large charge correlators in a double-scaling limit with $\varepsilon \ll 1$ and $\varepsilon Q = \text{fixed}$~\cite{Badel:2019oxl}. The double-scaling parameter~$\varepsilon Q \sim \mu^3$ controls, in particular, the gap of the radial mode, interpolating between the near-vacuum regime at $\varepsilon Q \ll 1$ and the conformal superfluid one for $\varepsilon Q \gg 1$. Results for EEC~\eqref{eq_EEC_intro} and other event shapes in this double-scaling limit could shed light on various aspects of event shapes in large charge states. In particular, it should be possible to analyze the microscopic structure of the sound jets discussed in Sec.~\ref{Sec:CollLimit}. Relatedly, the UV model would describe the transition between the light-ray OPE regime, which is trivial at leading order in $\varepsilon$ and is expected to govern the deeply collinear limit $\theta \ll 1/\mu$, and the EFT predictions discussed in this paper, similarly to the analysis of \cite{Chicherin:2023gxt} in planar $\mathcal{N}=4$ SYM. Additionally, a controlled microscopic model should allow matching the local counterterms discussed in Sec.~\ref{sec_EFT_singular} with the UV theory and, perhaps, also exploring more general detectors that cannot be obtained as integrals of local operators, in the spirit of~\cite{Caron-Huot:2022eqs}.

The techniques developed in this work are likely to prove useful also in the analysis of conformal collider observables of \emph{generic} heavy operators, that admit an EFT description as thermal states~\cite{Lashkari:2016vgj}. Real-time correlators at finite temperature are universally described by hydrodynamics~\cite{Kovtun:2012rj,Delacretaz:2020nit}, and present some new physical ingredients compared to the superfluid states analyzed here. For instance, hydrodynamic correlators also describe the propagation of a diffusion mode besides sound, and the Naivier-Stokes equations are dissipative.\footnote{We thank Z. Komargodski for useful discussions on this point.} 
The analysis of event shapes in such conformal fluid states might represent an important step forward toward a systematic application of EFT techniques to ECs in heavy-ion collisions.

Finite density and finite temperature states also admit interesting holographic duals. For instance, the conformal superfluid is dual to a superconducting boson star \cite{Hartnoll:2008kx,delaFuente:2020yua}, while generic heavy operators are described by black holes in AdS. Detector operators are dual to shock waves that perturb the AdS geometry \cite{Hofman:2008ar}, and it would be interesting to explore their behavior in these nontrivial backgrounds. The EFT results might provide useful guidance for the holographic analysis, perhaps along the lines of the fluid-gravity correspondence~\cite{Hubeny:2011hd}. 

It might also be interesting to study event shapes in a different class of semiclassical states, such as those corresponding to large spin operators in CFT \cite{Alday:2007mf,Komargodski:2012ek,Fitzpatrick:2012yx}, that also admit a nontrivial interplay with the large charge regime~\cite{Cuomo:2017vzg,Cuomo:2022kio,Choi:2025tql} and holography~\cite{Fitzpatrick:2014vua,Fardelli:2024heb}. For spinning operators, even one-point event shapes are not completely fixed by symmetries and contain information about the dynamics. It is also interesting to explore conformal collider observables at large charge in theories with different EFT descriptions, such as Fermi spheres~\cite{Komargodski:2021zzy,Dondi:2022zna,Delacretaz:2025ifh} or models with moduli spaces~\cite{Hellerman:2017veg,Grassi:2019txd,Caetano:2023zwe,Cuomo:2024fuy}.

Finally, some of the ideas developed in this work might find applications in different contexts. For instance, in Sec.~\ref{Sec:Generalized} we found it convenient to decompose the four-point correlator into an unusual basis, reminiscent of the complexification of spatial momentum that is commonly employed in the calculation of phase shifts as a function of the impact parameter~\cite{Camanho:2014apa}. It might be worthwhile exploring whether similar ideas could prove useful in more general studies of event shapes.

There are also some superficial similarities between the EFT setup analyzed in this paper and the calculation of event shapes in AdS/CFT. Intuitively, this is because semiclassical theories in AdS, as defined through a bottom-up approach or a truncated supergravity action, naturally possess a cutoff $\lesssim N^2$ on the spectrum of the dilation operator. As a result, also in holography one finds a breakdown of semiclassical gravity for EECs in the ultra-collinear limit $\theta\lesssim 1/N^2$~\cite{Hofman:2008ar,Chen:2024iuv}. Additionally, in the ``Effective Conformal Theory'' setup of \cite{Fitzpatrick:2010zm}, i.e. for a CFT dual to a bulk theory obtained integrating out heavy fields with dimension $\Delta_{\Lambda}\ll N^2$, generic event shapes are potentially ill-defined and need to be regulated, as in our analysis of generalized detectors in Sec.~\ref{Sec:Generalized}. It might be instructive to reformulate these issues from an EFT perspective, perhaps along the lines of Sec.~\ref{sec_EFT_singular}.

\section*{Acknowledgments}

We thank G.~Fardelli, Z.~Komargodski, Y.-Z.~Li, A.~Monin, I.~Moult, R.~Rattazzi, Z.~Sun and R.~Sundrum for useful discussions. We are especially grateful to M.~Walters and A. Zhiboedov for useful comments on a preliminary version of this manuscript. During this work, GC was supported by the Simons Foundation grant 994296 (Simons Collaboration on Confinement and QCD Strings) and by the BSF grant 2018068. 
EF and FN are partially supported by the Swiss National
Science Foundation under contract 200020-213104 and through the National Center of Competence in Research SwissMAP.
The work of LR is supported by NSF Grant No.~PHY-2210361 and by the Maryland Center for Fundamental Physics.

\appendix

\section{The large charge propagator}\label{app_Prop}

\subsection{Generalities and the short distance limit}\label{Sec:ShortDistanceProp}

Here we discuss some properties of the propagator for the Goldstone field $\pi$. It is convenient to start from Euclidean signature and then obtain the Wightman function $G_{\pi \pi}$ we are interested in by analytic continuation.
The Lagrangian in~\eqref{eq_cyl_action_fluct} in Euclidean signature reads 
\begin{equation}
    S^{(2)}_E=c(d-1)\mu^{d-2}\int d\tau d^{d-1}\hat{m}\,\left[
    \frac{1}{2}  \left(\pd_\tau\pi\right)^2
    +\frac{1}{2(d-1)}(\pd_i\pi )^2  \right]\,.
\end{equation}
Using rotational invariance the equation defining the Euclidean propagator $G_{\pi \pi}$ can be explicitly written as 
\begin{equation}\label{eq_prop_PDE_2}
- \partial_{\tau}^2G_{\pi\pi}^E(\tau,\hat{m}\cdot\hat{m}_0) -\frac{1}{d-1}\Delta^{(S^{d-1})} G_{\pi\pi}^E(\tau,\hat{m}\cdot\hat{m}_0)=\delta(\tau)\delta^{(S^{d-1})}\left(\hat{m}-\hat{m}_0\right)\,.
\end{equation}
The simplest and most natural option is to decompose $G_{\pi\pi}^E(\tau,x)$ into spherical harmonics and then solve the ODE for $\tau$. This yields the propagator as a sum of Gegenbauer polynomials
\begin{equation}\label{Eq:CylinderProp}
    G_{\pi\pi}^E(\tau,\hat{m}\cdot\hat{m}_0)=-\frac{1}{2\Omega_{d-1}}|\tau|+\sum_{\ell=1}^{\infty}\frac{2\ell+d-2}{(d-2)\Omega_{d-1}}\frac{e^{-\omega_{\ell}|\tau|}}{2\omega_{\ell}}C_{\ell}^{\left(\frac{d-2}{2}\right)}\left(\hat{m}\cdot\hat{m}_0\right)\,,
\end{equation}
where the $\omega_{\ell}$'s are given in~\eqref{eq_freq_phonon}.
This decomposition makes it manifest that the propagator decays exponentially at large Euclidean time distances $|\tau|\gg 1$. Upon Wick rotating $\tau \rightarrow it - \, \epsilon$ ($\epsilon>0$), this yields the Wightman function~\eqref{Eq:WightmanCyl}.

At short distances (but still within EFT) we can use a flat space approximation and we find
\begin{equation}\label{Eq:PropSmallAngle2}
G_{\pi\pi}^E(\tau,\cos\sigma)\stackrel{\tau,\sigma\rightarrow 0}{\simeq}\frac{(d-1)^{\frac{d-1}{2}}}{(d-2)\Omega_{d-1}\left[\tau^2+(d-1)\sigma^2\right]^{\frac{d-2}{2}}}\,.
\end{equation}
Expanding around~\eqref{Eq:PropSmallAngle2} we may further obtain subleading orders in the short-distance approximation.
Wick-rotating~\eqref{Eq:PropSmallAngle2}, we obtain the short-distance Wightman propagator~\eqref{Eq:PropSmallAngle}.

\subsection{An alternative decomposition of the propagator}\label{app_prop_dS}

It is also possible, and in fact convenient for our purposes, to proceed oppositely as what we did to obtain~\eqref{Eq:CylinderProp}. In other words, we can work in Fourier basis along the $\mathbb{R}$ direction and write
\begin{align}\label{eq_prop_dS2}
    G_{\pi\pi}^E(\tau, \hat{m}\cdot \hat{m}_0) = \int \frac{d \omega}{2 \pi} e^{i \omega \tau} F_{\omega}(\hat{m}\cdot \hat{m}_0)\,,
\end{align}
where $F_{\omega}(\hat{m}\cdot \hat{m}_0)$ solves the equation
\begin{align}
  \left[ \omega^2 - \frac{1}{d-1} \Delta^{(S^{d-1})} \right]F_{\omega}(\hat{m}\cdot \hat{m}_0) = \delta^{(S^{d-1})}(\hat{m}-\hat{m}_0)\,.
\end{align}
We thus recognize $F_{\omega}(\hat{m}\cdot\hat{m}_0)/(d-1)$ as the propagator of a scalar on the sphere with mass $m^2 = (d-1) \omega^2$ on $S^{d-1}$ (see for instance \cite{SalehiVaziri:2024joi}), namely
\begin{align}\label{eq_F_sphere}
    \frac{F_{\omega}(x)}{(d-1)} = \frac{\Gamma(\Delta_{\omega})\Gamma(d-2-\Delta_{\omega})}{(4 \pi)^{\frac{d-1}{2}}\Gamma(\frac{d-1}{2})} {}_2F_1\left(\Delta_{\omega} ,d-2-\Delta_{\omega} ;\frac{d-1}{2};\frac{1+x}{2} \right)\,,
\end{align}
where $x=\hat{m}\cdot\hat{m}_0$ and $\Delta_{\omega}$ is the solution of
\begin{align}\label{eq_Delta_m}
    \Delta_{\omega}(d-2-\Delta_{\omega}) = (d-1)\omega^2\,.
\end{align}
Note that \eqref{eq_F_sphere} is symmetric under $\Delta_{\omega}\leftrightarrow d-2-\Delta_{\omega}$ and hence we can always restrict to a single arbitrary solution of~\eqref{eq_Delta_m}.

Physically, the representation~\eqref{eq_prop_dS2} can be seen as the natural one that is obtained when quantizing the EFT on $dS_{d-1}\times \mathbb{R}_{\tau}$ and then Wick-rotating de Sitter back to the sphere.\footnote{This is therefore reminiscent of but different from angular or Rindler quantization, see e.g.~\cite{Agia:2022srj}. Rindler coordinates are also convenient in celestial holography, see e.g.~\cite{Cheung:2016iub}.}  Note, however, that when doing this the EFT ground state (or any primary state) has complex momentum along $\mathbb{R}$ and the natural unitarity conditions obeyed by the CFT are therefore not manifest in this quantization procedure. The convenience of the representation~\eqref{eq_prop_dS2} is that the sphere propagator decays exponentially for $\arccos(x)\gtrsim 1/\Delta_{\omega}$, similarly to the terms with $\omega_{\ell}|\tau|\gtrsim 1$ in \eqref{Eq:CylinderProp}.  Therefore the contribution from the region $|\omega|\gg 1/\arccos (x)$ is negligible in the integral~\eqref{eq_prop_dS} - which makes this representation suitable for numerical evaluation at finite angular distance. The same remains true when Wick-rotating to the Lorentzian cylinder $\tau\rightarrow i t$ as long as we consider space-like distances from the viewpoint of the sound-cone, i.e. $[\arccos (x)]^2(d-1)>t^2$, so that $e^{ \omega |t|} F_{\omega}(\hat{m}\cdot \hat{m}_0)\sim e^{-|\omega|\left(\sqrt{d-1}\arccos (x)-|t|\right)}$ is indeed small at large $\omega$.

To obtain a representation which is suitable for numerics, it is useful to consider the large $\omega$ limit of $F_{\omega}(x)$. We focus on $d=3$. Using $\Delta_{\omega}=\frac{1}{2}+i\left(\sqrt{2} \omega -\frac{1}{8 \sqrt{2} \omega }+\ldots\right)$ for $\omega\gg 1$ and that~\eqref{eq_F_sphere} simplifies to
\begin{equation}\label{eq_sphere_prop_2d}
\frac{F_\omega(x)}{2}\stackrel{d=3}{=}\frac{1}{4 \sin (\pi\Delta_{\omega})}P_{\Delta_{\omega} -1}(-x)\,,
\end{equation}
for positive $\omega$ we find the following asymptotic behavior: 
\begin{equation}\label{eq_F_asymp}
\begin{aligned}
F_\omega(x)=2e^{-\sqrt{2} \omega  \arccos(x)} &\left\{
\frac{\sqrt{\frac{1}{\omega }}}{2^{7/4} \sqrt{\pi } \sqrt[4]{1-x^2}}
-\frac{\left(\frac{1}{\omega }\right)^{3/2} \left[x\sqrt{1-x^2} -\left(1-x^2\right) \arccos(x)\right]}{32 \sqrt{\pi } \sqrt[4]{2-2 x^2} \left(1-x^2\right)}\right.\\
&\left.+O\left(\frac{1}{\omega^{5/2}}\right)\right\}
+O\left(e^{-2\sqrt{2}\pi\omega+\sqrt{2}\omega\arccos(x)}\right)\equiv F^{(asymp)}_{\omega}(x)\,.
\end{aligned}
\end{equation}
To obtain a numerical approximation to the Wightman propagator we thus truncate the integral in~\eqref{eq_prop_dS} at a cutoff $\Lambda\gtrsim [\sqrt{2}\arccos (x)-|t|]^{-1}$ and partially account for the remainder integrating~\eqref{eq_F_asymp}
\begin{equation}\label{eq_app_G_long}
G_{\pi \pi}(t, x)=G_{\pi\pi}^E(it, x) \simeq \int^{\Lambda}_{-\Lambda} \frac{d \omega}{2 \pi} e^{ \omega t} F_{\omega}(x)+
 \int^{\infty}_{\Lambda} \frac{d \omega}{2 \pi} (e^{ \omega t}+e^{-\omega t}) F_{\omega}^{(asymp)}(x)\,.
\end{equation}
The integral over the asymptotic term $F_{\omega}^{(asymp)}$ may be performed analytically. Truncating to the second nontrivial order as in~\eqref{eq_F_asymp}, we obtain an expression which is correct up to terms of order $\sim e^{\Lambda [|t|-\sqrt{2}\arccos (x)]}/\Lambda^{5/2}$. Analogous expressions can be derived in this way for derivatives of the propagator.

Let us finally comment on a subtlety. The sphere propagator~\eqref{eq_sphere_prop_2d}, on $S^2$, diverges as $\sim 1/\omega^2$ in the massless limit $\omega\rightarrow 0$. This makes the integrals~\eqref{eq_prop_dS2} and~\eqref{eq_app_G_long} divergent near $\omega=0$. This divergence may be regulated by removing from the integral a small integration region near $\omega=0$ and adding a suitable linear function of $a+b t$, with coefficients that diverge as the removed region shrinks to a point. Physically, this ambiguity is associated with the treatment of the zero-mode of the field. In practice, we shall only be interested in using expressions analogous to~\eqref{eq_app_G_long} to compute second derivatives of the propagator, for which the zero-mode contribution drops out. For these, the integral over $\omega$ is manifestly finite and receives only infinitesimal contributions from $\omega=0$. Hence we do not need to worry about this subtlety for our purposes.

\section{Disconnected contributions to event shapes}\label{App:Disconnected}

In this Appendix we compute the disconnected contributions to the $1/\Delta_Q$ correction to event shapes, that arise from the corrections to the saddle-point in the Fourier transform of the leading factorized result. For concreteness, we discuss in detail only the CCC, but the procedure generalizes straightforwardly to other event shapes. At the end of the section we also report the result for EEC and for event shapes involving generalized detectors constructed from scalar local operators as in~\eqref{Eq:GenDetector}.

We follow \cite{Firat:2023lbp}. The calculation consists of three simple steps. We first analytically continue the disconnected cylinder correlator~\eqref{eq_many_J_EFT} to Minkowski. To this aim it is easier to work out these correlators in Euclidean signature and then analytically continue the result to obtain the desired Lorentzian Wightman function. We then integrate over the the light-rays. Finally, we perform the Fourier transform of the source and sink operators. Note that this procedure, in some ways, is opposite compared to the discussion in Sec.~\ref{Sec:Detectors}, where we began from the Fourier transform. This is possible, and in fact convenient, due to the simplicity of the leading order correlator but, as discussed in the main text, this strategy is impractical at subleading orders.

\paragraph{Analytic continuation from Euclidean}

The semiclassical nature of the large charge EFT implies that correlations function of local operators, at leading order in $1/Q$, factorize into expectation values for the single operators. In particular for the $U(1)$ current we have  
\begin{align}
    &\begin{aligned}
    & \frac{ \langle \Onb(x_{fE})   J^{\mu}(x_{1E}) J^{\nu}(x_{2E}) \On(x_{iE})\rangle_E}{ \langle \Onb(x_{fE})  \On(x_{iE})\rangle_E} =\\
     &\qquad
     \frac{ \langle \Onb(x_{fE})   J^{\mu}(x_{1E})  \On(x_{iE})\rangle_E}{ \langle \Onb(x_{fE})  \On(x_{iE})\rangle_E}\cdot 
     \frac{ \langle \Onb(x_{fE})    J^{\nu}(x_{2E}) \On(x_{iE})\rangle_E}{ \langle \Onb(x_{fE})  \On(x_{iE})\rangle_E}
     \left(1+ O(Q^{-\frac{d}{d-1}}) \right)\,, \label{Eq:JJFactor} 
     \end{aligned}
\end{align}
where, with the subscript $_E$ we indicate that we are evaluating the previous correlators in Euclidean signature. The positions of the operators are also meant in Euclidean space. The three point functions on the RHS of the previous equations are completely fixed by conformal invariance and by the Ward identities
\begin{align}
\begin{aligned}
    \langle \bar{\mathcal{O}}_{n}(x_{fE})   J^{\mu}(x_{E})  \mathcal{O}_{n}(x_{iE})\rangle_E &= \frac{Q}{\Omega_{d-1}}\frac{V^{\mu}}{|x_{fE} - x_{iE}|^{2 \Delta_Q - d+2} |x_{E} - x_{iE}|^{d-2}|x_{fE} - x_{E}|^{d-2}}\,,\\
    V^{\mu} &= \frac{(x_{fE}-x_E)^{\mu}}{|x_{fE}-x_E|^2}+\frac{(x_{E}-x_{iE})^{\mu}}{|x_{E}-x_{iE}|^2}\,.
    \end{aligned}
\end{align}

We can now Wick rotate~\eqref{Eq:JJFactor} to Lorentzian signature using $J^0(x^0,x^i)=J^0_E(i x^0,x^i)$ and $J^j(x^0,x^i)=i\,J^j_E(i x^0,x^i)$, to guarantee that the current remains Hermitian. The following replacements
\begin{equation}
 x^0_{iE}\rightarrow i x^0_i-3 \epsilon\,,\quad x^0_{1E}\rightarrow i x^0_1-2\epsilon\,,\quad x^0_{2E}\rightarrow i x_2^0-2\epsilon   \,,\quad x^0_{fE}\rightarrow i x^0_f-\epsilon
\end{equation}
with $\epsilon>0$, guarantee the correct ordering since the two detectors commute. Note that the overall result has an $i$ upfront:
\begin{equation}\label{eq_prescrip2}
    \langle \bar{\mathcal{O}}_{n}(x_{f})   J^{\mu}(x)  \mathcal{O}_{n}(x_{i})\rangle =i\,
    \frac{Q}{\Omega_{d-1}}\frac{\frac{(x_f-x)^\mu}{-(x_f-x)^2}-\frac{(x_i-x)^\mu}{-(x_i-x)^2}}{(-x_{fi}^2)^{\Delta_Q-\frac{d-2}{2}}\left[-(x-x_i)^2\right]^{\frac{d-2}{2}}\left[-(x-x_f)^2\right]^{\frac{d-2}{2}}}\,.
\end{equation}

\paragraph{Light-ray integral}
From \eqref{Eq:JJFactor} it follows
\begin{align}\label{Eq:QQ2Factor}
    \langle \Onb(x_{f}) \Qno \Qnt \mO_Q(x_i) \rangle_{\text{disc}} &= \frac{\langle \Onb(x_{f}) \Qno \On(x_i) \rangle \langle \Onb(x_{f}) \Qnt \On(x_i) \rangle  }{\langle \Onb(x_{f})  \On(x_i) \rangle}\,, 
\end{align}
i.e. the matrix element of detector operators is the product of the expectation value of a single detector. Thus, we just need to singularly compute each of the factors in the previous equation. This is a fairly standard procedure (see, for instance, \cite{Hofman:2008ar}), but we review it here for completeness.

Starting from the definition \eqref{charge_detector}, we first take the limit $x_{\mathbf{n}_i}^+\rightarrow\infty$ of the three point function with the current 
\begin{equation}\label{eq_prescrip3}
\begin{aligned}
     \lim_{\xn^+ \rightarrow + \infty} (\xn^{+})^{d-2}&\langle \Onb (x_{f})   \bar{n} \cdot J(\xn) \On (x_{i}) \rangle =\\ 
     =i\frac{Q}{\Omega_{d-1}}&\frac{n\cdot x_{fi}}{(-x_{fi}^2)^{\Delta_{Q}-\frac{d-2}{2}}(
     n\cdot x_f-x^-_n-i\epsilon)^{\frac{d}{2}}
     (  n\cdot x_i-x^-_n+i\epsilon)^{\frac{d}{2}}}\,,
     \end{aligned}
\end{equation}
Then, we integrate over the light-ray 
\begin{equation}
 \langle \Onb (x_{f}) \Qno \On (x_i) \rangle
 =\frac{Q}{\Omega_{d-1}}2^{2-d}\pi \frac{\Gamma(d  -1)}{\Gamma(d/2)^2}\frac{i^{d-2} }{(-x_{fi}^2)^{\Delta_{Q}-\frac{d-2}{2}}(n\cdot x_{if}+i\epsilon)^{d-2}}
 \,,
\end{equation}
where we used
\begin{equation}
    \int_{-\infty}^{\infty} d \xn^- \frac{1}{(
     n\cdot x_f-x^-_n-i\epsilon)^{\alpha}
     (  n\cdot x_i-x^-_n+i\epsilon)^{\beta}} = \frac{\Gamma( \alpha + \beta -1)}{\Gamma(\alpha)\Gamma(\beta)}\frac{2 \pi \,i^{2\beta-1}}{(n\cdot x_{if}+i \epsilon)^{ \alpha +\beta -1}}\,.
\end{equation}

\paragraph{Fourier transform}
The last step is to take the Fourier transform. Namely we need
\begin{align}
    \langle \Qno \Qnt \rangle_{\text{disc}} =\frac{\int d^4x_{fi} e^{ip\cdot x_{fi}} \langle \Onb(x_f)\Qno \Qnt \On(x_i)\rangle}{\int d^4x_{fi} e^{ipx_{fi}} \langle \Onb(x_f)\On(x_i)\rangle}\,,
\end{align}
where the numerator can be read from \eqref{Eq:QQ2Factor}. We use the following Fourier transforms
\begin{align}
&\begin{aligned}\label{Eq:ExactFourier2Point}
\int d^d x \frac{e^{ip\cdot x}}{(-x^2)^{\Delta_Q}}
    =\frac{\pi^{\frac{d+2}{2}}(p^2)^{\Delta_Q-d/2}}{2^{2\Delta_Q-d-1}\Gamma(\Delta_Q)\Gamma\left(\Delta_Q-\frac{d-2}{2}\right)}\Theta(p^0)\Theta(p^2)
    \,,
\end{aligned}\\
&\begin{aligned}
\int d^dx \frac{e^{iE \, x^0}}{(-x^2)^{\alpha}(-n_2\cdot x+i\epsilon)^{\beta}(-n_1\cdot x+i\epsilon)^{\beta}}&=
\frac{i^{-2\beta}\pi^{\frac{d+2}{2}}E^{2\alpha+2\beta-d}}{2^{2\alpha+2\beta-d-1}
\Gamma(\alpha)\Gamma\left(\alpha+2\beta-\frac{d-2}{2}\right)
}\\
\label{eq_FT_4pt}
&\times
\, _2F_1\left(\beta ,\beta ;\alpha +2 \beta -\frac{d-2}{2} ;\frac{n_1\cdot n_2}{2}\right)
\Theta(E)\,, 
\end{aligned}
\end{align}
where we evaluated the second integral in the rest frame $p=(E,\mathbf{0})$. We obtain 
\begin{align}
    \langle \Qno \Qnt \rangle_{\text{disc}} &= \left(\frac{Q}{\Omega_{d-2}} \right)^2 \frac{\Gamma(\Delta_Q)\Gamma(\Delta_Q-\frac{d}{2}+1)}{\Gamma(\Delta_Q-d+2)\Gamma(\Delta_Q+\frac{d}{2}-1)}\times \notag \\&\hspace{4cm}\times\,_2F_1\left( d-2,d-2,\Delta_Q +\frac{d}{2}-1,\frac{n_1 \cdot n_2}{2}\right)\\
    &\simeq \left(\frac{Q}{\Omega_{d-2}} \right)^2 \left[ 1-\frac{(d-2)^2\,\mathbf{n}_1\cdot \mathbf{n}_2 }{2 \Delta_Q}+\ldots\right]\,.\notag
\end{align}
where we expanded for large $\Delta_Q$. 

With an analogous procedure we find the disconnected contribution to EEC. Using $T^{00}(x^0,x^i)=T^{00}_E(i x^0,x^i)$, $T^{0j}(x^0,x^i)=i\, T^{0j}_E(i x^0,x^i)$ and $T^{ij}(x^0,x^i)=-T^{ij}_E(i x^0,x^i)$, we find that the light-ray transform of the three-point function reads
\begin{equation}
    \langle\bar{\mO}_Q(x_f)\mathcal{E}(\mathbf{n})\mO_Q(x_i)\rangle=\lambda_{\bar{\mO}_QT\mO_Q}
    \frac{\Gamma(d+1)}{\Gamma\left(\frac{d+2}{2}\right)^2}\frac{2^{1-d}\pi\, i^{d-1}}{(-x_{fi}^2)^{\Delta_Q-\frac{d-2}{2}}(n\cdot x_{if}+i\epsilon)^{d-1}}\,,
\end{equation}
where the OPE coefficient is $\lambda_{\bar{\mO}_QT\mO_Q}=\frac{d\Delta_Q}{(d-1)\Omega_d}$. Using the formulas above for the Fourier transform, we obtain:
\begin{align}
    \langle \Eno \Ent \rangle_{\text{disc}} &= \left(\frac{E}{\Omega_{d-2}} \right)^2 \frac{\Delta_Q\,\Gamma(\Delta_Q+1)\Gamma(\Delta_Q-\frac{d}{2}+1)}{\Gamma(\Delta_Q-d+2)\Gamma(\Delta_Q+\frac{d}{2}+1)}\times \notag \\ &\hspace{4cm}\times\,_2F_1\left( d-1,d-1,\Delta_Q +\frac{d}{2}+1,\frac{n_1 \cdot n_2}{2}\right)\\
    &\simeq \left(\frac{E}{\Omega_{d-2}} \right)^2 \left[1-\frac{1+(d-1)^2\mathbf{n}_1\cdot \mathbf{n}_2}{2 \Delta_Q}+\ldots\right]\,. \notag
\end{align}

Finally, for a detector obtained from a scalar $\mathcal{O}$ with scaling dimension $\Delta$, we get
\begin{align}
    \langle \mathcal{D}_{\mathcal{O}}(\mathbf{n}_1) &\mathcal{D}_{\mathcal{O}}(\mathbf{n}_2) \rangle_{\text{disc}}=\frac{\pi ^2 4^{2-\Delta } \lambda_{\Onb \mathcal{O}\On}^2 \Gamma (\Delta -1)^2 \Gamma (\Delta_Q) \Gamma \left(-\frac{d}{2}+\Delta_Q+1\right)}{E^2 \Gamma \left(\frac{\Delta }{2}\right)^4 \Gamma (\Delta_Q-\Delta ) \Gamma \left(-\frac{d}{2}+\Delta +\Delta_Q-1\right)}\times \notag \\ &\hspace{4cm}\times\,_2F_1\left(\Delta-1,\Delta-1,\Delta +\Delta_Q-\frac{d}{2}-1 , \frac{n_1\cdot n_2}{2}\right)\\
    \notag
    \simeq &\left[
    \frac{\lambda_{\Onb \mathcal{O}\On}\sqrt{\pi }  \Gamma \left(\frac{\Delta -1}{2}\right) \Gamma (\Delta_Q) \Gamma \left(\Delta_Q+1-\frac{d}{2}\right)}{E\, \Gamma \left(\frac{\Delta }{2}\right) \Gamma \left(\Delta_Q-\frac{\Delta }{2}\right) \Gamma \left(\frac{\Delta -d}{2}+\Delta_Q\right)}\right]^2\left[1
    -\frac{1+(\Delta-1)^2\mathbf{n}_1\cdot \mathbf{n}_2}{2\Delta_Q}+\ldots
    \right]\,.
\end{align}

\section{Details on energy and charge correlators}\label{App:Connected}

In this Appendix we collect additional details on the EEC and CCC calculations presented in Sec.~\ref{Sec:EEC}. 

\subsection{Collinear limit}\label{App:AddColl}

In the collinear limit we can replace the Goldstone propagator $G_{\pi \pi}$ with the short distance expression in \eqref{Eq:PropSmallAngle}. It is convenient to define 
\begin{align}\label{Eq:DistanceOp}
    \delta t= (t_1 -t_2 -i \epsilon)\,,&& \cos \sigma = \hat{m}_1\cdot \hat{m}_2\,,
\end{align}
where we take $\epsilon>0$ for definiteness. The sign of $\epsilon$ is anyhow irrelevant since the two operators are everywhere spacelike separated. However it is important that a nonzero $\epsilon$ regulates the propagator at coincident points. Recall indeed that Lorentzian correlators are distributions obtained via a limit procedure from Euclidean ones; we will show below that retaining a nonzero $\epsilon$ is indeed crucial to obtain the distributional terms localized at $\theta=0$.

In 3 dimensions, solving~\eqref{eq_prop_PDE_2} with the boundary condition~\eqref{Eq:PropSmallAngle2} up to $O(\delta t^2,\sigma^2)$, we get 
\begin{align}\label{eq_prop_small_angle_3d}
G_{\pi\pi}(\delta t,\cos\sigma)&\stackrel{d=3}{=}\frac{1}{2\pi\left(2\sigma^2-\delta t^2\right)^{1/2}}\left(1-\frac{\sigma^2-\delta t^2}{12}\right)+\text{const.}+\ldots\,,
\end{align}
where the constant term is fixed from regularity at infinity, and is hence undetermined from this expansion only.

Provided the previous expression, the correlators of interest can be readily obtained via \eqref{eq_QQ_conn} and \eqref{eq_EE_conn}. The explicit calculation is straightforward but somewhat lengthy. Thus, in the following, we just provide the details of the derivation of the dominant contribution in the collinear limit, in $d=3$ for the CCC. This can be simply generalized to obtain the subleading term at small $\theta$ in \eqref{Eq:ShortDistanceQQ}, to EEC and to higher dimensions.

It is convenient to introduce the following coordinates: 
\begin{equation}\label{eq_tpm}  
\tilde{t}_{\pm} = \frac{t^-_1 \pm t^-_2}{2}\,.  
\end{equation}  
The distance between the operators in \eqref{Eq:DistanceOp} becomes  
\begin{align}  
\delta t = \tilde{t}_- - i \epsilon\,, && \cos\sigma &=  
\frac{1}{2} \left[(1+\cos\theta) \cos\tilde{t}_- - (1-\cos\theta) \cos\tilde{t}_+\right]\,,
\end{align}  
and we identify  
\begin{equation}
\tilde{t}_-^{\,2} \sim \theta^2 \ll 1  \,,
\end{equation}
as the integration region that yields the enhanced contribution. In this region, the previous angular distance and the measure in \eqref{eq_QQ_conn} can be expanded as  
\begin{align}\label{Eq:ExpAngDist}
    \sigma^2 = \tilde{t}_-^{\,2} + \frac{\theta ^2}{2} \left(1+ \cos \tilde{t}_+\right) + O(\theta^4,\theta^2 \tilde{t}_-^2)\,,
\end{align}  
and, using that the integrand is symmetric under $\tilde{t}_+\leftrightarrow-\tilde{t}_+$,  
\begin{align}
    \iint dt_1^- d t_2^- \cos\frac{t_1^-}{2} \cos \frac{t_2^-}{2} &&\longrightarrow && 4\int_{0}^{\pi} d\tilde{t}_+ \int_{- \pi+\tilde{t}_+}^{\pi-\tilde{t}_+} d\tilde{t}_- \left[ \frac{1+\cos\tilde{t}_+ }{2} + O(\tilde{t}_-^{\,2}) \right]\,. 
\end{align}  
Given \eqref{Eq:ExpAngDist}, we can immediately expand also the integrand to obtain  
\begin{align}\label{eq_app_short_int}
    \frac{\langle \Qno \Qnt \rangle^{(1)}_{\text{conn}}}{\langle \Qno\rangle\langle \Qnt\rangle} &\simeq -\frac{1}{6} \int_{0}^{\pi} d\tilde{t}_+ \int_{- \pi+\tilde{t}_+}^{\pi-\tilde{t}_+} d\tilde{t}_-  
    (1+\cos \tilde{t}_+)  
    \frac{\tilde{t}_-^{\,2} +2i\epsilon \,\tilde{t}_- 
    -5 \epsilon^2 + \theta^2 +\theta^2\cos \tilde{t}_+}
    {(\tilde{t}_-^{\,2}+ 2 i \epsilon \tilde{t}_- + \epsilon^2 +\theta^2+\theta^2\cos \tilde{t}_+)^{5/2}}\,,
\end{align}  
where, in the expansion, we considered \(\epsilon \sim \theta \sim \tilde{t}_-\). This is crucial to obtain the correct "Pf" distribution in \eqref{Eq:ShortDistanceQQ}. 

\begin{figure}[t]
    \centering
    \includegraphics[width=.5\linewidth]{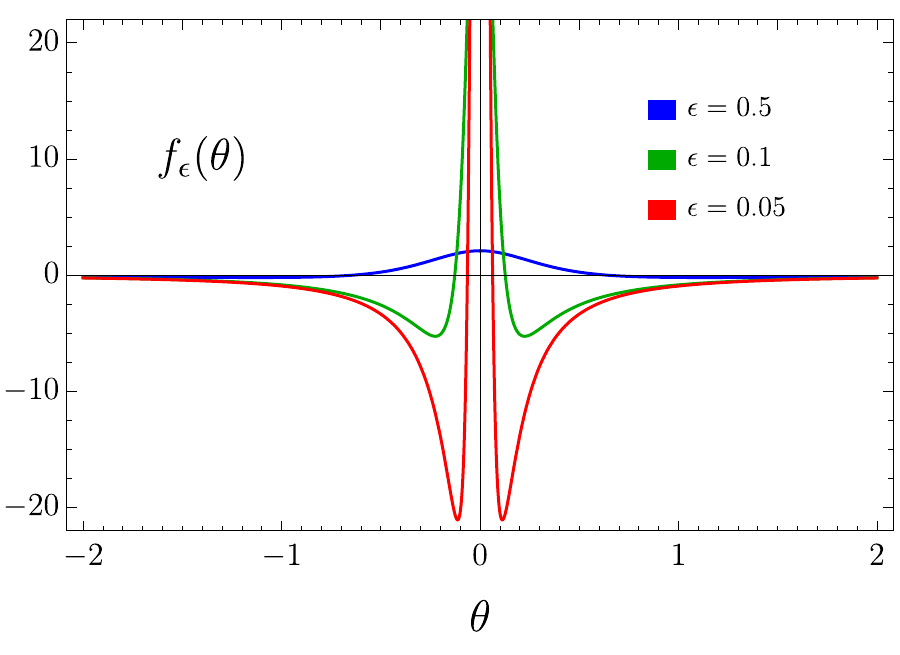}
    \caption{Collinear limit of charge-charge correlators for different values of the regulator $\epsilon$, see \eqref{eq_app_short_res} for more details. }
    \label{fig:Coll}
\end{figure}

Performing the integral~\eqref{eq_app_short_int} we get 
\begin{align}\label{eq_app_short_res}
    \frac{\langle \Qno \Qnt \rangle^{(1)}_{\text{conn}}}{\langle \Qno\rangle\langle \Qnt\rangle}\simeq -\frac{\pi}{3\theta^2} +\frac{\pi \,\epsilon\,(\epsilon^2+2\theta^2)}{3 \theta^2 (\epsilon^2 +\theta^2)^{3/2}}\equiv f_{\epsilon}(\theta)\,.
\end{align}
This function is plotted in Fig.~\ref{fig:Coll}: it decreases as $-\pi/(3\theta^2)$ until $\theta\sim \epsilon$, and then it has a sudden positive peak. In other words, the limit $\theta\rightarrow 0$ and $\epsilon\rightarrow 0$ do not commute:
\begin{equation}
    \lim_{\epsilon\rightarrow 0}f_{\epsilon}(\theta)=-\frac{\pi}{3\theta^2}\,,\qquad
    \lim_{\theta\rightarrow 0}f_{\epsilon}(\theta)=\frac{\pi}{6 \epsilon^2}\,.
\end{equation}
Therefore in the limit $\epsilon\rightarrow 0$ we obtain a nontrivial distribution. To identify which distribution, we simply need to integrate $f_{\epsilon}(\theta)$ against suitable test functions $g(\theta)$. We only consider test functions which are regular at $\theta=0$: $g(\theta)=g(0)+g'(0)\theta+\ldots$. Since $f_{0}(\theta)=-\pi/(3\theta^2)$ it suffices to check the integrals of $f_{\epsilon}(\theta)$ and $\theta f_{\epsilon}(\theta)$. Given arbitrary positive constants $B$ and $A$, we have
\begin{align}
\int^A_{-B} d\theta f_{\epsilon}(\theta)&=\frac{\pi  \left(1-\frac{\epsilon}{\sqrt{\theta ^2+\epsilon^2}}\right)}{3 \theta }\Bigg\vert_{-B}^A\xrightarrow{\epsilon\rightarrow 0}\frac{\pi}{3 A}+\frac{\pi}{3 B},,\\
 \int^A_{-B} d\theta \,\theta f_{\epsilon}(\theta)&=-\frac{ \pi }{3} \left[\frac{\epsilon }{\sqrt{\theta ^2+\epsilon ^2}}+\log \left(\sqrt{\theta ^2+\epsilon ^2}+\epsilon \right)\right]\Bigg\vert_{-B}^A\xrightarrow{\epsilon\rightarrow 0}\frac{\pi}{3}\log\frac{A}{B}\,.
\end{align}
In the limit $\epsilon\rightarrow 0^+$ the results agree with the ``Pf" distribution as defined in~\eqref{eq_Pf_def}. Note that, similarly to the principal value distribution, to integrate the finite part distribution in practice we may just apply the fundamental theorem of calculus neglecting the singularity at $\theta=0$.  The finite part distribution is equivalent to a second order plus distribution up to a regular contribution~\cite{Ebert:2018gsn} and, analogously to the relation between the plus distribution and its higher order generalizations, coincides with the derivative 
of the principal value distribution.

\subsection{Charge and momentum conservation}\label{App:Conservations}
$U(1)$ charge  and momentum conservation, following from Ward identities, directly imply \eqref{Eq:CCons} and \eqref{Eq:ECons}. In this Appendix we test the validity of these relations at Next-to-Leading order. 

First of all, we notice that the leading order results (see for instance \eqref{Eq:CCLO}) by themselves saturate~\eqref{Eq:CCons} and~\eqref{Eq:ECons}. This implies that
\begin{align}
  \int d^{d-2} \Omega_{\mathbf{n_1}} \langle \Qno \Qnt \rangle^{(1)}= 0 \quad&\implies\quad  
  \int_0^{2 \pi} d \theta \left.\langle \Qno \Qnt \rangle^{(1)}\right|_{\mathbf{n}_1\cdot \mathbf{n}_2 = \cos \theta} = 0\,,\\ \label{Eq:ConsEEC}
  \int d^{d-2} \Omega_{\mathbf{n_1}} \langle \Eno \Ent \rangle^{(1)}= 0 \quad &\implies\quad  
  \int_0^{2 \pi} d \theta \left.\langle \Eno \Ent \rangle^{(1)}\right|_{\mathbf{n}_1\cdot \mathbf{n}_2 = \cos \theta} = 0\,,\\ \label{Eq:ConsEEC1}
  \int d^{d-2} \Omega_{\mathbf{n_1}}\, \mathbf{n}_1 \langle \Eno \Ent \rangle^{(1)}= 0 \quad &\implies\quad  
  \int_0^{2 \pi} d \theta \cos \theta\,\left.\langle \Eno \Ent \rangle^{(1)}\right|_{\mathbf{n}_1\cdot \mathbf{n}_2 = \cos \theta} = 0\,.
\end{align}
The integrals can be performed over the position of one detector or, equivalently, over the detectors distance.

Let us start with the CCC. The disconnected contribution \eqref{Eq:DiscCC} trivially integrates to zero, therefore also the connected one must integrate to zero
\begin{align}
    \int_0^{2 \pi} d \theta_1 \langle \Qno \Qnt \rangle^{(1)}_{\text{disc}} = 0&&\Longrightarrow&&\int_0^{2 \pi} d \theta_1 \langle \Qno \Qnt \rangle^{(1)}_{\text{conn}} = 0\,.
\end{align}
How this is realized is nontrivial from the definition \eqref{Eq:CCCyl} and deserves some discussion. We focus for concreteness on $d=3$. Expanding the Legendre polynomial in terms of spherical harmonics, the propagator \eqref{Eq:WightmanCyl} can be conveniently rewritten as 
\begin{equation}\label{Eq:PropDec}
    G_{\pi\pi}(t_1-t_2,\hat{m}_1 \cdot \hat{m}_2)=
\sum_{\ell=1}^{\infty}\frac{e^{-i\omega_{\ell}(t_1-t_2)}}{2\omega_{\ell}}\sum_{m=-\ell}^{\ell}Y_{\ell m}(\sigma_1,\theta_1)(Y_{\ell m}(\sigma_2,0))^*
    -\frac{i}{8\pi}(t_1-t_2)
    \,,
\end{equation}
where we expressed $\hat{m}_i$ in spherical coordinates $(\sigma_i,\theta_i)$. We conveniently set $\theta_2 = 0$ so that $\theta_1$ coincides with the angular detectors distance ($\theta = \theta_1$). Given the explicit form of the connected correlator~\eqref{eq_QQ_conn}, we want to use~\eqref{Eq:PropDec} to show
\begin{align}\label{Eq:CConsProp}
   \int_0^{2 \pi} d \theta_1 \langle \Qno \Qnt \rangle^{(1)}_{\text{conn}} = \frac{Q^2}{48 \pi}\int_0^{2\pi} d\theta_1
     \iint dt^-_1 dt^-_2 
     \cos\frac{t^-_1}{2}\cos\frac{t^-_2}{2}\,
     \prod_{i=1,2}\left[2\partial_{t_i}-\partial_{\sigma_i}\right]G_{\pi \pi}=0\,.
\end{align}
It is clear that the integral over $\theta_1$ selects only the $m = 0$ terms in \eqref{Eq:PropDec} and the zero mode does not contribute. The surviving spherical harmonics in \eqref{Eq:PropDec} need to be integrated term by term in \eqref{Eq:CConsProp}. We were not able to perform this integral in general, but we checked analytically that the integral on $dt_1^-$ vanishes for the first ten terms $\ell=1,\ldots,10$ on \texttt{Mathematica}. Furthermore, we checked numerically that our numerical result for the CCC in Sec.~\ref{Sec:EEC}, integrated over the detector distance $\theta$ accounting for the $\text{Pf}$ distribution at small angles, vanishes at the percent level.

We now turn to EEC. The disconnected contribution now integrates to a non-zero value. Working again in 3 dimensions from \eqref{Eq:DiscEE} we get
\begin{align}\label{eq_app_EE_int_disc}
   \int_0^{2 \pi} d\theta_1 \langle \Eno \Ent \rangle^{(1)}_{\text{disc}}= -\left(\frac{E}{2 \pi}\right)^2 \pi\,,
\end{align}
and 
\begin{align}\label{eq_app_EE_int_disc2}
   \int_0^{2 \pi} d\theta_1 \cos \theta_1 \langle \Eno \Ent \rangle^{(1)}_{\text{disc}}= -\left(\frac{E}{2 \pi}\right)^2 2\pi\,,
\end{align}
that must be compensated from the connected in order to ensure \eqref{Eq:ConsEEC} and \eqref{Eq:ConsEEC1}. We start with the former, using \eqref{Eq:EECyl} and \eqref{Eq:PropDec} we get 
\begin{align}\label{Eq:EConsProp}
   \int_0^{2 \pi} d \theta_1 \langle \Eno \Ent \rangle^{(1)}_{\text{conn}}= \frac{3E^2}{64 \pi}\int_0^{2\pi} d\theta_1
     \iint dt^-_1 dt^-_2 
     \left(\cos\frac{t^-_1}{2}\cos\frac{t^-_2}{2}\right)^3\,
\prod_{i=1,2}\left[3\partial_{t_i}-2\partial_{\sigma_i}\right]G_{\pi \pi}.
\end{align}
Again, the integral over $\theta_1$ selects only the terms with $m=0$ in \eqref{Eq:CConsProp} and $\ell\geq 1$. We evaluated on \texttt{Mathematica} the integrals in \eqref{Eq:EConsProp} for the first 10 terms $\ell=1,\ldots,10$ and we found that only the $\ell=1$ mode contributes. Assuming that all higher $\ell$ contributions also vanish, the $\ell=1$ term exactly compensates for the disconnected term~\eqref{eq_app_EE_int_disc}:
\begin{align}\label{eq_EE_WI}
    \int_0^{2 \pi} d\theta_1 \langle \Eno \Ent \rangle^{(1)}_{\text{conn}}= \left(\frac{E}{2 \pi}\right)^2 \pi=-\int_0^{2 \pi} d\theta_1 \langle \Eno \Ent \rangle^{(1)}_{\text{disc}}\,,
\end{align}
as required by Energy conservation. In complete analogy we compute the connected contribution to \eqref{Eq:ConsEEC1}. Only the $\ell=1$ mode contributes, now with $m=\pm 1$. From that we consistently get
\begin{align}\label{eq_EE_WI2}
    \int_0^{2 \pi} d\theta_1 \cos \theta_1 \langle \Eno \Ent \rangle^{(1)}_{\text{conn}}= \left(\frac{E}{2 \pi}\right)^2 2\pi=-\int_0^{2 \pi} d\theta_1 \cos \theta_1 \langle \Eno \Ent \rangle^{(1)}_{\text{disc}}\,.
\end{align} 
Note that the $\ell=1$ mode has $\omega_1=1$ and corresponds to a descendant of the large charge ground state; it should therefore not be surprising that it plays a nontrivial role in Ward identities. We also checked that our numerical results for the connected contribution satisfies~\eqref{eq_EE_WI} and~\eqref{eq_EE_WI2} at the few percent level when one properly accounts for the distribution~\eqref{Eq:ShortDistanceEE} in the collinear limit. This is a non-trivial check of the consistency of our results.

\subsection{Back-to-back limit}\label{App:BacktoBack}

In this appendix we provide some details on the derivation of the results~\eqref{eq_back_QQ} and~\eqref{eq_back_EE} in $d=3$. We will discuss explicitly only the CCC, as the analysis of EEC is identical.

\begin{figure}[t]
    \centering
    \includegraphics[width=.5\linewidth]{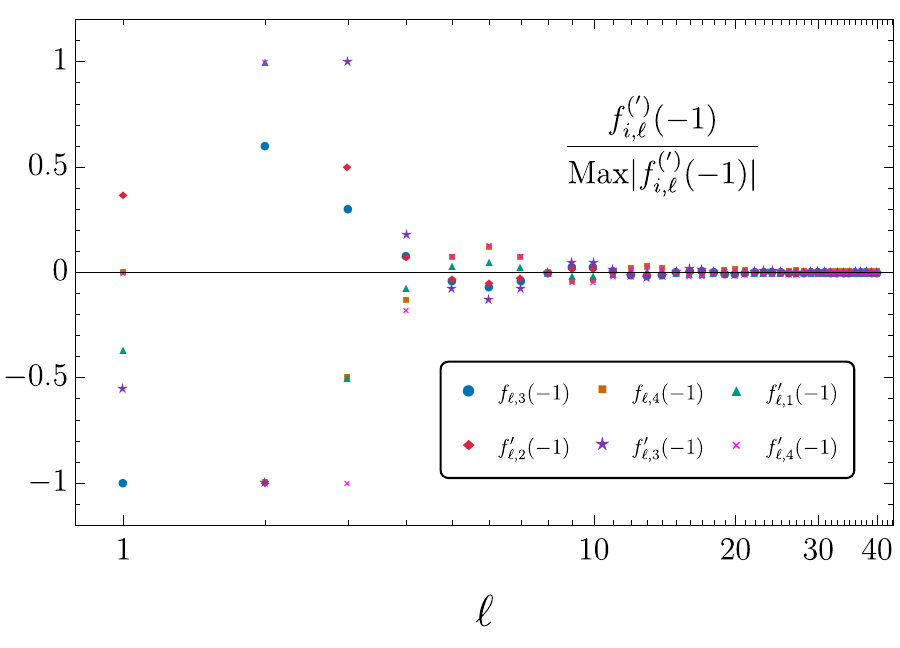}
    \caption{Various contributions to the back-to-back limit of charge-charge correlators for the first 40 harmonics ($\ell=1,\ldots, 40$). The plotted coefficients are defined in App.~\ref{App:BacktoBack}.}
    \label{fig:Back}
\end{figure}

\begingroup
\allowdisplaybreaks

It is convenient to define the summand in the decomposition~\eqref{Eq:WightmanCyl} as
\begin{equation}\label{eq_g_l}
    G_{\pi\pi}(t,x)=
    \sum_{\ell=1}^{\infty}g_{\ell}(t,x)-\frac{i}{4\pi}t\,,\qquad 
    g_{\ell}(t,x)=\frac{2\ell+1}{8\pi\omega_\ell}e^{-i\omega_{\ell}t}P_{\ell}\left(x\right)
    \,.
\end{equation}
We also introduce the following short-hand notation
\begin{equation}
 s_i=\sin\frac{t_i^-}{2}\,,\qquad c_i=\cos\frac{t_i^-}{2}\,.   
\end{equation}
Expanding the derivatives, we then recast~\eqref{eq_Fl_QQ} as
\begin{equation}\label{eq_F_l_QQ_app}
F^{(\ell)}_{QQ}(\cos\theta)=-\frac{\pi}{12}\left[
f_{1,\ell}(\cos\theta)+f_{2,\ell}(\cos\theta)+f_{3,\ell}(\cos\theta)+f_{4,\ell}(\cos\theta)\right]    \,,
\end{equation}
where
\begin{align}
    f_{1,\ell}(\cos\theta)&=4\iint dt_1^- dt_2^-c_1 c_2 \,g^{(2,0)}_{\ell}\,,\\
    f_{2,\ell}(\cos\theta)&=-2\iint dt_1^- dt_2^-c_1 c_2(s_1 c_2-c_2 s_1)(1+\cos\theta)g^{(1,1)}_{\ell}\,,\\
    f_{3,\ell}(\cos\theta)&=-\iint dt_1^- dt_2^-c_1 c_2 (s_1 s_2\cos\theta+c_1 c_2)g^{(0,1)}_{\ell}\,,\\
    f_{4,\ell}(\cos\theta)&=-\iint dt_1^- dt_2^-c_1 c_2 \left[c_1c_2s_1s_2(1+\cos^2\theta)-(c_1^2s_2^2+c_2^2s_1^2)\cos\theta\right]g^{(0,2)}_{\ell}\,.
\end{align}
At this point it is convenient to change coordinates as in~\eqref{eq_tpm}. Note that
\begin{equation}\label{eq_m1m2_app}
    \hat{m}_1\cdot \hat{m}_2\vert_{\theta=\pi}=-\cos\tilde{t}_+\,.
\end{equation}
Then expanding around $\theta=\pi$ we find
\begin{align}
    f_{1,\ell}(-1)=&8\int_0^\pi d\tilde{t}_+ \int_{-\pi+\tilde{t}_+}^{\pi-\tilde{t}_+} d\tilde{t}_-\left(\cos\tilde{t}_++\cos\tilde{t}_-\right)\,g^{(2,0)}_{\ell}(\tilde{t}_-,-\cos\tilde{t}_+)\,,\\
    f_{1,\ell}'(-1)=&2\int_0^\pi d\tilde{t}_+ \int_{-\pi+\tilde{t}_+}^{\pi-\tilde{t}_+} d\tilde{t}_-\left(\cos\tilde{t}_++\cos\tilde{t}_-\right)^2\,g^{(2,1)}_{\ell}(\tilde{t}_-,-\cos\tilde{t}_+)\,,\\
    f_{2,\ell}(-1)=&0\,,\\
    f_{2,\ell}'(-1)=&-2\int_0^\pi d\tilde{t}_+ \int_{-\pi+\tilde{t}_+}^{\pi-\tilde{t}_+} d\tilde{t}_-\sin\tilde{t}_-\left(\cos\tilde{t}_++\cos\tilde{t}_-\right)\,g^{(1,1)}_{\ell}(\tilde{t}_-,-\cos\tilde{t}_+)
    \,,\\
    f_{3,\ell}(-1)=&-2\int_0^\pi d\tilde{t}_+ \int_{-\pi+\tilde{t}_+}^{\pi-\tilde{t}_+} d\tilde{t}_-\cos\tilde{t}_+\left(\cos\tilde{t}_++\cos\tilde{t}_-\right)\,g^{(0,1)}_{\ell}(\tilde{t}_-,-\cos\tilde{t}_+)\,,\\
    \nonumber
f_{3,\ell}'(-1)=&\frac14\int_0^\pi d\tilde{t}_+ \int_{-\pi+\tilde{t}_+}^{\pi-\tilde{t}_+} d\tilde{t}_- \left[\cos (2 \tilde{t}_+)-\cos (2 \tilde{t}_-)\right] g^{(0,1)}_{\ell}(\tilde{t}_-,-\cos\tilde{t}_+)\\
    &-\frac{1}{2}\int_0^\pi d\tilde{t}_+ \int_{-\pi+\tilde{t}_+}^{\pi-\tilde{t}_+} d\tilde{t}_-  \cos \tilde{t}_+ \left(\cos \tilde{t}_-+\cos \tilde{t}_+\right)^2  g^{(0,2)}_{\ell}(\tilde{t}_-,-\cos\tilde{t}_+)
    \,,\\
    f_{4,\ell}(-1)=&
    -2\int_0^\pi d\tilde{t}_+ \int_{-\pi+\tilde{t}_+}^{\pi-\tilde{t}_+} d\tilde{t}_-\sin^2\tilde{t}_+\left(\cos\tilde{t}_++\cos\tilde{t}_-\right)\,g^{(0,2)}_{\ell} (\tilde{t}_-,-\cos\tilde{t}_+)   \,,\\
    f_{4,\ell}'(-1)=&\int_0^\pi d\tilde{t}_+ \int_{-\pi+\tilde{t}_+}^{\pi-\tilde{t}_+} d\tilde{t}_- \left(\cos \tilde{t}_++\cos  \tilde{t}_-\right)\sin^2\tilde{t}_+ g^{(0,2)}_{\ell}(\tilde{t}_-,-\cos\tilde{t}_+)\\
    &-\frac{1}{2}\int_0^\pi d\tilde{t}_+ \int_{-\pi+\tilde{t}_+}^{\pi-\tilde{t}_+} d\tilde{t}_-  \cos \tilde{t}_+ \sin^2\tilde{t}_+\left(\cos \tilde{t}_-+\cos \tilde{t}_+\right)^2  g^{(0,3)}_{\ell}(\tilde{t}_-,-\cos\tilde{t}_+)
    \,.
\end{align}
The integrals over $\tilde{t}_-$ are straightforward. Indeed by \eqref{eq_m1m2_app} the Legendre polynomial in~\eqref{eq_g_l} does not depend on $\tilde{t}_-$. The remaining integration over $\tilde{t}_+$ may then be carried over analytically expanding the Legendre polynomials into components, but in practice it is faster to do it numerically.

We checked up to high order in $\ell$ that
\begin{equation}
    f_{1,1}(-1)=\frac{8}{\pi}\,,\qquad f_{1,\ell>1}(-1)=0\,.
\end{equation}
The other terms instead yield a non-zero result for generic values of $\ell$, but oscillate and decay rapidly in absolute value with $\ell$. In Fig.~\ref{fig:Back} we plot our results for $\ell\in[1,40]$. Setting a cutoff $\ell_{max}=40$, we find that the sums give
\begin{align}
    &\sum_{\ell=1}^{\ell_{max}} \left[f_{\ell,1}(-1)+f_{\ell,2}(-1)+ f_{\ell,3}(-1)+f_{\ell,4}(-1)\right]\simeq 1.358\,,\\
    &\sum_{\ell=1}^{\ell_{max}}
    \left[f_{\ell,1}'(-1)+f_{\ell,2}'(-1)+ f_{\ell,3}'(-1)+f_{\ell,4}'(-1)\right]\simeq 0.220\,.
\end{align}
The value of the last few summands is at most of order $0.1\%$ of the total sums. Therefore, given the oscillatory nature of the sum we expect that the difference between the full series and the cutoff results is not much larger than $0.1\%$. Reinstating the prefactor in~\eqref{eq_F_l_QQ_app} we obtain the result for the charge correlator in~\eqref{eq_back_QQ}. The result~\eqref{eq_back_EE} for the energy correlator is derived analogously.

\endgroup

\subsection{Fourier decomposition of CCC and EEC}\label{app_Fourier_dec}

It is often convenient to decompose event shapes in Fourier basis. For CCC and EEC in $d=3$, such decomposition reads
\begin{equation}\label{eq_app_ell_dec}
    \langle \Qno\Qnt\rangle=\sum_{k=0}^\infty c_{\mathcal{Q\mathcal{Q}}}^{(k)}\cos(k\theta)\,,\qquad
    \langle \Eno\Ent\rangle=\sum_{k=0}^\infty c_{\mathcal{E\mathcal{E}}}^{(k)}\cos(k\theta)\,,
\end{equation}
where we used parity to exclude odd harmonics. Note that unitarity requires that the coefficients in~\eqref{eq_app_ell_dec} are non-negative \cite{Fox:1978vw}:
\begin{equation}\label{eq_cl_pos}
    c_{\mathcal{Q}\mathcal{Q}}^{(k)}\geq 0\,,\qquad c_{\mathcal{E\mathcal{E}}}^{(k)}\geq 0\,.
\end{equation}
Below we check this property and compute the first few coefficients at next-to-leading order.

The Fourier coefficients admit a large charge expansion analogous to~\eqref{eq_CC_exp}:
\begin{equation}
  c_{\mathcal{Q}\mathcal{Q}}^{(k)}= c_{\mathcal{Q}\mathcal{Q}}^{(k,0)} +\frac{1}{\Delta_Q}c_{\mathcal{Q}\mathcal{Q}}^{(k,1)}+\ldots\,,
\end{equation}
and similarly for $c_{\mathcal{E}\mathcal{E}}^{(k)}$. Given the factorized structure of the leading order results, we immediately find that only the homogeneous term is non-vanishing at leading order:
\begin{equation}
    c_{\mathcal{Q}\mathcal{Q}}^{(k,0)}=\left(\frac{Q}{2\pi}\right)^2\delta^k_0\,,\qquad
    c_{\mathcal{E}\mathcal{E}}^{(k,0)}=\left(\frac{E}{2\pi}\right)^2\delta^k_0\,.
\end{equation}
We can extract the subleading coefficients from the results in Sec.~\ref{Sec:EEC} using the completeness of the Fourier basis: 
\begin{align}\label{eq_app_cQQ_and_cEE}
    c_{\mathcal{Q}\mathcal{Q}}^{(k,1)}&=\frac{2}{\pi}\int_0^\pi d\theta\cos(k\theta)\langle \Qno \Qnt \rangle^{(1)}\,,\qquad
    c_{\mathcal{E}\mathcal{E}}^{(k,1)}=\frac{2}{\pi}\int_0^\pi d\theta\cos(k\theta)\langle \Eno \Ent \rangle^{(1)}\,.
\end{align}

As discussed and checked in detail in Sec.~\ref{App:Conservations}, Ward identities imply that the following coefficients vanish,
\begin{equation}
c_{\mathcal{Q}\mathcal{Q}}^{(0,1)}=c_{\mathcal{E}\mathcal{E}}^{(0,1)}=
    c_{\mathcal{E}\mathcal{E}}^{(1,1)}=0\,.
\end{equation}
The remaining coefficients are computed using~\eqref{Eq:PropDec} and recalling that the spherical harmonics explicitly read
\begin{equation}
    Y_{\ell m}(\sigma ,\theta)=N_{\ell}^m P_{\ell}^m(\cos\sigma)e^{im\theta}\,,\qquad
    N_{\ell}^m=\sqrt{\frac{(2\ell+1)}{4\pi}\frac{(\ell-m)!}{(\ell+m)!}}\,,
\end{equation}
where $P_{\ell}^m(x)$ is the associated Legendre polynomial. Since the disconnected terms in~\eqref{Eq:DiscCC} and~\eqref{Eq:DiscEE} only contribute to the $k=0,1$ harmonics, we find
\begin{align}\label{eq_app_cQQ1}
   c_{\mathcal{Q}\mathcal{Q}}^{(k\geq 1,1)}&=\left(\frac{Q}{2\pi}\right)^2\frac{\pi}{12}\left(
    \sum_{\ell=k}^{\infty}q_{\ell,k}-\frac12\delta^k_1\right)\,,\\
   c_{\mathcal{E}\mathcal{E}}^{(k\geq 2,1)}
   &=
   \left(\frac{E}{2\pi}\right)^2\frac{3\pi}{16}\sum_{\ell=k}^{\infty}\varepsilon_{\ell,k}\,,
   \label{eq_app_cEE1}
\end{align}
where $q_{\ell,k}$ and $\varepsilon_{\ell,k}$ are \emph{non-negative} coefficients defined as
\begin{align}\label{eq_app_qlk}
    q_{\ell,k}&=\frac{(N^k_{\ell})^2}{\omega_{\ell}}\left|\int dt^-\cos\frac{t^-}{2}\left[\left(2\pd_t-\pd_\sigma\right)e^{-i\omega_{\ell}t}P_{\ell}^k(\cos\sigma)\right]_{t+\sigma=\pi}\right|^2\geq 0\,, 
    \\
    \varepsilon_{\ell,k}&=\frac{(N^k_{\ell})^2}{\omega_{\ell}}\left|\int dt^-\left(\cos\frac{t^-}{2}\right)^3\left[\left(3\pd_t-2\pd_\sigma\right)e^{-i\omega_{\ell}t}P_{\ell}^k(\cos\sigma)\right]_{t+\sigma=\pi}\right|^2\geq 0\,.
    \label{eq_app_elk}
\end{align} 
We find that $q_{\ell,k}\sim 1/\ell^2$ and $ \varepsilon_{\ell,k}\sim 1/\ell^6$ for $\ell\gg k$, hence the sums in~\eqref{eq_app_cQQ1} and~\eqref{eq_app_cEE1} are convergent. From~\eqref{eq_app_cQQ1} and~\eqref{eq_app_cEE1} we immediately conclude that $c_{\mathcal{Q}\mathcal{Q}}^{(k\geq 2,1)}$ and $c_{\mathcal{E}\mathcal{E}}^{(k\geq 2,1)}$ are non-negative, as expected. We also checked numerically that $c_{\mathcal{Q}\mathcal{Q}}^{(1,1)}>0$. 

For any given $\ell$ and $k$ the integrals in~\eqref{eq_app_qlk} and~\eqref{eq_app_elk} can be performed analytically, but we did not find a convenient closed form for all $\ell$'s and $k$'s.\footnote{Since these integrals are highly oscilatory, numerical integration loses precision already at moderate values of $\ell$ and is thus not convenient.} In practice therefore we used \texttt{Mathematica} to evaluate such integrals up to a cutoff $\ell_{\text{max}}$ for $k\leq 5$, where $\ell_{\text{max}}=80$ for CCC and $\ell_{\text{max}}=50$ for EEC. We then obtained approximate results for $c_{\mathcal{Q}\mathcal{Q}}^{(k\geq 1,1)}$ and $c_{\mathcal{E}\mathcal{E}}^{(k\geq 2,1)}$ truncating the sums in~\eqref{eq_app_cQQ1} and~\eqref{eq_app_cEE1}. Since $q_{\ell,k}\sim 1/\ell^2$ and $\varepsilon_{\ell,k}\sim 1/\ell^6$, we estimated the error on this procedure as the average of the last five coefficients we computed times $x\,\ell_{\text{max}}$, where $x=1$ for $c_{\mathcal{Q}\mathcal{Q}}^{(k,1)}$ and $x=5$ for $c_{\mathcal{E}\mathcal{E}}^{(k,1)}$. Our results are shown in Tab.~\ref{tab:Fourier}. We checked that the values in Tab.~\ref{tab:Fourier} are compatible with those obtained by performing the integrals~\eqref{eq_app_cQQ_and_cEE} using the numerical result discussed in Sec.~\ref{Sec:Num}. Note also that the sound-jet singularity $\sim 1/\theta^2$ implies that the Fourier coefficients grow linearly with $\ell$ asymptotically.

\begin{table}[t]
    \centering
  \begin{tabular}{c|c|c|c|c|c|c}
   & $k=0$ & $k=1$ & $k=2$ & $k=3$ & $k=4$ & $k=5$  \\ \hline
  $c_{\mathcal{Q}\mathcal{Q}}^{(k,1)}$   & $0$ (exact) & $0.69(2)$ & $2.15(5)$ & $3.1(1)$ & $4.1(2)$ & $5.0(4)$  \\
   $c_{\mathcal{E}\mathcal{E}}^{(k,1)}$  & $0$ (exact) & $0$ (exact) & $2.58$ & $3.27$ & $4.04$ & $4.84$
\end{tabular}
    \caption{Results for the first Fourier coefficients. The parentheses represent the estimated error on the last displayed digit for $c_{\mathcal{Q}\mathcal{Q}}^{(k>1,1)}$, for $c_{\mathcal{E}\mathcal{E}}^{(k>1,1)}$ the error is negligible to the reported precision.}
    \label{tab:Fourier}
\end{table}

\bibliography{Biblio}

\providecommand{\href}[2]{#2}\begingroup\raggedright\begin{thebibliography}{10}

\bibitem{PhysRevD.17.2298}
C.~L. Basham, L.~S. Brown, S.~D. Ellis and S.~T. Love, \emph{Electron-positron
  annihilation energy pattern in quantum chromodynamics: Asymptotically free
  perturbation theory},
  \href{https://doi.org/10.1103/PhysRevD.17.2298}{\emph{Phys. Rev. D}
  {\bfseries 17} (1978) 2298}.

\bibitem{PhysRevLett.41.1585}
C.~L. Basham, L.~S. Brown, S.~D. Ellis and S.~T. Love, \emph{Energy
  correlations in electron-positron annihilation: Testing quantum
  chromodynamics},
  \href{https://doi.org/10.1103/PhysRevLett.41.1585}{\emph{Phys. Rev. Lett.}
  {\bfseries 41} (1978) 1585}.

\bibitem{PhysRevD.19.2018}
C.~L. Basham, L.~S. Brown, S.~D. Ellis and S.~T. Love, \emph{Energy
  correlations in electron-positron annihilation in quantum chromodynamics:
  Asymptotically free perturbation theory},
  \href{https://doi.org/10.1103/PhysRevD.19.2018}{\emph{Phys. Rev. D}
  {\bfseries 19} (1979) 2018}.

\bibitem{LOUISBASHAM1979297}
C.~L. Basham, L.~S. Brown, S.~D. Ellis and S.~T. Love, \emph{Energy
  correlations in perturbative quantum chromodynamics: A conjecture for all
  orders},
  \href{https://doi.org/https://doi.org/10.1016/0370-2693(79)90601-4}{\emph{Physics
  Letters B} {\bfseries 85} (1979) 297}.

\bibitem{Chen:2024nyc}
H.~Chen, P.~F. Monni, Z.~Xu and H.~X. Zhu, \emph{{Scaling Violation in Power
  Corrections to Energy Correlators from the Light-Ray Operator Product
  Expansion}},
  \href{https://doi.org/10.1103/PhysRevLett.133.231901}{\emph{Phys. Rev. Lett.}
  {\bfseries 133} (2024) 231901}
  [\href{https://arxiv.org/abs/2406.06668}{{\ttfamily 2406.06668}}].

\bibitem{Chen:2020vvp}
H.~Chen, I.~Moult, X.~Zhang and H.~X. Zhu, \emph{{Rethinking jets with energy
  correlators: Tracks, resummation, and analytic continuation}},
  \href{https://doi.org/10.1103/PhysRevD.102.054012}{\emph{Phys. Rev. D}
  {\bfseries 102} (2020) 054012}
  [\href{https://arxiv.org/abs/2004.11381}{{\ttfamily 2004.11381}}].

\bibitem{CMS:2024mlf}
{\scshape CMS} collaboration, A.~Hayrapetyan et~al., \emph{{Measurement of
  Energy Correlators inside Jets and Determination of the Strong Coupling
  \ensuremath{\alpha}S(mZ)}},
  \href{https://doi.org/10.1103/PhysRevLett.133.071903}{\emph{Phys. Rev. Lett.}
  {\bfseries 133} (2024) 071903}
  [\href{https://arxiv.org/abs/2402.13864}{{\ttfamily 2402.13864}}].

\bibitem{Komiske:2022enw}
P.~T. Komiske, I.~Moult, J.~Thaler and H.~X. Zhu, \emph{{Analyzing N-Point
  Energy Correlators inside Jets with CMS Open Data}},
  \href{https://doi.org/10.1103/PhysRevLett.130.051901}{\emph{Phys. Rev. Lett.}
  {\bfseries 130} (2023) 051901}
  [\href{https://arxiv.org/abs/2201.07800}{{\ttfamily 2201.07800}}].

\bibitem{Holguin:2022epo}
J.~Holguin, I.~Moult, A.~Pathak and M.~Procura, \emph{{New paradigm for
  precision top physics: Weighing the top with energy correlators}},
  \href{https://doi.org/10.1103/PhysRevD.107.114002}{\emph{Phys. Rev. D}
  {\bfseries 107} (2023) 114002}
  [\href{https://arxiv.org/abs/2201.08393}{{\ttfamily 2201.08393}}].

\bibitem{Riembau:2024tom}
M.~Riembau and M.~Son, \emph{{One-point correlators of conserved and
  nonconserved charges in QCD}},
  \href{https://doi.org/10.1103/PhysRevD.111.014004}{\emph{Phys. Rev. D}
  {\bfseries 111} (2025) 014004}
  [\href{https://arxiv.org/abs/2407.12082}{{\ttfamily 2407.12082}}].

\bibitem{Lee:2022uwt}
K.~Lee, B.~Me\c{c}aj and I.~Moult, \emph{{Conformal collider physics meets LHC
  data}}, \href{https://doi.org/10.1103/PhysRevD.111.L011502}{\emph{Phys. Rev.
  D} {\bfseries 111} (2025) L011502}
  [\href{https://arxiv.org/abs/2205.03414}{{\ttfamily 2205.03414}}].

\bibitem{Craft:2022kdo}
E.~Craft, K.~Lee, B.~Me\c{c}aj and I.~Moult, \emph{{Beautiful and Charming
  Energy Correlators}},  \href{https://arxiv.org/abs/2210.09311}{{\ttfamily
  2210.09311}}.

\bibitem{Ricci:2022htc}
L.~Ricci and M.~Riembau, \emph{{Energy correlators of hadronically decaying
  electroweak bosons}},
  \href{https://doi.org/10.1103/PhysRevD.106.114010}{\emph{Phys. Rev. D}
  {\bfseries 106} (2022) 114010}
  [\href{https://arxiv.org/abs/2207.03511}{{\ttfamily 2207.03511}}].

\bibitem{Caron-Huot:2017vep}
S.~Caron-Huot, \emph{{Analyticity in Spin in Conformal Theories}},
  \href{https://doi.org/10.1007/JHEP09(2017)078}{\emph{JHEP} {\bfseries 09}
  (2017) 078} [\href{https://arxiv.org/abs/1703.00278}{{\ttfamily
  1703.00278}}].

\bibitem{Kravchuk:2018htv}
P.~Kravchuk and D.~Simmons-Duffin, \emph{{Light-ray operators in conformal
  field theory}}, \href{https://doi.org/10.1007/JHEP11(2018)102}{\emph{JHEP}
  {\bfseries 11} (2018) 102}
  [\href{https://arxiv.org/abs/1805.00098}{{\ttfamily 1805.00098}}].

\bibitem{Hartman:2016lgu}
T.~Hartman, S.~Kundu and A.~Tajdini, \emph{{Averaged Null Energy Condition from
  Causality}}, \href{https://doi.org/10.1007/JHEP07(2017)066}{\emph{JHEP}
  {\bfseries 07} (2017) 066}
  [\href{https://arxiv.org/abs/1610.05308}{{\ttfamily 1610.05308}}].

\bibitem{Faulkner:2016mzt}
T.~Faulkner, R.~G. Leigh, O.~Parrikar and H.~Wang, \emph{{Modular Hamiltonians
  for Deformed Half-Spaces and the Averaged Null Energy Condition}},
  \href{https://doi.org/10.1007/JHEP09(2016)038}{\emph{JHEP} {\bfseries 09}
  (2016) 038} [\href{https://arxiv.org/abs/1605.08072}{{\ttfamily
  1605.08072}}].

\bibitem{Casini:2017roe}
H.~Casini, E.~Teste and G.~Torroba, \emph{{Modular Hamiltonians on the null
  plane and the Markov property of the vacuum state}},
  \href{https://doi.org/10.1088/1751-8121/aa7eaa}{\emph{J. Phys. A} {\bfseries
  50} (2017) 364001} [\href{https://arxiv.org/abs/1703.10656}{{\ttfamily
  1703.10656}}].

\bibitem{Ceyhan:2018zfg}
F.~Ceyhan and T.~Faulkner, \emph{{Recovering the QNEC from the ANEC}},
  \href{https://doi.org/10.1007/s00220-020-03751-y}{\emph{Commun. Math. Phys.}
  {\bfseries 377} (2020) 999}
  [\href{https://arxiv.org/abs/1812.04683}{{\ttfamily 1812.04683}}].

\bibitem{Kologlu:2019mfz}
M.~Kologlu, P.~Kravchuk, D.~Simmons-Duffin and A.~Zhiboedov, \emph{{The
  light-ray OPE and conformal colliders}},
  \href{https://doi.org/10.1007/JHEP01(2021)128}{\emph{JHEP} {\bfseries 01}
  (2021) 128} [\href{https://arxiv.org/abs/1905.01311}{{\ttfamily
  1905.01311}}].

\bibitem{Chang:2020qpj}
C.-H. Chang, M.~Kologlu, P.~Kravchuk, D.~Simmons-Duffin and A.~Zhiboedov,
  \emph{{Transverse spin in the light-ray OPE}},
  \href{https://doi.org/10.1007/JHEP05(2022)059}{\emph{JHEP} {\bfseries 05}
  (2022) 059} [\href{https://arxiv.org/abs/2010.04726}{{\ttfamily
  2010.04726}}].

\bibitem{Belin:2020lsr}
A.~Belin, D.~M. Hofman, G.~Mathys and M.~T. Walters, \emph{{On the stress
  tensor light-ray operator algebra}},
  \href{https://doi.org/10.1007/JHEP05(2021)033}{\emph{JHEP} {\bfseries 05}
  (2021) 033} [\href{https://arxiv.org/abs/2011.13862}{{\ttfamily
  2011.13862}}].

\bibitem{Korchemsky:2021htm}
G.~P. Korchemsky and A.~Zhiboedov, \emph{{On the light-ray algebra in conformal
  field theories}}, \href{https://doi.org/10.1007/JHEP02(2022)140}{\emph{JHEP}
  {\bfseries 02} (2022) 140}
  [\href{https://arxiv.org/abs/2109.13269}{{\ttfamily 2109.13269}}].

\bibitem{Caron-Huot:2022eqs}
S.~Caron-Huot, M.~Kologlu, P.~Kravchuk, D.~Meltzer and D.~Simmons-Duffin,
  \emph{{Detectors in weakly-coupled field theories}},
  \href{https://doi.org/10.1007/JHEP04(2023)014}{\emph{JHEP} {\bfseries 04}
  (2023) 014} [\href{https://arxiv.org/abs/2209.00008}{{\ttfamily
  2209.00008}}].

\bibitem{Hofman:2008ar}
D.~M. Hofman and J.~Maldacena, \emph{{Conformal collider physics: Energy and
  charge correlations}},
  \href{https://doi.org/10.1088/1126-6708/2008/05/012}{\emph{JHEP} {\bfseries
  05} (2008) 012} [\href{https://arxiv.org/abs/0803.1467}{{\ttfamily
  0803.1467}}].

\bibitem{Kologlu:2019bco}
M.~Kologlu, P.~Kravchuk, D.~Simmons-Duffin and A.~Zhiboedov, \emph{{Shocks,
  Superconvergence, and a Stringy Equivalence Principle}},
  \href{https://doi.org/10.1007/JHEP11(2020)096}{\emph{JHEP} {\bfseries 11}
  (2020) 096} [\href{https://arxiv.org/abs/1904.05905}{{\ttfamily
  1904.05905}}].

\bibitem{Chen:2024iuv}
H.~Chen, R.~Karlsson and A.~Zhiboedov, \emph{{Energy correlations and Planckian
  collisions}},  \href{https://arxiv.org/abs/2404.15056}{{\ttfamily
  2404.15056}}.

\bibitem{Cordova:2018ygx}
C.~C\'ordova and S.-H. Shao, \emph{{Light-ray Operators and the BMS Algebra}},
  \href{https://doi.org/10.1103/PhysRevD.98.125015}{\emph{Phys. Rev. D}
  {\bfseries 98} (2018) 125015}
  [\href{https://arxiv.org/abs/1810.05706}{{\ttfamily 1810.05706}}].

\bibitem{Hu:2022txx}
Y.~Hu and S.~Pasterski, \emph{{Celestial conformal colliders}},
  \href{https://doi.org/10.1007/JHEP02(2023)243}{\emph{JHEP} {\bfseries 02}
  (2023) 243} [\href{https://arxiv.org/abs/2211.14287}{{\ttfamily
  2211.14287}}].

\bibitem{Belitsky:2013bja}
A.~V. Belitsky, S.~Hohenegger, G.~P. Korchemsky, E.~Sokatchev and A.~Zhiboedov,
  \emph{{Event shapes in $\mathcal{N} = 4$ super-Yang-Mills theory}},
  \href{https://doi.org/10.1016/j.nuclphysb.2014.04.019}{\emph{Nucl. Phys. B}
  {\bfseries 884} (2014) 206}
  [\href{https://arxiv.org/abs/1309.1424}{{\ttfamily 1309.1424}}].

\bibitem{Dixon:2018qgp}
L.~J. Dixon, M.-X. Luo, V.~Shtabovenko, T.-Z. Yang and H.~X. Zhu,
  \emph{{Analytical Computation of Energy-Energy Correlation at Next-to-Leading
  Order in QCD}},
  \href{https://doi.org/10.1103/PhysRevLett.120.102001}{\emph{Phys. Rev. Lett.}
  {\bfseries 120} (2018) 102001}
  [\href{https://arxiv.org/abs/1801.03219}{{\ttfamily 1801.03219}}].

\bibitem{Dixon:2019uzg}
L.~J. Dixon, I.~Moult and H.~X. Zhu, \emph{{Collinear limit of the
  energy-energy correlator}},
  \href{https://doi.org/10.1103/PhysRevD.100.014009}{\emph{Phys. Rev. D}
  {\bfseries 100} (2019) 014009}
  [\href{https://arxiv.org/abs/1905.01310}{{\ttfamily 1905.01310}}].

\bibitem{Herrmann:2024yai}
E.~Herrmann, M.~Kologlu and I.~Moult, \emph{{Energy Correlators in Perturbative
  Quantum Gravity}},  \href{https://arxiv.org/abs/2412.05384}{{\ttfamily
  2412.05384}}.

\bibitem{Delacretaz:2018cfk}
L.~V. Delacr\'etaz, T.~Hartman, S.~A. Hartnoll and A.~Lewkowycz,
  \emph{{Thermalization, Viscosity and the Averaged Null Energy Condition}},
  \href{https://doi.org/10.1007/JHEP10(2018)028}{\emph{JHEP} {\bfseries 10}
  (2018) 028} [\href{https://arxiv.org/abs/1805.04194}{{\ttfamily
  1805.04194}}].

\bibitem{Andres:2022ovj}
C.~Andres, F.~Dominguez, R.~Kunnawalkam~Elayavalli, J.~Holguin, C.~Marquet and
  I.~Moult, \emph{{Resolving the Scales of the Quark-Gluon Plasma with Energy
  Correlators}},
  \href{https://doi.org/10.1103/PhysRevLett.130.262301}{\emph{Phys. Rev. Lett.}
  {\bfseries 130} (2023) 262301}
  [\href{https://arxiv.org/abs/2209.11236}{{\ttfamily 2209.11236}}].

\bibitem{Andres:2024ksi}
C.~Andres, F.~Dominguez, J.~Holguin, C.~Marquet and I.~Moult, \emph{{Towards an
  interpretation of the first measurements of energy correlators in the
  quark-gluon plasma}},
  \href{https://doi.org/10.1007/JHEP03(2025)166}{\emph{JHEP} {\bfseries 03}
  (2025) 166} [\href{https://arxiv.org/abs/2407.07936}{{\ttfamily
  2407.07936}}].

\bibitem{Bossi:2024qho}
H.~Bossi, A.~S. Kudinoor, I.~Moult, D.~Pablos, A.~Rai and K.~Rajagopal,
  \emph{{Imaging the wakes of jets with energy-energy-energy correlators}},
  \href{https://doi.org/10.1007/JHEP12(2024)073}{\emph{JHEP} {\bfseries 12}
  (2024) 073} [\href{https://arxiv.org/abs/2407.13818}{{\ttfamily
  2407.13818}}].

\bibitem{Alfimov:2014bwa}
M.~Alfimov, N.~Gromov and V.~Kazakov, \emph{{QCD Pomeron from AdS/CFT Quantum
  Spectral Curve}}, \href{https://doi.org/10.1007/JHEP07(2015)164}{\emph{JHEP}
  {\bfseries 07} (2015) 164} [\href{https://arxiv.org/abs/1408.2530}{{\ttfamily
  1408.2530}}].

\bibitem{Homrich:2022cfq}
A.~Homrich, D.~Simmons-Duffin and P.~Vieira, \emph{{Multitwist Trajectories and
  Decoupling Zeros in Conformal Field Theory}},
  \href{https://doi.org/10.1103/PhysRevLett.134.011602}{\emph{Phys. Rev. Lett.}
  {\bfseries 134} (2025) 011602}
  [\href{https://arxiv.org/abs/2211.13754}{{\ttfamily 2211.13754}}].

\bibitem{Homrich:2024nwc}
A.~Homrich, D.~Simmons-Duffin and P.~Vieira, \emph{{Light-ray wave functions
  and integrability}},
  \href{https://doi.org/10.1007/JHEP10(2024)125}{\emph{JHEP} {\bfseries 10}
  (2024) 125} [\href{https://arxiv.org/abs/2409.02160}{{\ttfamily
  2409.02160}}].

\bibitem{Hellerman:2015nra}
S.~Hellerman, D.~Orlando, S.~Reffert and M.~Watanabe, \emph{{On the CFT
  Operator Spectrum at Large Global Charge}},
  \href{https://doi.org/10.1007/JHEP12(2015)071}{\emph{JHEP} {\bfseries 12}
  (2015) 071} [\href{https://arxiv.org/abs/1505.01537}{{\ttfamily
  1505.01537}}].

\bibitem{Monin:2016jmo}
A.~Monin, D.~Pirtskhalava, R.~Rattazzi and F.~K. Seibold, \emph{{Semiclassics,
  Goldstone Bosons and CFT data}},
  \href{https://doi.org/10.1007/JHEP06(2017)011}{\emph{JHEP} {\bfseries 06}
  (2017) 011} [\href{https://arxiv.org/abs/1611.02912}{{\ttfamily
  1611.02912}}].

\bibitem{Banerjee:2017fcx}
D.~Banerjee, S.~Chandrasekharan and D.~Orlando, \emph{{Conformal dimensions via
  large charge expansion}},
  \href{https://doi.org/10.1103/PhysRevLett.120.061603}{\emph{Phys. Rev. Lett.}
  {\bfseries 120} (2018) 061603}
  [\href{https://arxiv.org/abs/1707.00711}{{\ttfamily 1707.00711}}].

\bibitem{Cuomo:2023mxg}
G.~Cuomo, J.~M. V.~P. Lopes, J.~Matos, J.~Oliveira and J.~Penedones,
  \emph{{Numerical tests of the large charge expansion}},
  \href{https://doi.org/10.1007/JHEP05(2024)161}{\emph{JHEP} {\bfseries 05}
  (2024) 161} [\href{https://arxiv.org/abs/2305.00499}{{\ttfamily
  2305.00499}}].

\bibitem{Firat:2023lbp}
E.~Firat, A.~Monin, R.~Rattazzi and M.~T. Walters, \emph{{Flux correlators and
  semiclassics}}, \href{https://doi.org/10.1007/JHEP03(2024)067}{\emph{JHEP}
  {\bfseries 03} (2024) 067}
  [\href{https://arxiv.org/abs/2309.14428}{{\ttfamily 2309.14428}}].

\bibitem{Chicherin:2023gxt}
D.~Chicherin, G.~P. Korchemsky, E.~Sokatchev and A.~Zhiboedov, \emph{{Energy
  correlations in heavy states}},
  \href{https://doi.org/10.1007/JHEP11(2023)134}{\emph{JHEP} {\bfseries 11}
  (2023) 134} [\href{https://arxiv.org/abs/2306.14330}{{\ttfamily
  2306.14330}}].

\bibitem{Lashkari:2016vgj}
N.~Lashkari, A.~Dymarsky and H.~Liu, \emph{{Eigenstate Thermalization
  Hypothesis in Conformal Field Theory}},
  \href{https://doi.org/10.1088/1742-5468/aab020}{\emph{J. Stat. Mech.}
  {\bfseries 1803} (2018) 033101}
  [\href{https://arxiv.org/abs/1610.00302}{{\ttfamily 1610.00302}}].

\bibitem{Delacretaz:2020nit}
L.~V. Delacretaz, \emph{{Heavy Operators and Hydrodynamic Tails}},
  \href{https://doi.org/10.21468/SciPostPhys.9.3.034}{\emph{SciPost Phys.}
  {\bfseries 9} (2020) 034} [\href{https://arxiv.org/abs/2006.01139}{{\ttfamily
  2006.01139}}].

\bibitem{Nicolis:2015sra}
A.~Nicolis, R.~Penco, F.~Piazza and R.~Rattazzi, \emph{{Zoology of condensed
  matter: Framids, ordinary stuff, extra-ordinary stuff}},
  \href{https://doi.org/10.1007/JHEP06(2015)155}{\emph{JHEP} {\bfseries 06}
  (2015) 155} [\href{https://arxiv.org/abs/1501.03845}{{\ttfamily
  1501.03845}}].

\bibitem{Rychkov:2016iqz}
S.~Rychkov, \emph{{EPFL Lectures on Conformal Field Theory in D\ensuremath{>}=
  3 Dimensions}}, SpringerBriefs in Physics. 1, 2016,
  \href{https://doi.org/10.1007/978-3-319-43626-5}{10.1007/978-3-319-43626-5},
  [\href{https://arxiv.org/abs/1601.05000}{{\ttfamily 1601.05000}}].

\bibitem{Camanho:2014apa}
X.~O. Camanho, J.~D. Edelstein, J.~Maldacena and A.~Zhiboedov, \emph{{Causality
  Constraints on Corrections to the Graviton Three-Point Coupling}},
  \href{https://doi.org/10.1007/JHEP02(2016)020}{\emph{JHEP} {\bfseries 02}
  (2016) 020} [\href{https://arxiv.org/abs/1407.5597}{{\ttfamily 1407.5597}}].

\bibitem{Komargodski:2021zzy}
Z.~Komargodski, M.~Mezei, S.~Pal and A.~Raviv-Moshe, \emph{{Spontaneously
  broken boosts in CFTs}},
  \href{https://doi.org/10.1007/JHEP09(2021)064}{\emph{JHEP} {\bfseries 09}
  (2021) 064} [\href{https://arxiv.org/abs/2102.12583}{{\ttfamily
  2102.12583}}].

\bibitem{Dondi:2022zna}
N.~Dondi, S.~Hellerman, I.~Kalogerakis, R.~Moser, D.~Orlando and S.~Reffert,
  \emph{{Fermionic CFTs at large charge and large N}},
  \href{https://doi.org/10.1007/JHEP08(2023)180}{\emph{JHEP} {\bfseries 08}
  (2023) 180} [\href{https://arxiv.org/abs/2211.15318}{{\ttfamily
  2211.15318}}].

\bibitem{Delacretaz:2025ifh}
L.~V. Delacr\'etaz, S.~D. Chowdhury and U.~Mehta, \emph{{Symmetry and causality
  constraints on Fermi liquids}},
  \href{https://arxiv.org/abs/2501.02073}{{\ttfamily 2501.02073}}.

\bibitem{Hellerman:2017veg}
S.~Hellerman, S.~Maeda and M.~Watanabe, \emph{{Operator Dimensions from
  Moduli}}, \href{https://doi.org/10.1007/JHEP10(2017)089}{\emph{JHEP}
  {\bfseries 10} (2017) 089}
  [\href{https://arxiv.org/abs/1706.05743}{{\ttfamily 1706.05743}}].

\bibitem{Grassi:2019txd}
A.~Grassi, Z.~Komargodski and L.~Tizzano, \emph{{Extremal correlators and
  random matrix theory}},
  \href{https://doi.org/10.1007/JHEP04(2021)214}{\emph{JHEP} {\bfseries 04}
  (2021) 214} [\href{https://arxiv.org/abs/1908.10306}{{\ttfamily
  1908.10306}}].

\bibitem{Caetano:2023zwe}
J.~a. Caetano, S.~Komatsu and Y.~Wang, \emph{{Large charge \textquoteright{}t
  Hooft limit of $ \mathcal{N} $ = 4 super-Yang-Mills}},
  \href{https://doi.org/10.1007/JHEP02(2024)047}{\emph{JHEP} {\bfseries 02}
  (2024) 047} [\href{https://arxiv.org/abs/2306.00929}{{\ttfamily
  2306.00929}}].

\bibitem{Cuomo:2024fuy}
G.~Cuomo, L.~Rastelli and A.~Sharon, \emph{{Moduli spaces in CFT: large charge
  operators}}, \href{https://doi.org/10.1007/JHEP09(2024)185}{\emph{JHEP}
  {\bfseries 09} (2024) 185}
  [\href{https://arxiv.org/abs/2406.19441}{{\ttfamily 2406.19441}}].

\bibitem{Alvarez-Gaume:2019biu}
L.~Alvarez-Gaume, D.~Orlando and S.~Reffert, \emph{{Large charge at large N}},
  \href{https://doi.org/10.1007/JHEP12(2019)142}{\emph{JHEP} {\bfseries 12}
  (2019) 142} [\href{https://arxiv.org/abs/1909.02571}{{\ttfamily
  1909.02571}}].

\bibitem{Badel:2019oxl}
G.~Badel, G.~Cuomo, A.~Monin and R.~Rattazzi, \emph{{The Epsilon Expansion
  Meets Semiclassics}},
  \href{https://doi.org/10.1007/JHEP11(2019)110}{\emph{JHEP} {\bfseries 11}
  (2019) 110} [\href{https://arxiv.org/abs/1909.01269}{{\ttfamily
  1909.01269}}].

\bibitem{Son:2002zn}
D.~T. Son, \emph{{Low-energy quantum effective action for relativistic
  superfluids}},  \href{https://arxiv.org/abs/hep-ph/0204199}{{\ttfamily
  hep-ph/0204199}}.

\bibitem{Cuomo:2020rgt}
G.~Cuomo, \emph{{A note on the large charge expansion in 4d CFT}},
  \href{https://doi.org/10.1016/j.physletb.2020.136014}{\emph{Phys. Lett. B}
  {\bfseries 812} (2021) 136014}
  [\href{https://arxiv.org/abs/2010.00407}{{\ttfamily 2010.00407}}].

\bibitem{Jafferis:2017zna}
D.~Jafferis, B.~Mukhametzhanov and A.~Zhiboedov, \emph{{Conformal Bootstrap At
  Large Charge}}, \href{https://doi.org/10.1007/JHEP05(2018)043}{\emph{JHEP}
  {\bfseries 05} (2018) 043}
  [\href{https://arxiv.org/abs/1710.11161}{{\ttfamily 1710.11161}}].

\bibitem{Belitsky:2013xxa}
A.~V. Belitsky, S.~Hohenegger, G.~P. Korchemsky, E.~Sokatchev and A.~Zhiboedov,
  \emph{{From correlation functions to event shapes}},
  \href{https://doi.org/10.1016/j.nuclphysb.2014.04.020}{\emph{Nucl. Phys. B}
  {\bfseries 884} (2014) 305}
  [\href{https://arxiv.org/abs/1309.0769}{{\ttfamily 1309.0769}}].

\bibitem{Korchemsky:2021okt}
G.~P. Korchemsky, E.~Sokatchev and A.~Zhiboedov, \emph{{Generalizing event
  shapes: in search of lost collider time}},
  \href{https://doi.org/10.1007/JHEP08(2022)188}{\emph{JHEP} {\bfseries 08}
  (2022) 188} [\href{https://arxiv.org/abs/2106.14899}{{\ttfamily
  2106.14899}}].

\bibitem{Ebert:2018gsn}
M.~A. Ebert, I.~Moult, I.~W. Stewart, F.~J. Tackmann, G.~Vita and H.~X. Zhu,
  \emph{{Subleading power rapidity divergences and power corrections for
  q$_{T}$}}, \href{https://doi.org/10.1007/JHEP04(2019)123}{\emph{JHEP}
  {\bfseries 04} (2019) 123}
  [\href{https://arxiv.org/abs/1812.08189}{{\ttfamily 1812.08189}}].

\bibitem{Agia:2022srj}
N.~Agia and D.~L. Jafferis, \emph{{Angular Quantization in CFT}},
  \href{https://arxiv.org/abs/2204.11872}{{\ttfamily 2204.11872}}.

\bibitem{Fox:1978vu}
G.~C. Fox and S.~Wolfram, \emph{{Observables for the Analysis of Event Shapes
  in e+ e- Annihilation and Other Processes}},
  \href{https://doi.org/10.1103/PhysRevLett.41.1581}{\emph{Phys. Rev. Lett.}
  {\bfseries 41} (1978) 1581}.

\bibitem{Kovtun:2012rj}
P.~Kovtun, \emph{{Lectures on hydrodynamic fluctuations in relativistic
  theories}}, \href{https://doi.org/10.1088/1751-8113/45/47/473001}{\emph{J.
  Phys. A} {\bfseries 45} (2012) 473001}
  [\href{https://arxiv.org/abs/1205.5040}{{\ttfamily 1205.5040}}].

\bibitem{Hartnoll:2008kx}
S.~A. Hartnoll, C.~P. Herzog and G.~T. Horowitz, \emph{{Holographic
  Superconductors}},
  \href{https://doi.org/10.1088/1126-6708/2008/12/015}{\emph{JHEP} {\bfseries
  12} (2008) 015} [\href{https://arxiv.org/abs/0810.1563}{{\ttfamily
  0810.1563}}].

\bibitem{delaFuente:2020yua}
A.~de~la Fuente and J.~Zosso, \emph{{The large charge expansion and AdS/CFT}},
  \href{https://doi.org/10.1007/JHEP06(2020)178}{\emph{JHEP} {\bfseries 06}
  (2020) 178} [\href{https://arxiv.org/abs/2005.06169}{{\ttfamily
  2005.06169}}].

\bibitem{Hubeny:2011hd}
V.~E. Hubeny, S.~Minwalla and M.~Rangamani, \emph{{The fluid/gravity
  correspondence}},  in \emph{{Theoretical Advanced Study Institute in
  Elementary Particle Physics}: {String theory and its Applications: From meV
  to the Planck Scale}}, pp.~348--383, 2012,
  \href{https://arxiv.org/abs/1107.5780}{{\ttfamily 1107.5780}}.

\bibitem{Alday:2007mf}
L.~F. Alday and J.~M. Maldacena, \emph{{Comments on operators with large
  spin}}, \href{https://doi.org/10.1088/1126-6708/2007/11/019}{\emph{JHEP}
  {\bfseries 11} (2007) 019} [\href{https://arxiv.org/abs/0708.0672}{{\ttfamily
  0708.0672}}].

\bibitem{Komargodski:2012ek}
Z.~Komargodski and A.~Zhiboedov, \emph{{Convexity and Liberation at Large
  Spin}}, \href{https://doi.org/10.1007/JHEP11(2013)140}{\emph{JHEP} {\bfseries
  11} (2013) 140} [\href{https://arxiv.org/abs/1212.4103}{{\ttfamily
  1212.4103}}].

\bibitem{Fitzpatrick:2012yx}
A.~L. Fitzpatrick, J.~Kaplan, D.~Poland and D.~Simmons-Duffin, \emph{{The
  Analytic Bootstrap and AdS Superhorizon Locality}},
  \href{https://doi.org/10.1007/JHEP12(2013)004}{\emph{JHEP} {\bfseries 12}
  (2013) 004} [\href{https://arxiv.org/abs/1212.3616}{{\ttfamily 1212.3616}}].

\bibitem{Cuomo:2017vzg}
G.~Cuomo, A.~de~la Fuente, A.~Monin, D.~Pirtskhalava and R.~Rattazzi,
  \emph{{Rotating superfluids and spinning charged operators in conformal field
  theory}}, \href{https://doi.org/10.1103/PhysRevD.97.045012}{\emph{Phys. Rev.
  D} {\bfseries 97} (2018) 045012}
  [\href{https://arxiv.org/abs/1711.02108}{{\ttfamily 1711.02108}}].

\bibitem{Cuomo:2022kio}
G.~Cuomo and Z.~Komargodski, \emph{{Giant Vortices and the Regge Limit}},
  \href{https://doi.org/10.1007/JHEP01(2023)006}{\emph{JHEP} {\bfseries 01}
  (2023) 006} [\href{https://arxiv.org/abs/2210.15694}{{\ttfamily
  2210.15694}}].

\bibitem{Choi:2025tql}
J.~Choi and E.~Lee, \emph{{Large charge operators at large spin from
  relativistically rotating vortices}},
  \href{https://arxiv.org/abs/2501.07198}{{\ttfamily 2501.07198}}.

\bibitem{Fitzpatrick:2014vua}
A.~L. Fitzpatrick, J.~Kaplan and M.~T. Walters, \emph{{Universality of
  Long-Distance AdS Physics from the CFT Bootstrap}},
  \href{https://doi.org/10.1007/JHEP08(2014)145}{\emph{JHEP} {\bfseries 08}
  (2014) 145} [\href{https://arxiv.org/abs/1403.6829}{{\ttfamily 1403.6829}}].

\bibitem{Fardelli:2024heb}
G.~Fardelli, A.~L. Fitzpatrick and W.~Li, \emph{{Holography and Regge phases
  with U(1) charge}},
  \href{https://doi.org/10.1007/JHEP08(2024)202}{\emph{JHEP} {\bfseries 08}
  (2024) 202} [\href{https://arxiv.org/abs/2403.07079}{{\ttfamily
  2403.07079}}].

\bibitem{Fitzpatrick:2010zm}
A.~L. Fitzpatrick, E.~Katz, D.~Poland and D.~Simmons-Duffin, \emph{{Effective
  Conformal Theory and the Flat-Space Limit of AdS}},
  \href{https://doi.org/10.1007/JHEP07(2011)023}{\emph{JHEP} {\bfseries 07}
  (2011) 023} [\href{https://arxiv.org/abs/1007.2412}{{\ttfamily 1007.2412}}].

\bibitem{SalehiVaziri:2024joi}
K.~Salehi~Vaziri, \emph{{A non-perturbative construction of the de Sitter
  late-time boundary}},  \href{https://arxiv.org/abs/2412.00183}{{\ttfamily
  2412.00183}}.

\bibitem{Cheung:2016iub}
C.~Cheung, A.~de~la Fuente and R.~Sundrum, \emph{{4D scattering amplitudes and
  asymptotic symmetries from 2D CFT}},
  \href{https://doi.org/10.1007/JHEP01(2017)112}{\emph{JHEP} {\bfseries 01}
  (2017) 112} [\href{https://arxiv.org/abs/1609.00732}{{\ttfamily
  1609.00732}}].

\bibitem{Fox:1978vw}
G.~C. Fox and S.~Wolfram, \emph{{Event Shapes in e+ e- Annihilation}},
  \href{https://doi.org/10.1016/0550-3213(79)90120-2}{\emph{Nucl. Phys. B}
  {\bfseries 149} (1979) 413}.

\end{thebibliography}\endgroup
\bibliographystyle{JHEP.bst}

\end{document}